\documentclass[aps,prd,preprint,showpacs,eqsecnum,nofootinbib]{revtex4}
\usepackage{graphicx,epsf,bm,subfigure,epstopdf,amsmath,xfrac,amssymb}
\usepackage{accents,placeins}
\newif\ifdraft
\drafttrue
\newif\ifpreprint
\preprinttrue

\def\fig#1{fig.~{\ref{#1}}}
\def\Fig#1{Fig.~{\ref{#1}}}

\def\sect#1{section~{\ref{#1}}}
\def\sects#1#2{sections~{\ref{#1}} and~{\ref{#2}}}

\def\Tab#1{Table~{\ref{#1}}}

\def\spa#1.#2{\left\langle#1\,#2\right\rangle}
\def\spb#1.#2{\left[#1\,#2\right]}
\def\spash#1.#2{\spa{\smash{#1}}.{\smash{#2}}}
\def\spbsh#1.#2{\spb{\smash{#1}}.{\smash{#2}}}
\def\lor#1.#2{\left(#1\,#2\right)}
\def\sand#1.#2.#3{%
\left\langle\smash{#1^{-}}{\vphantom1}\right|{#2}%
\left|\smash{#3^{-}}{\vphantom1}\right\rangle}
\def\sandpp#1.#2.#3{%
\left\langle\smash{#1^{+}}{\vphantom1}\right|{#2}%
\left|\smash{#3^{+}}{\vphantom1}\right\rangle}
\def\sandpm#1.#2.#3{%
\left\langle\smash{#1^{+}}\vphantom1\right|{#2}%
\left|\smash{#3^{-}}\vphantom1\right\rangle}
\def\sandmp#1.#2.#3{%
\left\langle\smash{#1^{-}}\vphantom1\right|{#2}%
\left|\smash{#3^{+}}{\vphantom1}\right\rangle}
\def\sandmppm#1.#2.#3{%
\left\langle\smash{#1^{\mp}}\vphantom1\right|{#2}%
\l
eft|\smash{#3^{\pm}}{\vphantom1}\right\rangle}
\def\sandnn#1.#2.#3{%
\left\langle\smash{#1}\vphantom1\right|{#2}%
\left|\smash{#3}{\vphantom1}\right\rangle}
\def\sandmn#1.#2.#3{%
\left\langle\smash{#1^{-}}\vphantom1\right|{#2}%
\left|\smash{#3}{\vphantom1}\right\rangle}
\def\sandnm#1.#2.#3{%
\left\langle\smash{#1}\vphantom1\right|{#2}%
\left|\smash{#3^{-}}{\vphantom1}\right\rangle}

\def\e{\epsilon}

\def\eqn#1{eq.~(\ref{#1})}
\def\Eqn#1{Equation~(\ref{#1})}
\def\eqns#1#2{eqs.~(\ref{#1}) and~(\ref{#2})}

\def\be{\begin{equation}}
\def\ee{\end{equation}}

\def\onehalf{\frac12}

\def\Ord{{\cal O}}
\def\MB{{\textsf{MB\/}}}
\def\MBresolve{{\textsf{MBresolve\/}}}
\newbox\charbox
\newbox\slabox
\def\s#1{{      
        \setbox\charbox=\hbox{$#1$}
        \setbox\slabox=\hbox{$/$}
        \dimen\charbox=\ht\slabox
        \advance\dimen\charbox by -\dp\slabox
        \advance\dimen\charbox by -\ht\charbox
        \advance\dimen\charbox by \dp\charbox
        \divide\dimen\charbox by 2
        \raise-\dimen\charbox\hbox to \wd\charbox{\hss/\hss}
        \llap{$#1$} }}

\def\simasO^#1{\stackrel{#1}{\sim}}
\def\simas^#1{\hskip -3pt%
\stackrel{#1}{\lower 5pt\hbox{$\widetilde{\hphantom{#1}}$}}%
\hskip -3pt}

\def\negonehalf{-\frac12}
\def\stdparam{-\frac1{20}}

\newcommand{\asin}{\mathop{\rm asin}\nolimits}
\newcommand{\atan}{\mathop{\rm atan}\nolimits}
\newcommand{\asinh}{\mathop{\rm asinh}\nolimits}
\renewcommand{\Re}{\mathop{\rm Re}\nolimits}
\renewcommand{\Im}{\mathop{\rm Im}\nolimits}
\newcommand{\sign}{\mathop{\rm sign}\nolimits}

\newcommand\gatop[2]{\genfrac{}{}{0pt}{}{#1}{#2}}

\begin{document}

\title{Efficient Evaluation of Massive Mellin--Barnes Integrals}

\author{Janusz Gluza}\affiliation{
Department of Field Theory and Particle Physics, Institute of Physics,
University of Silesia, Uniwersytecka 4, PL--40-007 Katowice, Poland
}
\author{Tomasz Jeli\'nski}\affiliation{
	Department of Field Theory and Particle Physics, Institute of Physics,
	University of Silesia, Uniwersytecka 4, PL--40-007 Katowice, Poland
}
\author{David A. Kosower}\affiliation{
School of Natural Sciences, Institute for Advanced Study,
1 Einstein Drive, Princeton, New Jersey 08540\\
{\rm and}\\
Institut de Physique Th\'eorique, CEA, CNRS, Universit\'e Paris--Saclay,
          F--91191 Gif-sur-Yvette cedex, France}

\begin{abstract}
We show how to evaluate one-dimensional
Minkowski-region Mellin--Barnes representations arising from 
massive loop integrals, by modifying the contours
of integration.  
We implement an exact solution to the differential equation
determining the contours of stationary phase.  We also present several
simple approximations to these contours.
Our approach
points the way to more efficient computations of 
massless and massive Mellin--Barnes
integrals in both Euclidean and Minkowski regions.
\end{abstract}


\maketitle

\section{Introduction}
\label{IntroSection}

The Mellin--Barnes approach has proven
to be a versatile and successful approach
to evaluating higher-loop integrals, 
both analytically and numerically~\cite{SmirnovBook,FreitasReview}.
Its early successes included the analytic computation of
the planar~\cite{SmirnovDoubleBox} and nonplanar~\cite{Tausk}
 two-loop double-box integrals.
In this approach, one first introduces a Feynman parametrization
into loop integrals, performs the loop integrals, and then uses
Mellin--Barnes representations for the integrands to allow the
Feynman parameter integrals to be computed.  The integrals are typically
infrared-divergent, and may have ultraviolet divergences as well.  
These divergences are usually regulated dimensionally; the resulting 
singularities are hidden inside
the integrands of the Mellin--Barnes integrals.  One can
move the contours to make these singularities manifest, yielding
a representation in which the poles in the regulator $\e$
are manifest, and in which the coefficients are finite Mellin--Barnes
integrals which can be computed analytically or numerically.
Czakon's \MB{} package~\cite{CzakonMB} automated the process of moving contours
to resolve singularities; a related algorithm was later implemented
in \MBresolve{} by Smirnov and Smirnov~\cite{MBresolve}.  Other publicly
available packages connected with the Mellin--Barnes evaluation of
Feynman integrals are available on the \textsf{MBtools\/} webpage~\cite{MBtools}:
\textsf{AMBRE}~\cite{AMBRE,Dubovyk:2016ocz}, which assists in creating
Mellin--Barnes representations; \textsf{MBasymptotics}~\cite{MBasymptotics},
which performs parametric expansions of Mellin--Barnes integrals;
and \textsf{barnesroutines}~\cite{BarnesRoutines}, which automates the
application of Barnes lemmas.

The Mellin--Barnes approach has been used extensively for numerical
cross-checks of analytic results in the Euclidean region, including
in two-loop massive Bhabha scattering in QED~\cite{Bhabha2L}; in three-loop
massless form factors~\cite{FF3} and static potentials~\cite{Smirnov:2009fh};
in massive two-loop QCD form-factors~\cite{Gluza:2009yy}; in $B$-physics 
studies~\cite{Bphysics};
in hadronic top-quark physics~\cite{Baernreuther:2013caa}; 
and for angular integrations in phase-space
integrals~\cite{PhaseSpaceIntegrals}.
It has also been used to obtain direct numerical results in computations
in supersymmetric Yang-Mills theories: for the four-loop cusp anomalous 
dimension~\cite{Bern:2006ew} and two-loop five-point amplitudes~\cite{Bern:2006vw}; 
as well as in ${\cal N}=6$ Chern--Simons theory at six loops and 
beyond~\cite{Bak:2009tq}.  Very recently, it has been applied to integrals
in chiral perturbation theory~\cite{Ananthanarayan:2016pos}.

The \MB{} and \MBresolve{} packages yield numerically
convergent integrals in the Euclidean region
for integrals arising from a mixture of massless and massive propagators.
The same integrals are typically only conditionally convergent in
the Minkowski region, and hence fail to converge numerically.
For Feynman integrals that arise at one loop, one can ultimately
perform the integrations analytically,
and assemble these results into numerical software libraries~\cite{vanOldenborgh:1990yc,%
vanHameren:2010cp,Ellis:2007qk,Denner:2016kdg,Fleischer:2012et}.
The convergence failure is then an annoyance, and prevents use of
numerical approaches as cross checks, but it is not a critical problem.

At two loops and beyond, not all desired integrals are available analytically,
and the obstruction is of greater importance.
A variety of other techniques can be applied to the numerical calculation
of Feynman integrals in the Minkowski region.  These include sector
decomposition~\cite{SectorDecomposition} 
(as implemented, for example, in 
\textsf{SecDec~3}~\cite{Borowka:2012qfa,Borowka:2012yc,Borowka:2015mxa}
 and \textsf{Fiesta~4\/}~\cite{Smirnov:2015mct});
numerical subtraction of singularities in loop-momentum
space~\cite{tka_as,loosub,kreimeruv,xloops,kreimerir,Freitas:2012iu},
along with appropriate complex contour deformations of the Feynman-parameter
integrations.  (See refs.~\cite{ns1,Dittmaier:2003bc,becker1,becker2,loosub} for
earlier one-loop results.)

The Mellin--Barnes
integrals produced by the \MB{} and \MBresolve{} packages
use standard contours, parallel to the imaginary axis.
The representation
was re-examined by Freitas and Huang~\cite{FreitasHuang}, who pointed
out that using tilted contours of integration
different from the textbook contours chosen by \MB{} can make
Minkowski-region massive integrals convergent.  These authors
did not specify
exactly how these tilted linear contours should be chosen.
A recent Mellin--Barnes based numerical 
package, \textsf{MBnumerics\/}~\cite{mbnum},
takes a different approach, shifting and rotating contours, re-mapping
integrands, and dropping small contributions to compute multi-dimensional
integrals with multiple scales in the Minkowski region~\cite{Dubovyk:2016ocz}.
 It has been applied to
the two-loop bosonic contributions to 
$Z\rightarrow b\bar b$~\cite{Dubovyk:2016aqv}.

In this article, we re-examine the choice of contours, and show how
to detemine contours that are in a certain sense close approximations
to optimal contours of integration.  We study one-dimensional 
Mellin--Barnes integrals both in the Euclidean and Minkowski regions.
Our contours improve the numerical efficiency of computation in the
former region, and provide an efficient and
 convergent representation in the latter.  We also present a
 connection to the tilted contours suggested by Freitas and Huang~\cite{FreitasHuang}.
 While
these contours are not computationally optimal, they do have the virtue
of simplicity.  We believe that the approach described here will generalize
to higher-dimensional Mellin--Barnes integrals; but the generalization
is not trivial nor completely straightforward, and accordingly we postpone
any discussion of it to future work.

This article is organized as follows.  In the next section, we study
an example of a one-dimensional massive integral in the Euclidean
region.  We also
discuss the
differential equation drawn from the mathematics literature which 
determines an exact contour of stationary phase.  In 
\sect{ContourApproximationSection}, we present several approximations
to the exact contour, for parameter values in the Euclidean region.
In \sect{AsymptoticFormSection}, we show how to match contours to
their asymptotic forms, and in \sect{OtherIntegralsSection}, we give
examples of various use cases exemplifying the utility of
using approximate contours and matching to their asymptotic forms.
We use as examples integrals that may arise in Feynman diagrams.
In \sect{MinkowskiSection}, we study 
approximations to contours of stationary phase
for parameters in the Minkowski
region.  In \sect{NoExtremaSection}, we examine one important
special case, of integrals with no stationary point along the real
axis even for parameters in the Euclidean region.  In \sect{EvaluationSection},
we examine briefly the evaluation of integrals using the various
contour approximations discussed in earlier sections.  We
give some concluding remarks in \sect{Conclusions}.

\section{A Euclidean Integral}
\label{DifferentialEquationSection}

Let us begin by studying Mellin--Barnes integrals in the Euclidean region.
In this region, the standard contours used in the definition of
the integrals --- as well as by the \MB{} package --- are usually suitable
for numerical integration as well.  But we can improve upon them, and
the improvements offer a stepping stone to the modifications required
to obtain a numerically convergent form
in the Minkowski region.

We begin with a one-dimensional integral,
\begin{equation}
I_1(s) = \frac{1}{2\pi i} \int_{c_0-i\infty}^{c_0+i\infty} dz\; F_1(z,s)\,,
\label{OneDimensionalExample}
\end{equation}
where
\begin{equation}
F_1(z,s) = (-s)^{-z} \frac{\Gamma^3(-z)\Gamma(1+z)}{\Gamma(-2z)}\,.
\label{OneDimensionalExampleIntegrand}
\end{equation}
This integral was considered in refs.~\cite{CzakonMB,Dubovyk:2016ocz}.  
As mentioned there, it can be
evaluated analytically, with result,
\begin{equation}
\frac{4}{\sqrt{\frac4{-s}+1}} \asinh\sqrt{-\frac{s}4}\,,
\label{Int1result}
\end{equation}
in the Euclidean region $s<0$;
\begin{equation}
- \frac{4}{\sqrt{\frac4s-1}} \asin\sqrt{\frac{s}4}\,,
\label{Int1belowthreshold}
\end{equation}
below threshold ($s\in [0,4]$) in the Minkowski region;
and,
\begin{equation}
\frac{4}{\sqrt{1-\frac4s}}
\biggl[-\frac{i\pi}2\sign\Im s 
       + \ln\biggl(\sqrt{\frac{s}4}+\sqrt{\frac{s}4-1}\biggr)
\biggr]\,,
\label{Int1abovethreshold}
\end{equation}
above threshold.  (The forms in \eqns{Int1belowthreshold}{Int1abovethreshold}
correspond to \eqn{Int1result} when $s$ is given a small imaginary part.)

\begin{figure}[t]
\includegraphics[clip,scale=0.66]{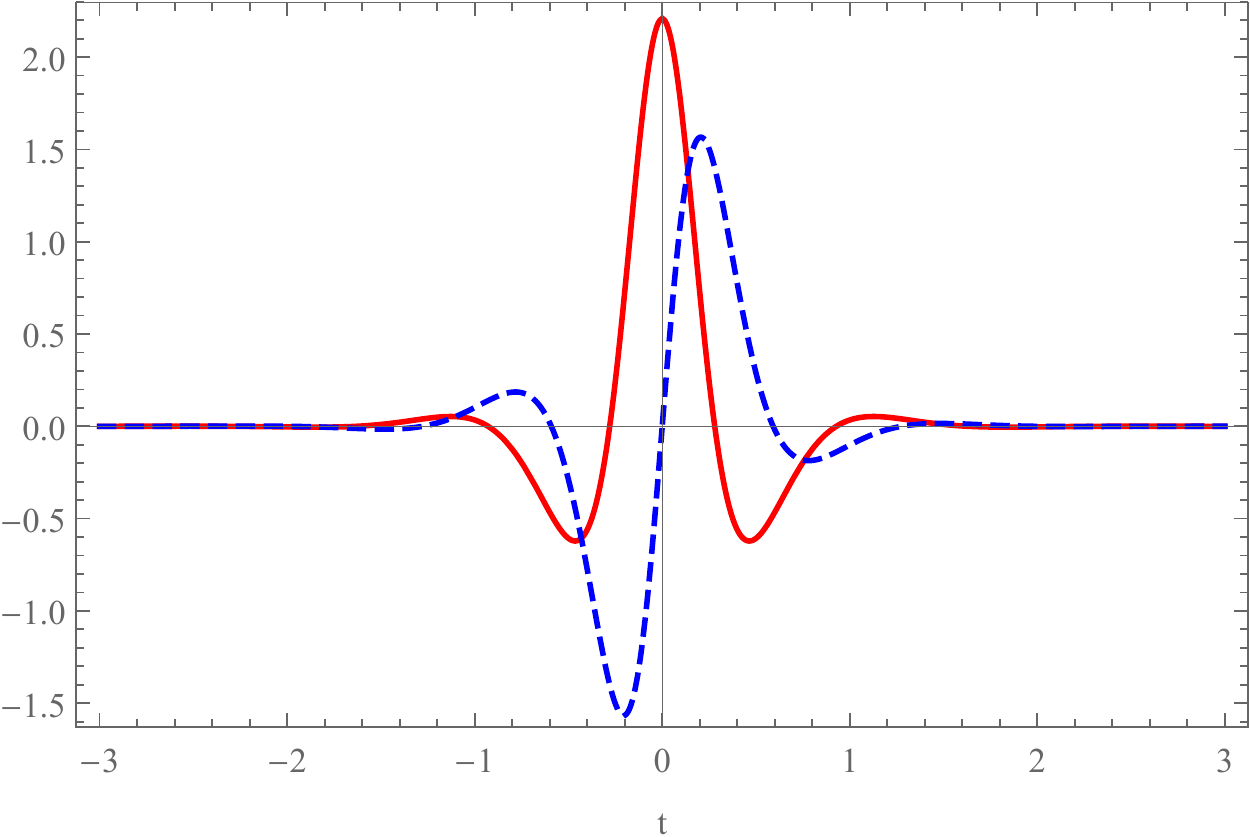}
\caption{The real (red) and imaginary (dashed blue) parts of
the integrand $F_1(z,s)$ of \eqn{OneDimensionalExampleIntegrand} for 
$s=\stdparam$ along the
`textbook' (and \MB{}) contour, $\Re z = c_0 = \negonehalf$.}
\label{OneDimContourIntegrandA}
\end{figure}

In the Euclidean region, $s<0$, and hence the integrand is real
for real $z$.  The reflection symmetry $z\leftrightarrow \bar z$ then
ensures the integral is real as well.  
  The \MB{} package, left to its own devices, will choose
$c_0= \negonehalf$.
Let us consider the integral
for $s=\stdparam$.
The real and
imaginary parts of the integrand along the contour
are shown in \fig{OneDimContourIntegrandA}.  Both oscillate around zero,
though the oscillations are damped as one moves away from the real
axis, and the resulting numerical integral converges nicely.  
(Obtaining the vanishing result for the imaginary part
does require
non-trivial cancellations in a numerical integration, of course.)

Nonetheless, let us ask: why choose this particular contour?  Or,
more pointedly: can we do any better?  Is there a more efficient contour?

\begin{figure}[ht]
\includegraphics[clip,scale=0.66]{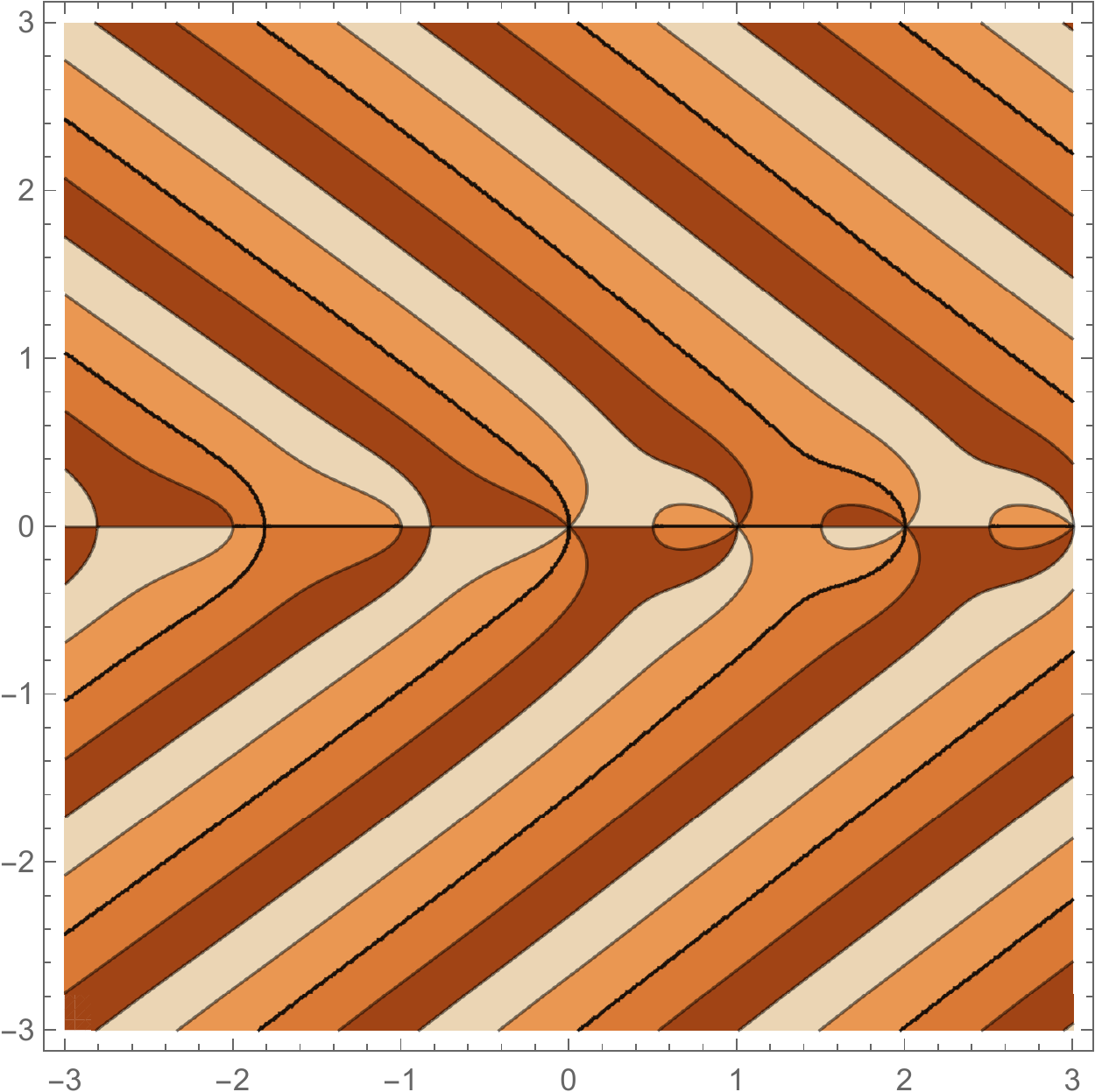}
\caption{A contour plot of phases
	for the integrand $F_1(z,s)$ of \eqn{OneDimensionalExampleIntegrand}
	with $s=\stdparam$, with contours for phases $-\pi/2$,
	$0$, $\pi/2$, and $\pi$ also shown.
	The contour with phase 0 is the boundary
	between the darkest and lightest shade.}
\label{OneDimensionalAllContours}
\end{figure}

Complex-analysis textbooks would tell us that the answer is yes: we
should choose the contour of steepest descent.  This will yield the
most-rapidly convergent integral, because it is also a contour
of stationary phase and hence minimizes oscillations in the integrand.
\Fig{OneDimensionalAllContours} shows a variety of contours
of stationary phase for the integrand $F_1(z,s=\stdparam)$; in the
case at hand, we should pick the contour of phase 0.
  As in the case of parton-distribution
evolution~\cite{PDFEvolution}, we must face the questions of whether
we can find a good approximation to this contour without undue 
computational effort; and whether we can easily adapt the contour
to different values of the parameter $s$.  As we shall see, we can
provide affirmative answers to both questions.

Our first task in finding the contour of stationary phase in this
case is to find a local minimum along the real axis.  The minimum closest
to $z=\negonehalf$ is given by the solution to,
\begin{equation}
-\ln (-s) + 2\psi(-2 z)-3\psi(-z)+\psi(z+1) = 0
\label{StationaryPointEquationExample}
\end{equation}
which is at $z_s \simeq -0.825618$ for $s=\stdparam$.  Because we
have the analytic form of the integrand, we can easily obtain this
equation analytically, and then use an efficient algorithm (e.g.
Newton--Raphson) to find a numerical solution.

Before looking at a sequence of approximations to the desired
contour, let us examine the exact contour $z(t)$, anchored
at the above minimum.  In the present
case, $\Im F_1(z,s)$ will vanish exactly along the contour.  We
seek contours described by a meromorphic function; and $F_1$ itself
is meromorphic as well.  The contour then satisfies a differential
equation~\cite{ContourDifferentialEquation},
\begin{equation}
\frac{dz}{dt} = -\overline{\frac{\partial \ln F_1(z,s)}{\partial z}}\,.
\label{DifferentialEquation}
\end{equation}
The phase of $F_1$ is given by $\Im\ln F_1$; along
a contour satisfying \eqn{DifferentialEquation}, the phase does
not vary,
\def\zbar{\overline{z}}
\def\Fbar{\overline{F}}
\begin{equation}
\begin{aligned}
\frac{d\textrm{phase}}{dt}
&=\frac{d\Im\ln F_1}{dt}
=\frac1{2i} \biggl[
\frac{d\ln F_1}{dt}
-\frac{d\ln \Fbar_1}{dt}
\biggr]
\\& = \frac1{2i} \biggl[
\frac{\partial\ln F_1}{\partial z}\frac{dz}{dt}
-\overline{\frac{\partial\ln F_1}{\partial z}}\frac{d\zbar}{dt}
\biggr]
\\& =-\frac1{2i} \biggl[
\frac{\partial\ln F_1}{\partial z}
   \overline{\frac{\partial \ln F_1}{\partial z}}
-\overline{\frac{\partial\ln F_1}{\partial z}}
   \frac{\partial \ln F_1}{\partial z}
\biggr]
\\& = 0\,.
\end{aligned}
\end{equation}
Furthermore,
\begin{equation}
\begin{aligned}
\frac{d|F_1|^2}{dt} &= 
 |F_1|^2 \biggl[
   \frac{d\ln F_1}{dt}
   +\frac{d\ln \Fbar_1}{dt}
\biggr]
\\& = |F_1|^2 \biggl[
\frac{\partial\ln F_1}{\partial z}\frac{dz}{dt}
+\overline{\frac{d\ln F_1}{\partial z}}\frac{d\zbar}{dt}
\biggr]
\\& = -|F_1|^2 \biggl[
\frac{\partial\ln F_1}{\partial z}
   \overline{\frac{\partial \ln F_1}{\partial z}}
+\overline{\frac{d\ln F_1}{\partial z}}
   \frac{\partial \ln F_1}{\partial z}
\biggr]
\\& = -2|F_1|^2 \biggl|
\frac{\partial\ln F_1}{\partial z}\biggr|^2 < 0\,,
\end{aligned}
\end{equation}
so that as expected it is a contour of steepest descent.

The stationary point $z_s$ is also a stationary point of this
equation; a solution which starts at $z_s$ will stay there for
all $t$.  As boundary data for the differential equation, we
must therefore choose a different point.  A suitable choice is
given by perturbing away from $z_s$ along one of the two directions
of steepest descent.  In general, one can find these by finding
the eigenvectors of the Hessian of the integrand; in this case,
the required directions are parallel to the imaginary axis,
in either the positive and negative direction.  One can then
solve the equation numerically; one must do so separately in
the upper- and lower-half planes.  (Alternatively, in the Euclidean
region the lower half-plane contour will be the complex conjugate
of the upper half-plane contour.)  The pair of contours together
is called the Lefschetz thimble ${\cal J}(z_s)$ associated to the
stationary point $z_s$.

In the case at hand, one starts with $z_s$ as given by the solution
to \eqn{StationaryPointEquationExample}, and looks for the tangent
to it.  For Euclidean values of $s$, 
the line $\Re z = z_s$ will be that tangent, because a minimum along 
the real axis is
a saddle point of the integrand in the complex plane.  We can
perturb away from the stationary point along the tangent,
\begin{equation}
z_s(\delta) = z_s + i\delta\,,
\end{equation}
in order to obtain a suitable starting point for the differential equation.
(The smaller $\delta$, the more accurate the solution will be.)  
We make use of the {\sl Mathematica\/} routine {\sf NDSolve\/} to
solve the differential equation~\eqn{DifferentialEquation} with
starting points $z_s(\pm\delta)$.  This yields a numerical representation
of the exact contour of stationary phase.  We show
examples of exact contours in following sections.

In general, however, solving the differential 
equation~(\ref{DifferentialEquation}), and then using the
solution repeatedly in a numerical integration, may be computationally
expensive. Furthermore, we may encounter integrals for which the exact
contour of stationary phase is not optimal for numerical
integration, and where the MB integral would require special
treatment with an exact contour.  
This motivates us to seek approximations to the exact
contour of stationary phase, which we consider in the next section.

\section{Contour Approximations}
\label{ContourApproximationSection}

\begin{figure}[t]
\includegraphics[clip,scale=0.66]{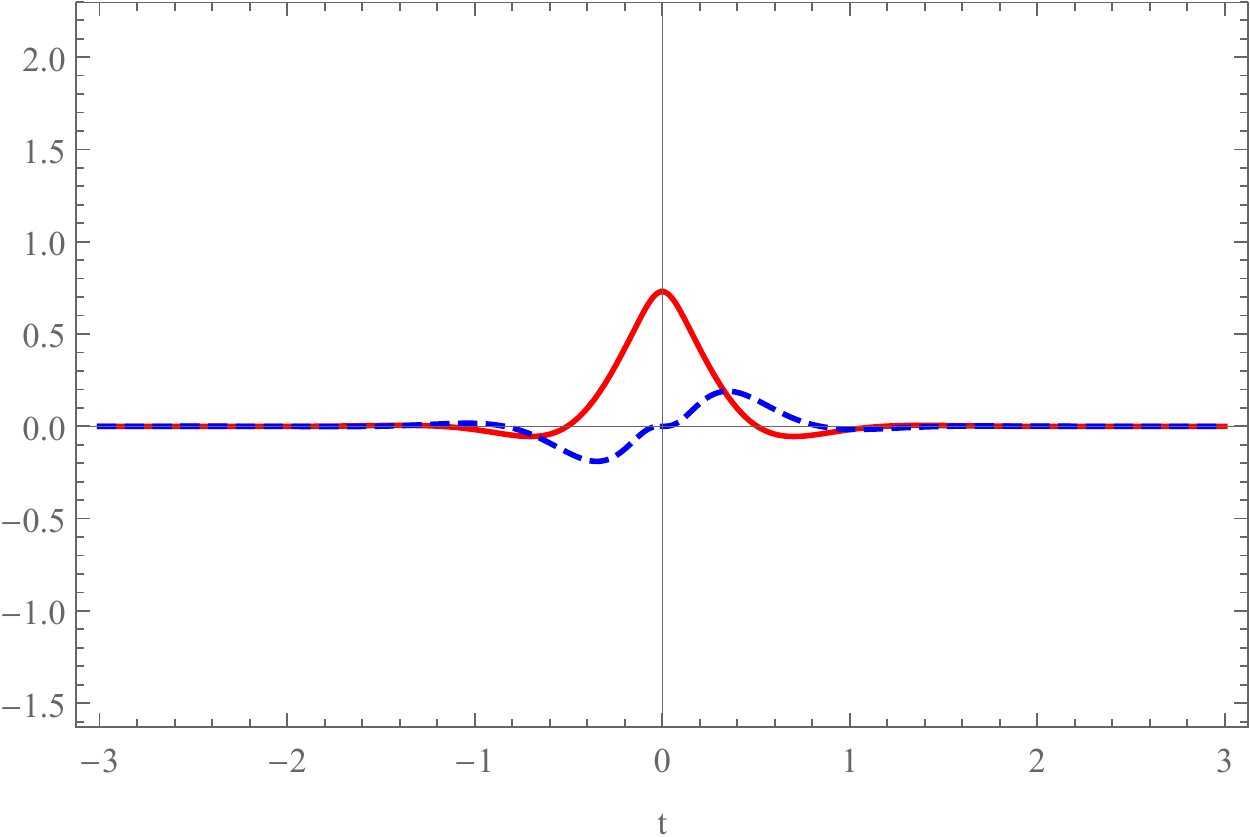}
\caption{The real (red) and imaginary (dashed blue) parts of
the integrand $F_1(z,s)$ of \eqn{OneDimensionalExampleIntegrand} for 
$s=\stdparam$ along the contour $\Re z = -0.825618$,
tangent to the contour of steepest descent.}
\label{OneDimContourIntegrandB}
\end{figure}

The simplest approximation to the exact contour of stationary
phase is given by a straight line tangent to it.
As noted in the previous section,
the line $\Re z = z_s$ will be that tangent.
If we take our contour to be this line, we see
that while the integrand still oscillates (see \fig{OneDimContourIntegrandB}),
the oscillations are damped more quickly than along the original contour.

We can improve on the tangent approximation; to
find a better approximation, we follow the same
procedure as in ref.~\cite{PDFEvolution}.  Parametrize $z(t) = x(t) + i y(t)$, choosing
\begin{equation}
x(t_0 = 0) = z_s,\qquad 
y(0) = 0,\qquad x, y {\rm\ real}.
\end{equation}
If we require the contour to be symmetric under reflection in the real
axis (as is desirable for numerical evaluation), $x$ will be an even
function, and $y$ an odd one.  We can rescale $t$ to make $y'(0)=1$.
Taking the contour to be smooth, we will also have $x'(0)=0$.
The expansion of the integrand around $z=z_s$ then takes the form,
\begin{equation}
\begin{aligned}
F_1(z(t)) &\sim F_1(z_s) + \frac{F_1''(z_s)}2 (x'(0)^2- y'(0)^2 + 2 i x'(0) y'(0)) t^2 + \cdots\\
&= F_1(z_s) - \frac{F_1''(z_s)}2 t^2 + \cdots
\end{aligned}
\end{equation}
where we drop the $s$ argument for brevity.
As all derivatives of $F_1(x)$ are real, the equation $\Im F_1(z(t))=0$ is
satisfied to this order; and as $F_1''(z_s)$ is positive, the integrand
decreases with $t$.  However, the contour will not continue parallel to
the imaginary axis; to see where it goes, we must expand to higher order.
Consider the expansion to $\Ord(t^3)$,
\begin{equation}
F_1(z(t)) \sim F_1(z_s) - \frac{F_1''(z_s)}2 t^2
+\frac16 \biggl( -i F_1^{(3)}(z_s) + 3 i F_1''(z_s) x''(0)
         \biggr) t^3 + \cdots
\end{equation}
To this order, the stationary-phase condition ($\Im F_1(z(t)) = 0$) requires,
\begin{equation}
x''(0) = \frac{F_1^{(3)}(z_s)}{3 F_1''(z_s)}\,.
\end{equation}
In the neighborhood of $z_s$, the approximate contour then has the form,
\begin{equation}
z_q(t) = z_s + i t + c_2 t^2\,,
\label{QuadraticContour}
\end{equation}
where $c_2$ is real, and given by,
\begin{equation}
c_2 = \frac{F_1^{(3)}(z_s)}{6 F_1''(z_s)}\,.
\label{QuadraticCoefficient}
\end{equation}
Because $x^{(3)}(0)$ vanishes, the terms of $\Ord(t^4)$ are automatically
real, and only at $\Ord(t^5)$ do imaginary terms now appear in the
expansion of $F(z(t))$.

\begin{figure}[t]
\begin{minipage}[b]{1.03\linewidth}
\begin{tabular}{cc}
\hskip -6mm
\includegraphics[clip,scale=0.60]{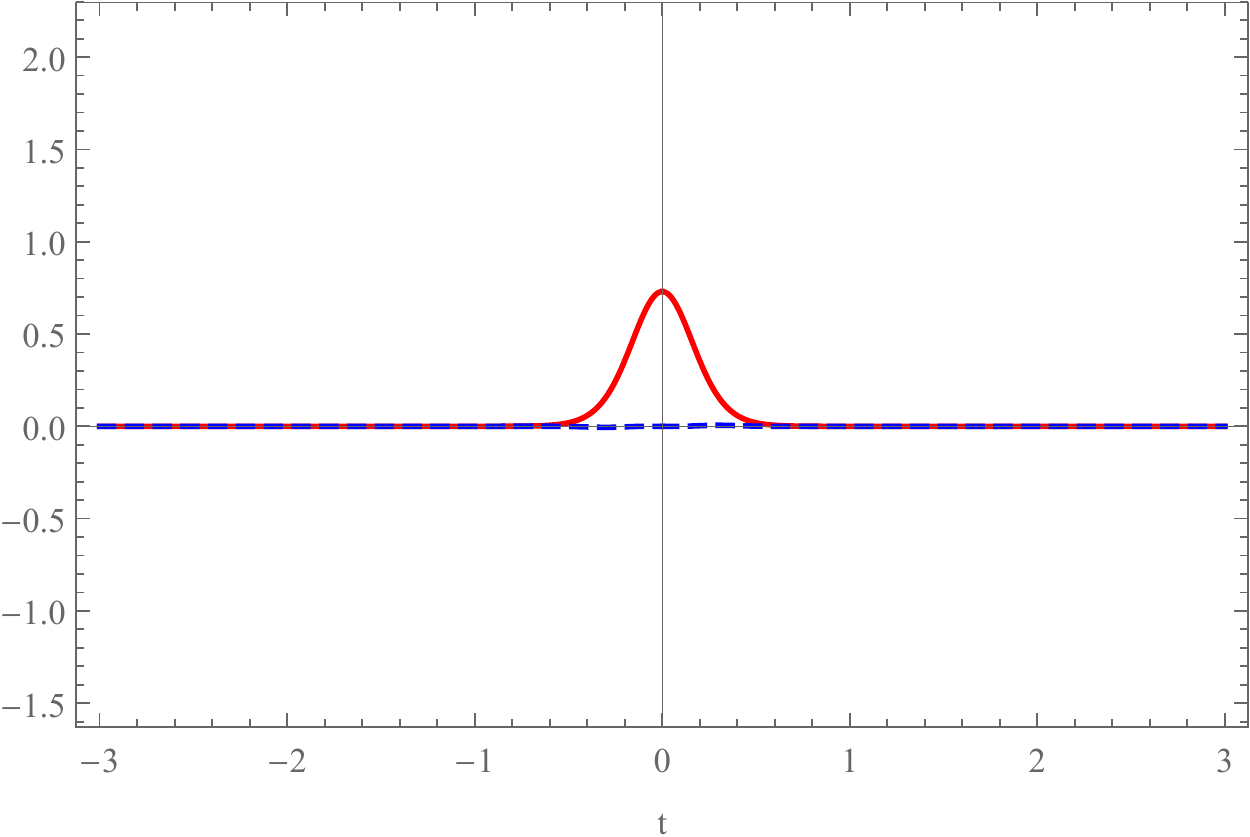}
\hskip5mm
&\includegraphics[clip,scale=0.645]{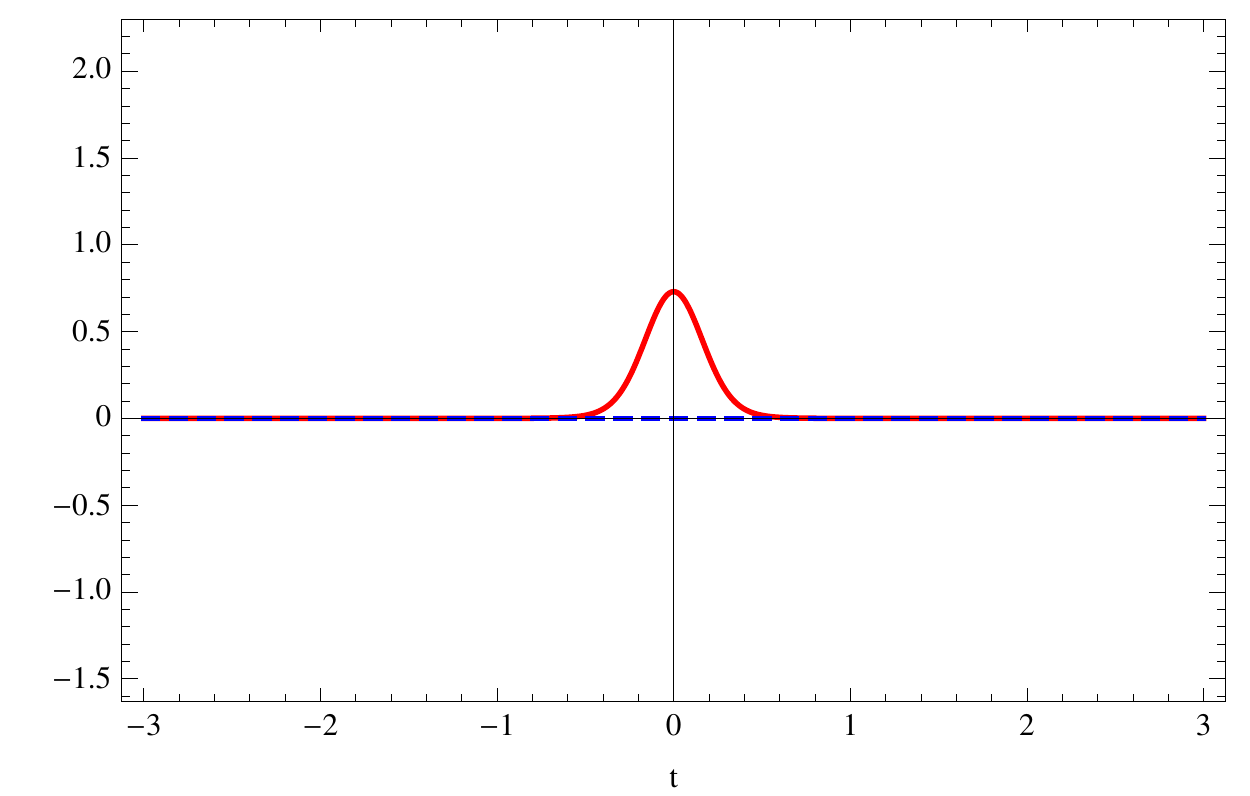}
\\[3mm]
(a)&(b)\\[3mm]
\end{tabular}
\end{minipage}
\caption{The real (red) and imaginary (dashed blue) parts of
the integrand $F_1(z,s)$ of \eqn{OneDimensionalExampleIntegrand} for 
$s=\stdparam$ along (a) the quadratic approximation $z_q(t)$
to the contour of stationary phase (b) the exact contour.}
\label{OneDimContourIntegrandC}
\end{figure}

In the example at hand ($s=\stdparam$), the quadratic contour is,
\begin{equation}
z_q(t) = -0.825618 + i t - 1.65358 t^2\,.
\label{QuadraticContourExample}
\end{equation}
The value of the integrand along the contour is shown 
in~\fig{OneDimContourIntegrandC}(a).  The imaginary part is essentially zero, 
and the 
real part is free of oscillations in the region which gives the bulk of the
contributions to the result.  (The parametrized
integrand will still have an imaginary part, because of the $z'(t)$ factor,
but there is no need to compute it.)
For comparison, in~\fig{OneDimContourIntegrandC}(b)
we show the integrand along the exact contour of stationary
phase, computed using the differential equation as described in
\sect{DifferentialEquationSection}.  The imaginary part
of the contour is chosen to be $t$ in both figures.

\begin{figure}[ht]
\includegraphics[clip,scale=0.66]{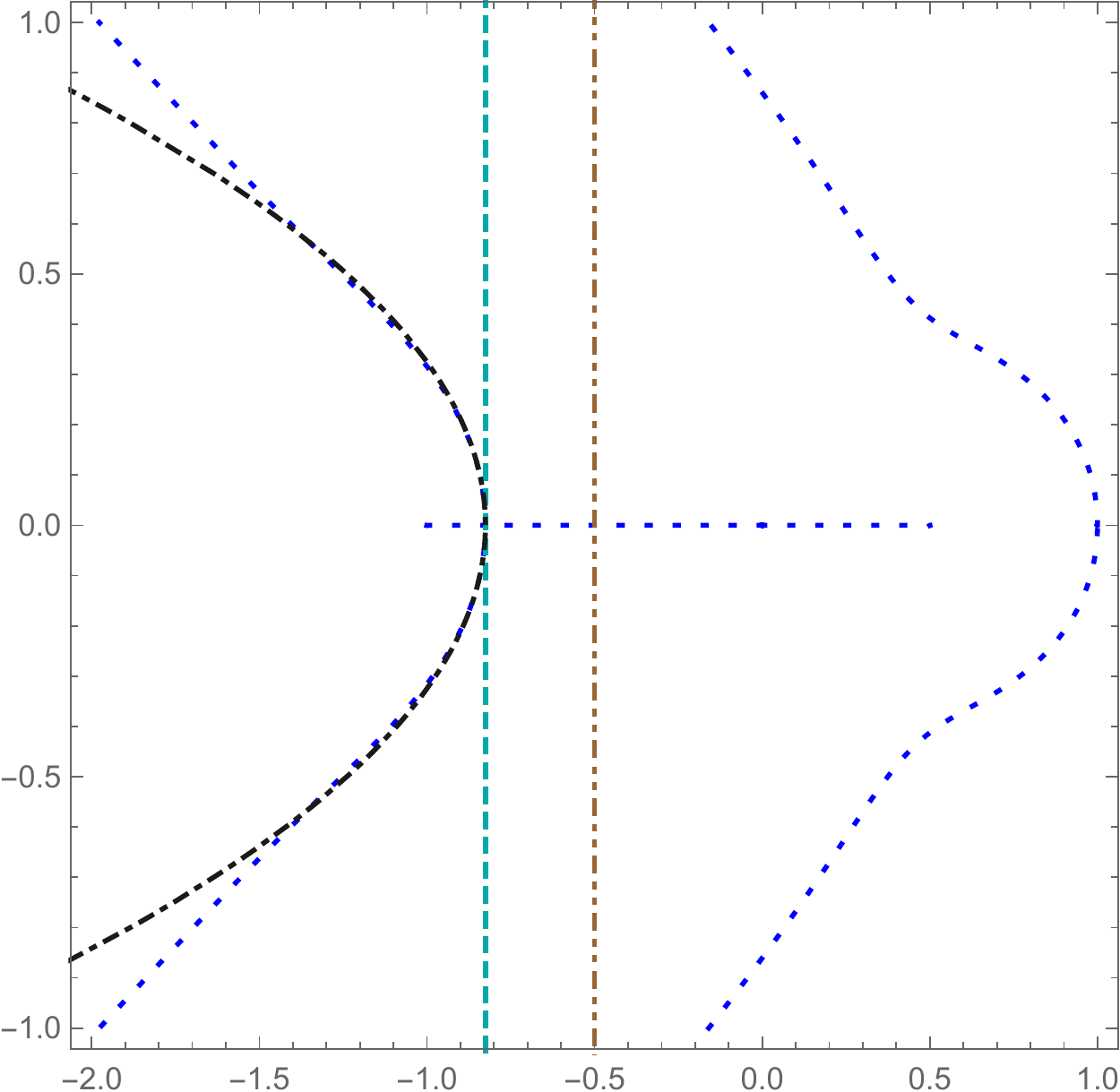}
\caption{The original linear (double-dot dashed brown),
tangent (dot-dashed dark turquoise), and quadratic (dot-dashed dark gray)
contours for the integrand $F_1(z,s)$ of
\eqn{OneDimensionalExampleIntegrand} with 
$s=\stdparam$, shown along with the exact contours of zero phase 
(dotted blue).}
\label{OneDimensionalSimpleContours}
\end{figure}

The two linear contours (original and tangent) and the quadratic
contour are shown along
with the exact contours of stationary phase
in \fig{OneDimensionalSimpleContours}.

\def\th{\theta}
\def\thinf{\theta_\infty}
\def\zinf{z_{\infty}}
\def\zinfh{z_0^{(\sharp)}}
\def\zinfp{z_0^{(+\infty)}}
\def\zinfm{z_0^{(-\infty)}}
There is another improvement we can make to the contour. 
Notice that the contours of constant phase, shown in
\fig{OneDimensionalAllContours}, 
are all asymptotically straight lines 
as $z\rightarrow \infty$ (so long as we stay away from the real axis).  
We can see this analytically
by using the 
asymptotic expansion of the gamma function,
\begin{equation}
\Gamma(z) \simas^{z\rightarrow \infty} 
\sqrt{\frac{2\pi}{z}} z^z e^{-z}\,,
\label{AsymptoticGamma}
\end{equation}
in \eqn{OneDimensionalExampleIntegrand}
to obtain an asymptotic form for the integrand (with $s<0$),
\begin{equation}
F \sim {\rm const}
\,4^z (-s)^{-z} \frac{(1 + z)^{1/2 + z}}{(-z)^{1+ z}}\,;
\end{equation}
paying careful attention to the branch cuts,
we can further simplify this expression to obtain,
\begin{equation}
F\sim {\rm const}
  \,4^z (-s)^{-z} \frac{(-1+i\delta\sign\Im z)^{z+1/2}}{\sqrt{-z}}
\,,
\label{Integrand1AsymptoticExpansion}
\end{equation}
where $\delta$ is an infinitesimal positive number.

We can compute the phase of this expression via,
\begin{equation}
\arg(z) = -i \ln(z/|z|)\,,
\end{equation}
to obtain,
\begin{equation}
\arg F = \frac{\pi}2-\frac12\arg(-z) 
  +\pi \Re z\sign\Im z+\Im z \ln\biggl(-\frac4s\biggr)\,
\label{IntegrandArg}
\end{equation}
(implicitly taken ${\rm mod}\ 2\pi$),
which indeed is a linear equation as $z\rightarrow\infty$.

Let us write the asymptotic form of the
contour in the following form,
\begin{equation}
z_\infty(t) = \left\{\begin{array}{l}
\zinf + i r e^{i\thinf} t\,,\quad t>0\,,\\
\zinf + i r e^{-i\thinf} t\,,\quad t<0\,,\\
\end{array}\right.
\label{InfinityContour}
\end{equation}
where $r$ is real.  In this parametrization, 
$\thinf\in[-\frac{\pi}2,\frac{\pi}2]$, with $\thinf=0$ corresponding
to a line parallel to the imaginary axis.
To fix the parameters $\zinf$ and $\thinf$, substitute
this form into \eqn{IntegrandArg}, and expand as $t\rightarrow \infty$.
Requiring the coefficient of
$t$ to vanish yields an equation for $\thinf$,
\begin{equation}
\thinf = \atan\biggl[\frac1{\pi}\ln(-4/s)\biggr]\,.
\end{equation}
Taking the limit $t\rightarrow \infty$ in $\arg(-z)$,
but setting $t=0$ elsewhere then allows us to solve for $\zinf$,
\begin{equation}
\zinf = -\frac34+\frac{\thinf}{2\pi}\,.
\end{equation}
Of course, there are many contours of zero phase, as seen in
\fig{OneDimensionalAllContours}; all asymptotic lines will 
share the same $\thinf$, but each will have a different 
$\zinf$.  The different $\zinf$ values will be separated by
even integers; the one chosen here is the one lying in the original
interval of interest $(-1,0)$.

\def\qp{{q^+}}
\def\qs{{q^\sharp}}

We will give a more general discussion of the asymptotic forms in the next
section, where we consider a more general 
ratio of gamma functions, and additional factors of polygamma functions.
Such additional factors
 will not affect the leading terms in 
the phase, and hence will leave the angle $\thinf$ unchanged; they will
however affect the position of the intercept $\zinf$.
 
\begin{figure}[ht]
\includegraphics[clip,scale=0.66]{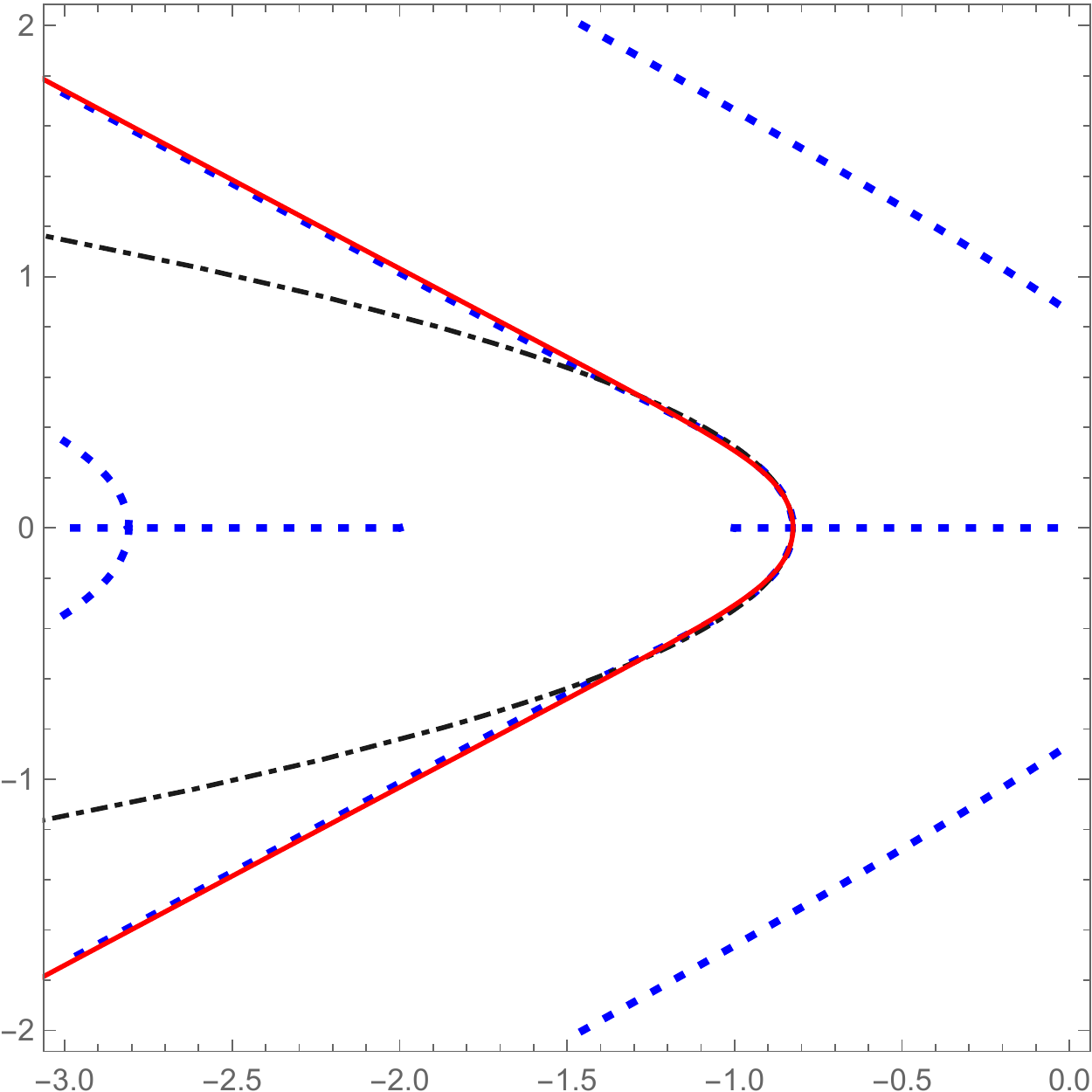}
\caption{The quadratic (dot-dashed dark gray) and [3/2] Pad\'e
 (solid red)
contours for the integrand $F_1(z,s)$ of
\eqn{OneDimensionalExampleIntegrand} with
$s=\stdparam$, shown against the exact contours of zero phase (dotted blue).}
\label{OneDimensionalPadeContour}
\end{figure}

To combine the quadratic contour with the asymptotic contour, 
we can replace the quadratic form
in \eqn{QuadraticContour} with a Pad\'e approximation.
We need a sufficient number of coefficients to fix two angles (the
tangents at the stationary point and at infinity), two intercepts
(with corresponding complex parts set to zero), and the quadratic 
behavior of the curve (again with vanishing imaginary part).  
This corresponds to eight real degrees of freedom.
While a [2/1] Pade does allow for four complex or eight 
real coefficients, we can remove one real parameter by rescaling the curve
parameter $t$, which leaves us with too few coefficients to fix
in order to
match the behaviors at both small and large $t$.  Instead we use
a [3/2] Pad\'e approximation.
For convenience we write it in the following form,
\begin{equation}
z_p(t) = \left\{
\begin{array}{l}
\displaystyle z_s + i t + 
  \frac{t^2 (a_2 + i b_2 a_3 t)}{1+i b_1 t + b_2 t^2}\,,\quad t>0\,,\\
\displaystyle z_s + i t + 
  \frac{t^2 (a_2^* + i b_2^* a_3^* t)}{1+i b_1^* t + b_2^* t^2}\,,\quad t<0\,,\\
\end{array}\right.
\label{PadeContour}
\end{equation}
which ensures the correct symmetry under reflection through the real
axis.
Matching coefficients as $t\rightarrow 0$, we find that,
\begin{equation}
a_2 = c_2\,.
\label{PadeContourCoefficientsI}
\end{equation}
Defining,
\begin{equation}
\begin{aligned}
\tau_E &= 1-e^{i\thinf} \rho\,,\\
d_E &= \tau_E^2 + c_2 (z_s-\zinf)\,,
\end{aligned}
\label{PadeContourCoefficientsII}
\end{equation}
and also matching the leading coefficient as $t\rightarrow \infty$, 
we find that,
\begin{equation}
a_3 = -\tau_E\,.
\label{PadeContourCoefficientsIII}
\end{equation}
The $\rho$ parameter corresponds to the magnitude of the coefficient
in the leading $t\rightarrow\infty$ coefficient; in general, it does
not appear possible to use it to improve the contour for practical
purposes beyond the constraints described below, and so we simply
set it to 1 here and in all following equations for the parameters
in \eqn{PadeContour}.  Matching the next-to-leading coefficient
as $t\rightarrow\infty$, we obtain,
\begin{equation}
b_1 = \frac{c_2+b_2 (z_s-\zinf)}{\tau_E}\,,
\label{PadeContourCoefficientsIV}
\end{equation}
where $b_2$ will be given below.
Matching only through $\Ord(t^2)$ at small $t$
 would leave one complex parameter completely
unfixed.  We can choose it so that the integrand is real
through $\Ord(t^5)$. The quartic order gives a linear equation
which can be solved for the real part of $b_2$ in terms of 
its imaginary part,
\begin{equation}
\Re b_2 = \frac1{\Re d_E-2 (\Re \tau_E)^2}\bigl[
(-c_2^2+\Im b_2 \Re \tau_E (\Re d_E+2 (\Im \tau_E)^2)/\Im \tau_E\bigr]
\end{equation}
while the quintic order then gives a quadratic equation for
the imaginary part,
\begin{equation}
q_2 (\Im b_2)^2 + q_1 \Im b_2 + q_0 = 0\,.
\label{PadeContourCoefficientsV}
\end{equation}
Define,
\begin{equation}
\begin{aligned}
c_j &= \frac{F_1^{(j+1)}(z_s)}{(j+1)!\,F_1''(z_s)}\,,\\
f_E &= (\Re d_E)^2 + 4(\Im \tau_E\,\Re\tau_E)^2\,;
\end{aligned}
\end{equation}
the coefficients in \eqn{PadeContourCoefficientsIV} are then,
\begin{equation}
\begin{aligned}
q_0 &= c_2^4 (\Im\tau_E)^2 \Re d_E (1-\Re\tau_E)
\\&\hphantom{=}
       + c_2 (3 c_2^3-4 c_2 c_3+c_4)
             (\Im\tau_E)^2 \Re d_E (\Re d_E-4(\Re\tau_E)^2)
\\&\hphantom{=}
       -4 c_2 (\Im\tau_E\Re\tau_E)^2 (c_2^3-2 c_2^3\Re\tau_E
                           -(3 c_2^3-4 c_2 c_3+c_4)(\Re\tau_E)^2)\,,\\
q_1 &= 2 c_2^2 \Im\tau_E\,f_E\bigl[(\Re\tau_E-2(\Re\tau_E)^2+\Re d_E\bigr]\,,\\
q_2 &= -\Re d_E\,f_E \bigl[(\Im\tau_E)^2-(\Re\tau_E)^2+\Re d_E\bigr]\,.\\
\end{aligned}
\label{PadeContourCoefficientsVI}
\end{equation}
A reasonable heuristic is to take the smaller of two positive
solutions; to take the positive solution if one is negative; and
to take the solution of smaller magnitude if both are negative. 
(If the solutions are complex, take the common real part.)

\begin{figure}[ht]
\begin{minipage}[b]{1.03\linewidth}
\begin{tabular}{cc}
\hskip -6mm
\includegraphics[clip,scale=0.5]{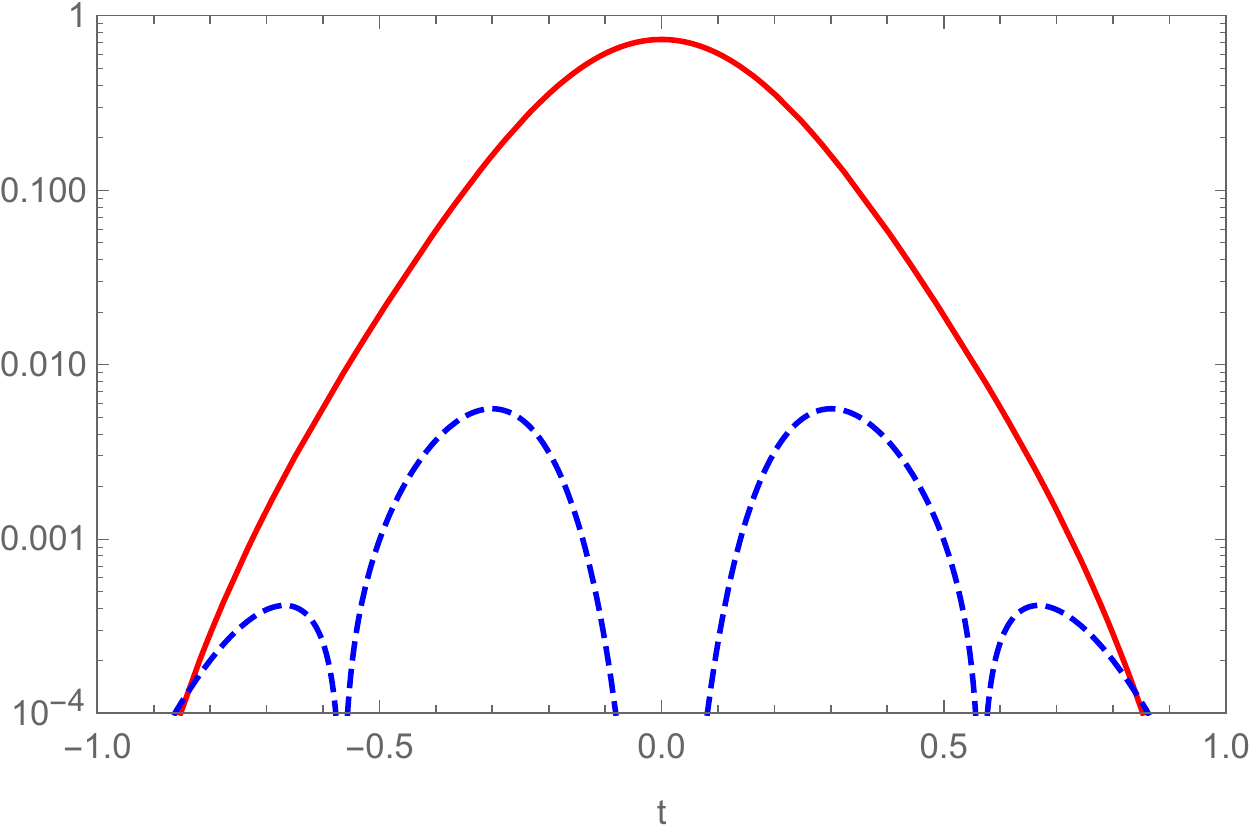}
\hskip5mm
&\includegraphics[clip,scale=0.5]{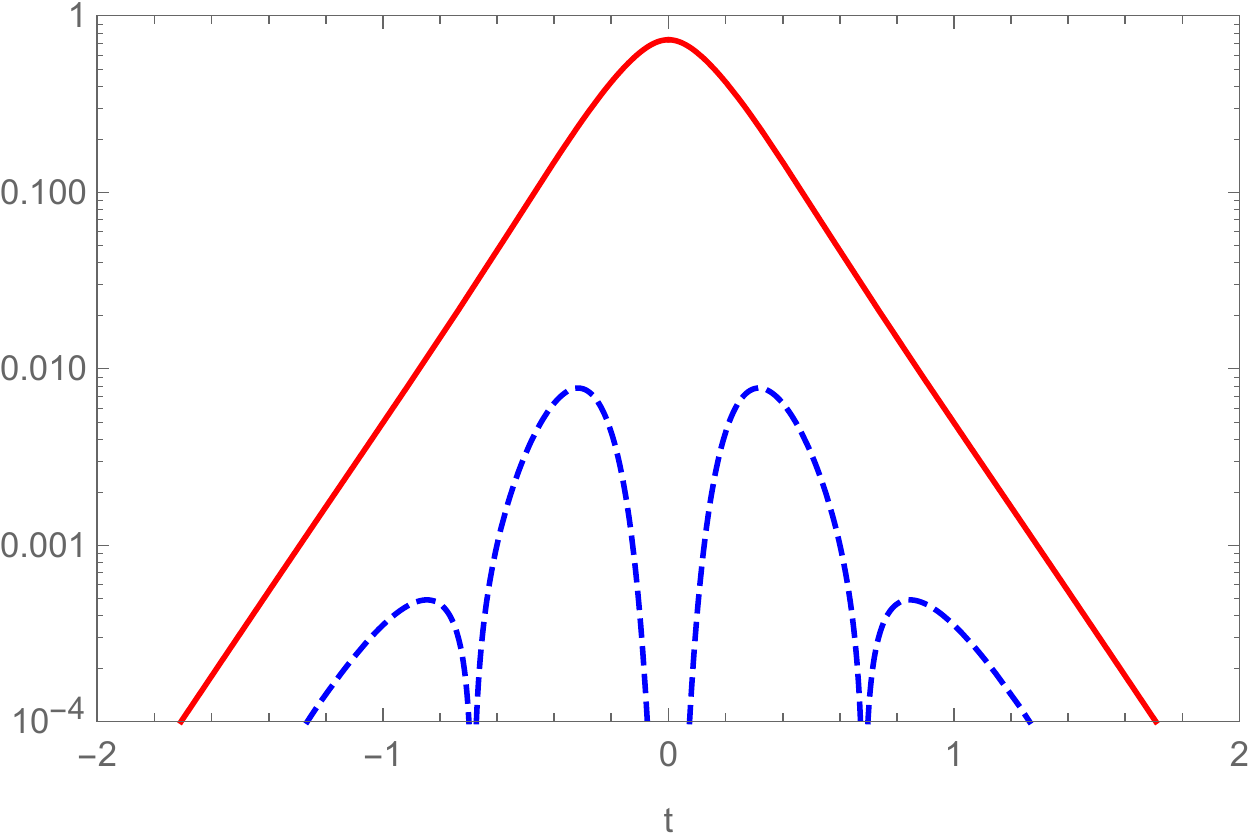}
\\[3mm]
(a)&(b)\\[3mm]
\end{tabular}
\end{minipage}
\caption{The absolute values of the
real (red) and imaginary (dashed blue) parts of
the integrand $F_1(z,s)$ of \eqn{OneDimensionalExampleIntegrand} for 
$s=\stdparam$ on (a) the quadratic contour of \eqn{QuadraticContourExample}
(b) the Pad\'e contour of \eqn{PadeContourExample}.}
\label{OneDimContourLogIntegrandC}
\end{figure}

\begin{figure}[ht]
\includegraphics[clip,scale=0.52]{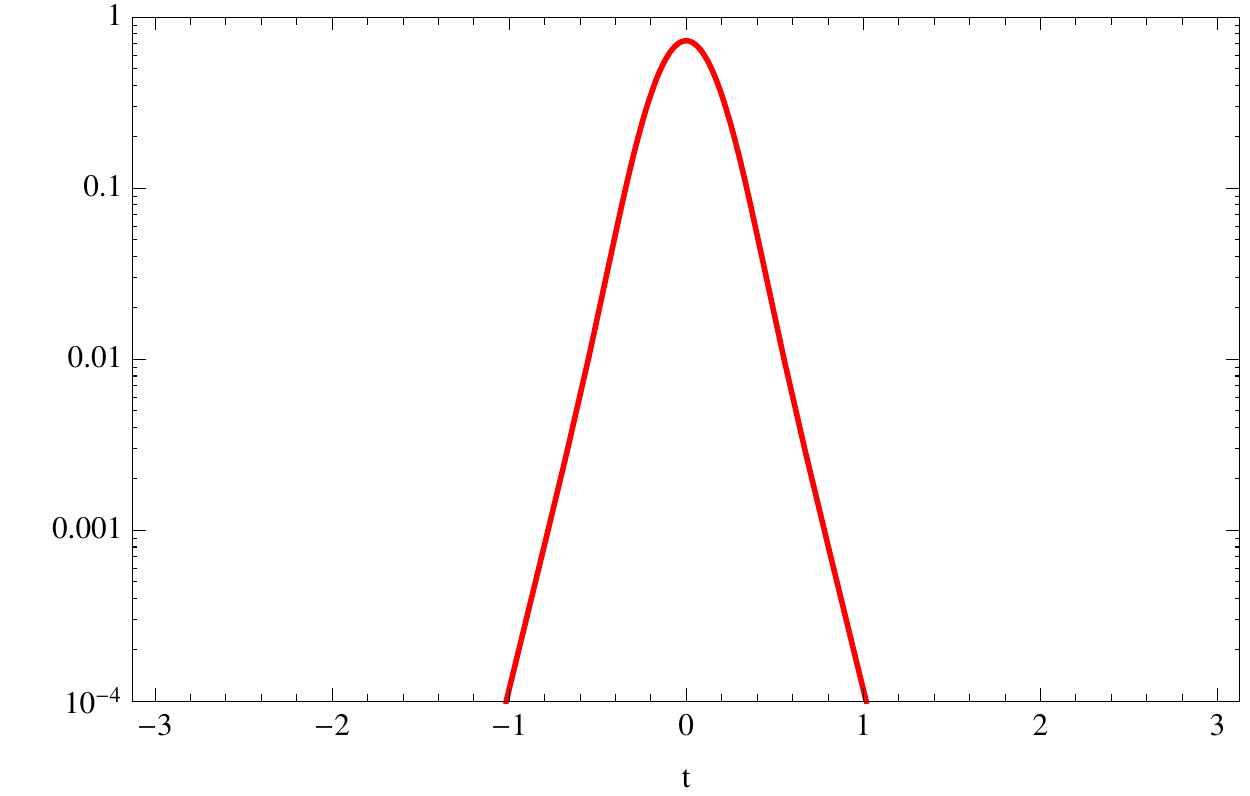}
\caption{The absolute values of the
real (red) and imaginary (dashed blue) parts of
the integrand $F_1(z,s)$ of \eqn{OneDimensionalExampleIntegrand} for 
$s=\stdparam$ on the exact contour of stationary phase.}
\label{OneDimExactContourLogIntegrandC}
\end{figure}

In the example at hand, this curve is,
\begin{equation}
-0.825618 + i t + 
\frac{\bigl(-1.65358 - (7.03903 + 6.94838 i\sign t)\, |t|\bigr)\, t^2}
{1 + (4.25685 - 1.28081 i\sign t)\, |t| 
  + (10.32798 + 3.24601 i\sign t) \,t^2}
\label{PadeContourExample}
\end{equation}
This contour is shown in \fig{OneDimensionalPadeContour},
along with the quadratic and exact contours.
The shape of the integrand along this contour 
is again very similar to
that shown in \fig{OneDimContourIntegrandC} for the quadratic
contour.  The differences
are noticeable only on a logarithmic scale, shown side-by-side in
\fig{OneDimContourLogIntegrandC}.
In \fig{OneDimExactContourLogIntegrandC}, we show for comparison
the integrand along the exact contour of stationary phase running
through the stationary point, as computed using the 
differential equation. 
The difference in shape is due to the different parametrization
of the curve (for the exact contour, we take the imaginary part to
simply be $t$).  The principal improvement along the exact
contour is the complete
absence of the imaginary part.

We can of course construct higher-order curves, to serve as closer
approximations to the true stationary-phase contour in the small-$t$
region, if desired.  For example, a quartic contour would be given by,
\begin{equation}
z_4(t) = z_s + i t + c_2 t^2 + g_4 t^4\,,
\label{QuarticContour}
\end{equation}
where 
\begin{equation}
g_4 = -\frac{F^{(5)}(z_s)}{120 F''(z_s)}
+\frac{F^{(4)}(z_s) F^{(3)}(z_s)}{36 (F''(z_s))^2}
-\frac1{72} \left(\frac{F^{(3)}(z_s)}{F''(z_s)}\right)^3\,.
\end{equation}
The formula for $g_4$ is obtained by requiring that the imaginary
part of $F(z(t))$ vanish to $\Ord(t^5)$; in the Euclidean region,
it will then automatically be real to $\Ord(t^6)$.

\section{Matching to Asymptotic Forms}
\label{AsymptoticFormSection}

\def\numG{{\gatop{\rm numer\ }{\Gamma{\rm s}}}}
\def\denG{{\gatop{\rm denom\ }{\Gamma{\rm s}}}}
\def\PG{{\psi{\rm s}}}
We are interested in one-dimensional Mellin--Barnes
integrals that arise from Feynman integrals with massive propagators.
Labeling the integration variable of
each Mellin--Barnes integral by $z$, 
the integrands contain gamma functions
and their derivatives, with arguments of the form $n\pm z$ and 
$n\pm 2 z$, where $n$ is an integer (positive, negative, or zero).
In general, the integrand is a sum of terms, where each term is a
pure product of gamma functions, their derivatives, and inverses
of gamma functions.  Let us focus on each
term separately, or equivalently restrict attention to
integrands of the form,
\begin{equation}
(-s)^{-z} \frac{\prod_{j\in{\rm numer}} \Gamma(a_j+n_j z)
	\prod_{j\in{\rm numer}} \psi^{(d_j)}(b_j+m_j z)
	 }{\prod_{j\in{\rm denom}} \Gamma(a_j+n_j z)}\,.
\label{GenericTerm}
\end{equation}  This makes it possible to write down general
formul\ae{} for the asymptotic behavior of the integrand, and
the corresponding parameters governing the contours of stationary
phase.

 The difference of numerator and denominator gamma
function arguments is independent of $z$,
\begin{equation}
\frac{d}{dz} \left[\sum_{\numG} {\rm argument(z)}
-\sum_{\denG} {\rm argument(z)}
                  \right] = 0\,.
\label{ArgumentIdentity0}
\end{equation}
Denoting the coefficient of $z$ in the argument to the $j$th
gamma function by $n_j$, as in \eqn{GenericTerm} we can rewrite 
this identity as,
\begin{equation}
\sum_{\numG} {n_j}-\sum_{\denG} {n_j}
 = 0\,.
\label{ArgumentIdentity}
\end{equation}

We can use this feature to derive a formula for the critical parameter $s_0$,
as well as for the behavior of the integrand at large $z$.  Considering
only the exponential terms in the asymptotic form for the gamma
function~(\ref{AsymptoticGamma}), we see that the gamma function factors
in the integrand (or
a single term if the integrand is a sum of terms) behave for large $z$
(away from the real axis) as,
\begin{equation}
\begin{aligned}
&\exp\biggl[-\biggl({\displaystyle\sum_{j\in\numG}} n_j
                     -{\displaystyle\sum_{j\in\denG}} n_j\biggr) z\biggr]
\\
&\times \exp\biggl[
         \biggl(\sum_{j\in\numG} n_j-\sum_{j\in\denG} n_j\biggr) z\ln z\biggr]
\\
&\times \exp\biggl[
       \biggl(\sum_{j\in\numG} n_j\ln |n_j|-\sum_{j\in\denG}n_j \ln |n_j|\biggr) 
            z\biggr]
\\
&\times \exp\biggl[\biggl(
       \sum_{j\in\numG} n_j\ln \bigl(\sign n_j-i\delta\sign\Im z\bigr)
       -\sum_{j\in\denG}n_j \ln \bigl(\sign n_j-i\delta\sign\Im z\bigr)\biggr) 
            z\biggr]
\end{aligned}
\end{equation}
where we drop overall constants.
 Using the above identity~(\ref{ArgumentIdentity}), 
this asymptotic form simplifies to,
\begin{equation}
\begin{aligned}
&\exp\biggl[\biggl(
       \sum_{j\in\numG} n_j\ln |n_j|-\sum_{j\in\denG}n_j \ln |n_j|\biggr) 
            z\biggr]\\
&\times\exp\biggl[\biggl(
       \sum_{j\in\left.\numG\right|n_j<0} |n_j|
       -\sum_{j\in\left.\denG\right|n_j<0}|n_j| \biggr) 
            i\pi z\sign\Im z\biggr]\,.
\end{aligned}
\label{GeneralAsymptoticBehavior}
\end{equation}
The exponential part of the integrand's behavior is not modified by possible
polygamma function factors, as these have logarithmic or power-like
asymptotic behavior.
The critical value of $s$, which determines in which direction the
contour must bend in order to ensure convergence as the contour parameter
$t\rightarrow\pm\infty$, is then given by,
\begin{equation}
s_0 = \frac{\prod_{j\in\numG} |n_j|^{n_j}}{\prod_{j\in\denG} |n_j|^{n_j}}\,.
\label{Threshold}
\end{equation}
\def\Nm{N_{-}}
\def\Sp{S_{+}}
\def\Sm{S_{-}}
Assuming the argument $s$ appears in the integrand as $(-s)^{-z}$,
for $|s|<s_0$, the contour must bend left, towards negative
values of $\Re z$, while for $|s|>s_0$, it must bend right, towards 
positive values of $\Re z$. 
The integer offset in the argument of the $j$th gamma function
(that is, the integer value obtained by setting $z=0$)
in \eqn{GenericTerm} is denoted by $a_j$.
It will be convenient to define,
\begin{equation}
\Nm \equiv 
\sum_{j\in\left.\numG\right|n_j<0} |n_j|
       -\sum_{j\in\left.\denG\right|n_j<0}|n_j| \,,
\label{NminusDef}
\end{equation}
\begin{equation}
S_{+} \equiv 
\sum_{j\in\left.\numG\right|n_j>0} 1
       -\sum_{j\in\left.\denG\right|n_j>0}1 \,,
\label{SplusDef}
\end{equation}
\begin{equation}
S_{-} \equiv 
\sum_{j\in\left.\numG\right|n_j<0} 1
       -\sum_{j\in\left.\denG\right|n_j<0}1 \,,
\label{SminusDef}
\end{equation}
\begin{equation}
A_{+} \equiv 
\sum_{j\in\left.\numG\right|n_j>0} a_j
       -\sum_{j\in\left.\denG\right|n_j>0} a_j  \,,
\label{AplusDef}
\end{equation}
and 
\begin{equation}
A_{-} \equiv 
\sum_{j\in\left.\numG\right|n_j<0} a_j
       -\sum_{j\in\left.\denG\right|n_j<0} a_j \,.
\label{AminusDef}
\end{equation}
(We do not need $N_{+}$, corresponding to $n_j>0$, 
thanks to \eqn{ArgumentIdentity}.)
\def\SP{S^{(\psi)}}
\def\DP{D^{(\psi)}}
Denoting the number of derivatives 
of the $j$th polygamma function $\psi^{(d)}(z)$ 
(always in the numerator) by
$d_j$, with $m_j$ the coefficient of $z$ in the argument 
let us also define,
\begin{equation}
\begin{array}{l}
\displaystyle \DP_{+} \equiv 
\sum_{j\in\left.\PG\right|m_j>0} d_j \,,\\
\displaystyle \DP_{-} \equiv 
\sum_{j\in\left.\PG\right|m_j<0} d_j \,,\\
\displaystyle \SP \equiv 
\sum_{j\in\PG} 1 \,,
\label{PGDef}
\end{array}
\end{equation}
where the sums are not taken over the basic polygamma function $\psi(z)$
but only over its derivatives $\psi^{(d)}(z)$.

Using them, we can express the remaining square-root factors in
\eqn{AsymptoticGamma} in a compact form, so that the asymptotic
behavior of the integrand as a whole is,
\begin{equation}
\begin{aligned}
F &\sim {\rm const} \biggl(\frac{s_0}{-s}\biggr)^{z} 
e^{i\pi z \Nm \sign\Im z} (-1)^{\DP_{+}+\DP_{-}+\SP}
z^{-S_{+}/2+A_{+}-\DP_{+}} (-z)^{-S_{-}/2+A_{-}-\DP_{-}}
\\
&={\rm const} \biggl(\frac{s_0}{-s}\biggr)^{z} e^{i\pi z \Nm\sign\Im z}
(-1+i\delta\sign\Im z)^{-S_{+}/2+A_{+}+\DP_{-}} \\
&\hskip 10mm\times
(-z)^{-S_{+}/2-S_{-}/2+A_{+}+A_{-}-\DP_{+}-\DP_{-}}\,.
\end{aligned}
\end{equation}
where we have rewritten $z^a = [(-1+i\delta \sign\Im z)(-z)]^a$,
and used the fact that $\DP_{\pm}$ and $\SP$ are integers, along with
$(-1)^n = (-1+i\delta \sign\Im z)^n$ for integer $n$.

The phase of this expression is,
\begin{equation}
\begin{aligned}
\arg F &= 
\biggl[\Im z\ln\biggl(\frac{s_0}{-s}\biggr)+\pi \Re z \Nm\sign\Im z
\\&\hphantom{=}
-\frac{\pi}2 \bigl((S_{+}-2 A_{+}-2\DP_{-}-2\SP) \bmod 4\bigr) \sign\Im z
\\&\hphantom{=}
+\bigl(-S_{+}/2-S_{-}/2+A_{+}+A_{-}-\DP_{+}-\DP_{-}\bigr) \arg(-z)\biggr] \bmod 2\pi\,,
\end{aligned}
\label{GeneralPhaseExpression}
\end{equation}
generalizing \eqn{IntegrandArg}. (Here, $n\bmod m$ is understood to mean
$(\sign n) (|n|\bmod m)$, and the $\bmod\,2\pi$ is understood to reduce
the variable to the range $(-\pi,\pi]$.)

\def\Step{\Theta}
Substituting the form in \eqn{InfinityContour}, expanding in $t$,
and setting the coefficient of the $\Ord(t)$ term to zero,
and that of the $\Ord(t^0)$ term
to $\phi_s$ allows us to obtain general formul\ae{} for $\thinf$ and
$\zinf$,
\begin{equation}
\begin{aligned}
\thinf &= \atan\biggl[\frac1{\pi \Nm}\ln\biggl(\frac{s_0}{-s}\biggr)
\biggr]\,,\qquad \Nm \neq 0\,,\\
\thinf &= \sign\ln\biggl(\frac{s_0}{-s}\biggr)\,\frac{\pi}2\,,
\qquad \Nm = 0\,,\\
\zinf &= \frac{\phi_s \sign\Im z_s}{\pi \Nm}
+\frac{(S_{+}-2 A_{+}-2\DP_{-}-2\SP)\bmod 4}{2 \Nm}\\
& -\frac1{\Nm \pi}
   \biggl[\bigl(A_{+}+A_{-}-S_{+}/2-S_{-}/2\\
   &\hskip 35mm-\DP_{+}-\DP_{-}\bigr) 
         \biggl(\thinf-\frac{\pi}2\biggr)
 \bmod 2\pi\biggr]
\,,\qquad \Nm \neq 0\,,\\
\zinfp &=
i\bigg|\frac{\phi_s}{\ln(s_0/(-s))}
  +\frac{\pi}2\frac{(S_{+}-2 A_{+}-2\DP_{-}-2\SP_{+})\bmod 4}{\ln(s_0/(-s))}\\
&
-\frac1{\ln(s_0/(-s))}
   \biggl[\bigl(A_{+}+A_{-}-S_{+}/2-S_{-}/2\\
   &\hskip 35mm-\DP_{+}-\DP_{-}\bigr) 
         \biggl(\thinf-\frac{\pi}2\biggr)
 \bmod 2\pi\biggr]\bigg|
\,,\qquad \Nm = 0\,,\\
\end{aligned}
\label{AsymptoticThetaEuclidean}
\end{equation}
valid in the Euclidean region (for $s\neq -s_0$).  
Factors of the polygamma $\psi(z)$ 
(with no derivatives) will correct the large-$t$ behavior by terms
of $\Ord(1/\ln t)$, which are noticeable visually on contour plots,
but have no practical importance in computing the integral.  In
the generic case, $\phi_s = 0$ or~$\pi$, but the formul\ae{} are
valid more generally; we take $\sign\Im 0$ to be 1.
(We have implicitly assumed that $\thinf\in [-\frac{\pi}2,\frac{\pi}2]$
in deriving these results.)

In these formul\ae{},
 $\phi_s=0$ if $F(z_s)\ge 0$, while
$\phi_s=\pi$ if $F(z_s)<0$.
For $\Nm\neq 0$,
the value for $\zinf$ may be shifted
by an integer multiple of $2/\Nm$.  (A good heuristic is
to choose $\zinf$ in the original interval of interest.)
For $\Nm = 0$, the asymptotes are
parallel to the real axis, and so we cannot take $\zinf$ to be real; we
must replace $\zinf$ by $\zinfp\Step(t)-\zinfp\Step(-t)$ in
\eqn{PadeContourCoefficientsII}, where
$\Step(t)$ is the usual Heaviside step function.
For this value of $\Nm$,
we can shift $\zinfp$ by a multiple of $2\pi i/\ln(s_0/(-s))$.
(In some cases, it may be appropriate to shift by half this
amount. A good heuristic here is to choose $\zinfp$ to lie in the
interval $i[1/c_2-\pi/\ln(s_0/(-s)),1/c_2+\pi/\ln(s_0/(-s))]$,
where $c_2$ is the quadratic coefficient given 
in \eqn{QuadraticCoefficient}.)

These expressions for $\thinf$ and $\zinfp$, along with that for $c_2$
given in \eqn{QuadraticCoefficient}, allow us to compute
the coefficients of the Pad\'e approximation~(\ref{PadeContour}) 
in the Euclidean region via 
eqs.~(\ref{PadeContourCoefficientsI}--\ref{PadeContourCoefficientsVI},%
\ref{NminusDef}--\ref{PGDef},\ref{AsymptoticThetaEuclidean}) for a 
generic one-dimensional
Mellin--Barnes integrand arising from
Feynman diagrams.

\section{Other Integrands}
\label{OtherIntegralsSection}

\def\sln{\mathop{\rm sln}}
\begin{figure}[ht]
\includegraphics[clip,scale=0.66]{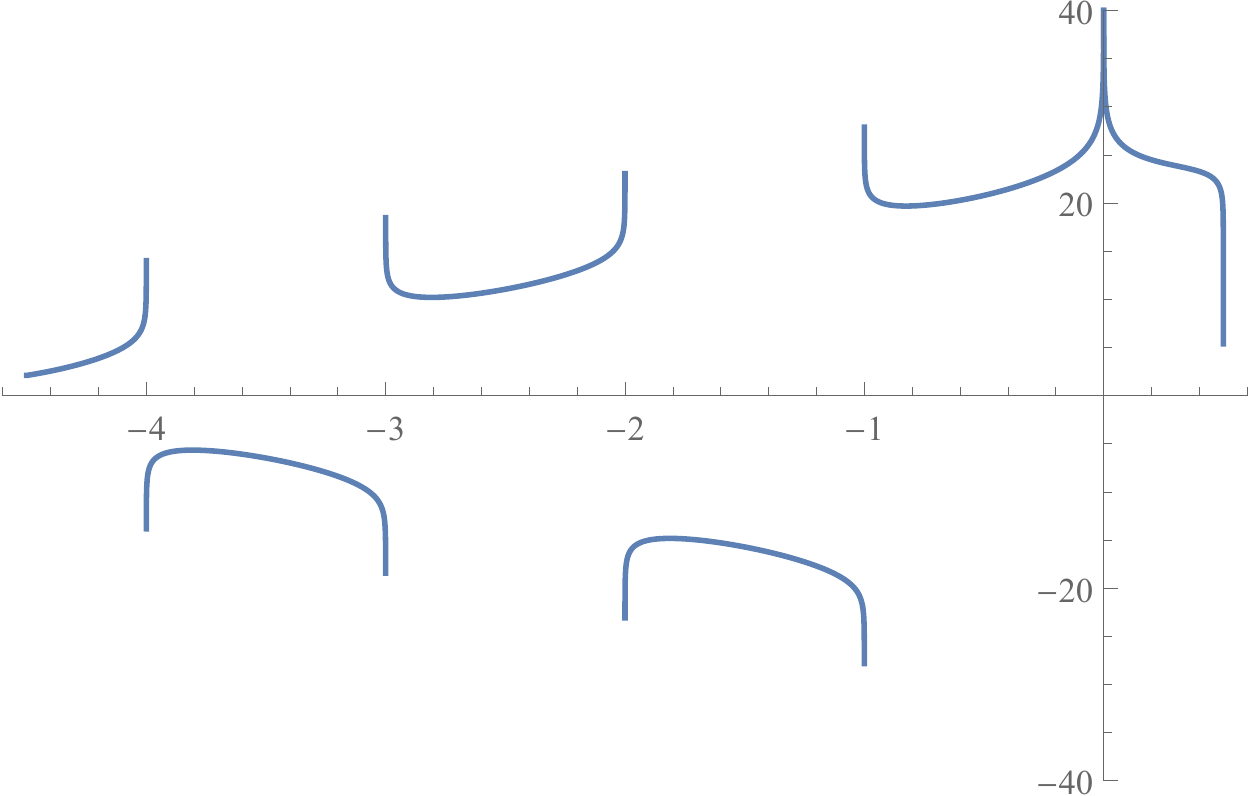}
\caption{The behavior of the integrand $F_1(z,s)$ 
in \eqn{OneDimensionalExampleIntegrand},
along the real axis.  The plot displays $\sln_{20}({\rm integrand})$,
with $s=\stdparam$.}
\label{OneDimIntegrand}
\end{figure}

The integrands in which we are interested
have poles at most integer values; may have zeros or
poles at half-integer values, and may have additional zeros along the
real axis.  The integrand considered in the previous section is
generic, but its properties are not universal: it has poles
at every integer value, and zeros at every positive half-integer
value.  At the ends of each integer interval $(n,n+1)$ with $n<0$, it blows
up with the same sign, so that each integer interval contains an
extremum.  This is shown in \fig{OneDimIntegrand}, with the aid
of a function designed to compress the vertical scale,
\begin{equation}
\sln_m x \equiv \sign x\,\ln(1+ |x| e^m)\,.
\end{equation}
For purposes of drawing contours of stationary phase and approximations
thereto,
it is not essential that the extremum be a minimum, of course; if it is a local
maximum, one can convert it to a minimum by considering the negative
of the integrand.

What other classes of integrands should we consider?  In this section,
we discuss a few less-generic but possible forms of integrands, and
discuss how the contours in the previous section are modified.

\subsection{Intervals Without Extrema}

\begin{figure}[ht]
\includegraphics[clip,scale=0.66]{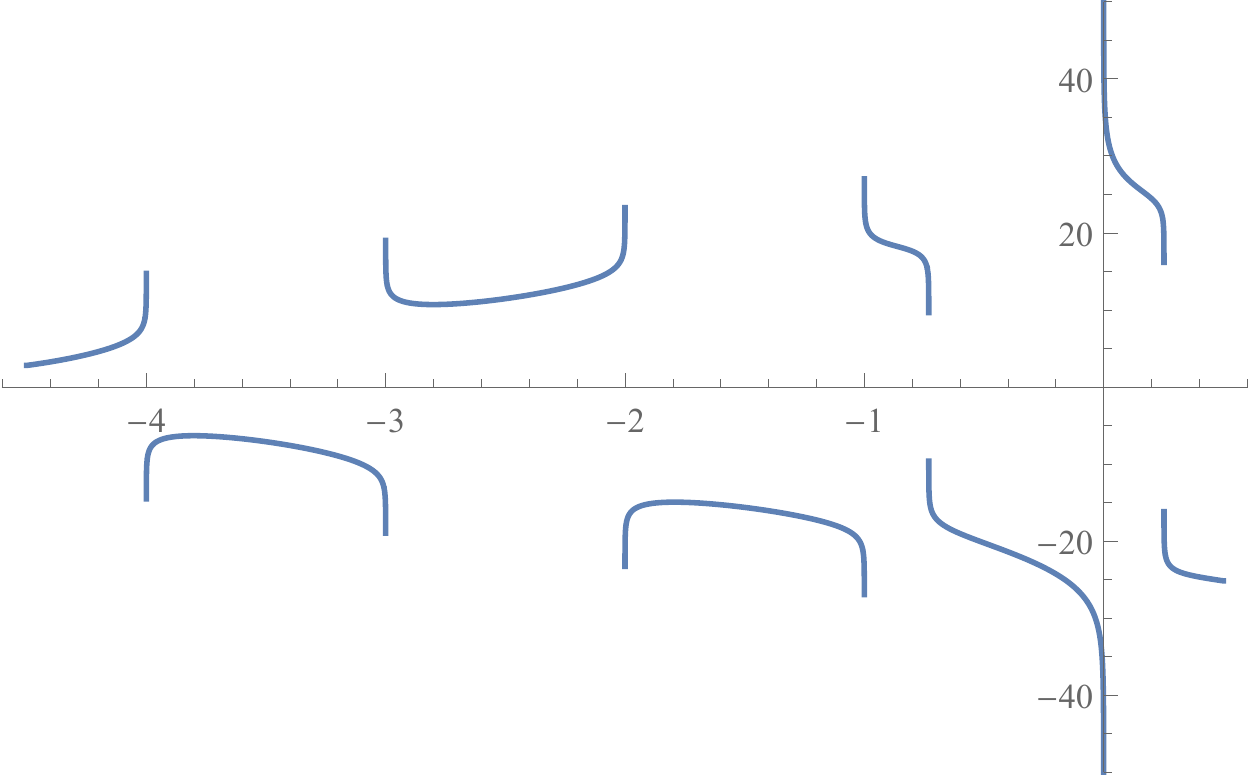}
\caption{The behavior of the integrand $F_2(z,s)$ of \eqn{Integrand2},
along the real axis.  The plot displays $\sln_{20}({\rm integrand})$,
with $s=\stdparam$.}
\label{OneDimIntegrand2}
\end{figure}

Not all integrands that arise in calculations of interest share the
nice feature of the one in \eqn{OneDimensionalExampleIntegrand}, namely
that the integrand blows up with the same sign at both ends of
for integer (or half-integer)
intervals of interest.  For example, consider the integrand
obtained by multiplying that of \eqn{OneDimensionalExampleIntegrand} by 
a polygamma function,
\begin{equation}
F_2(z,s)=
  (-s)^{-z} \frac{\Gamma^3(-z)\Gamma(1+z) \psi(-2z)}{\Gamma(-2z)}\,.
\label{Integrand2}
\end{equation}
If we are interested in a deforming the contour passing through 
$\Re z=-\onehalf$ slightly to obtain a contour of stationary phase,
we see from \fig{OneDimIntegrand2} that this is not possible, because
the integrand has no extremum in the interval $(-1,0)$.  Indeed,
the lone stationary point on the real axis is replaced
by a complex-conjugate pair of stationary points.  In addition, the
function necessarily has at least one zero in the interval.
These features require
a substantial modification of the contours discussed in the previous section,
a point to which we shall return in~\sect{NoExtremaSection}.

In the case of $F_2$, however, we can simply shift the contour 
to the interval $(-2,-1)$.  We pick up
an additional contribution from the residue at $-1$, so that
\begin{equation}
I_2(s) = \frac{1}{2\pi i} \int_{c_0-i\infty}^{c_0+i\infty} dz
\;F_2(z,s) = (-1+\gamma_E) +s
\frac{1}{2\pi i} \int_{c_1-i\infty}^{c_1+i\infty} dz
\;F_2(z,s)\,,
\end{equation}
where $c_1 = -3/2$.
We can apply the approach of the previous section to the
second integral, as its integrand does have 
a stationary point on the real axis in the new interval.

It can happen that there is {\it no\/} interval which has a local
extremum.  As mentioned
above, we will return to a consideration of such integrands in
 \sect{NoExtremaSection}.

\subsection{Wrong-Direction Quadratic Contour}

\begin{figure}[ht]
\includegraphics[clip,scale=0.66]{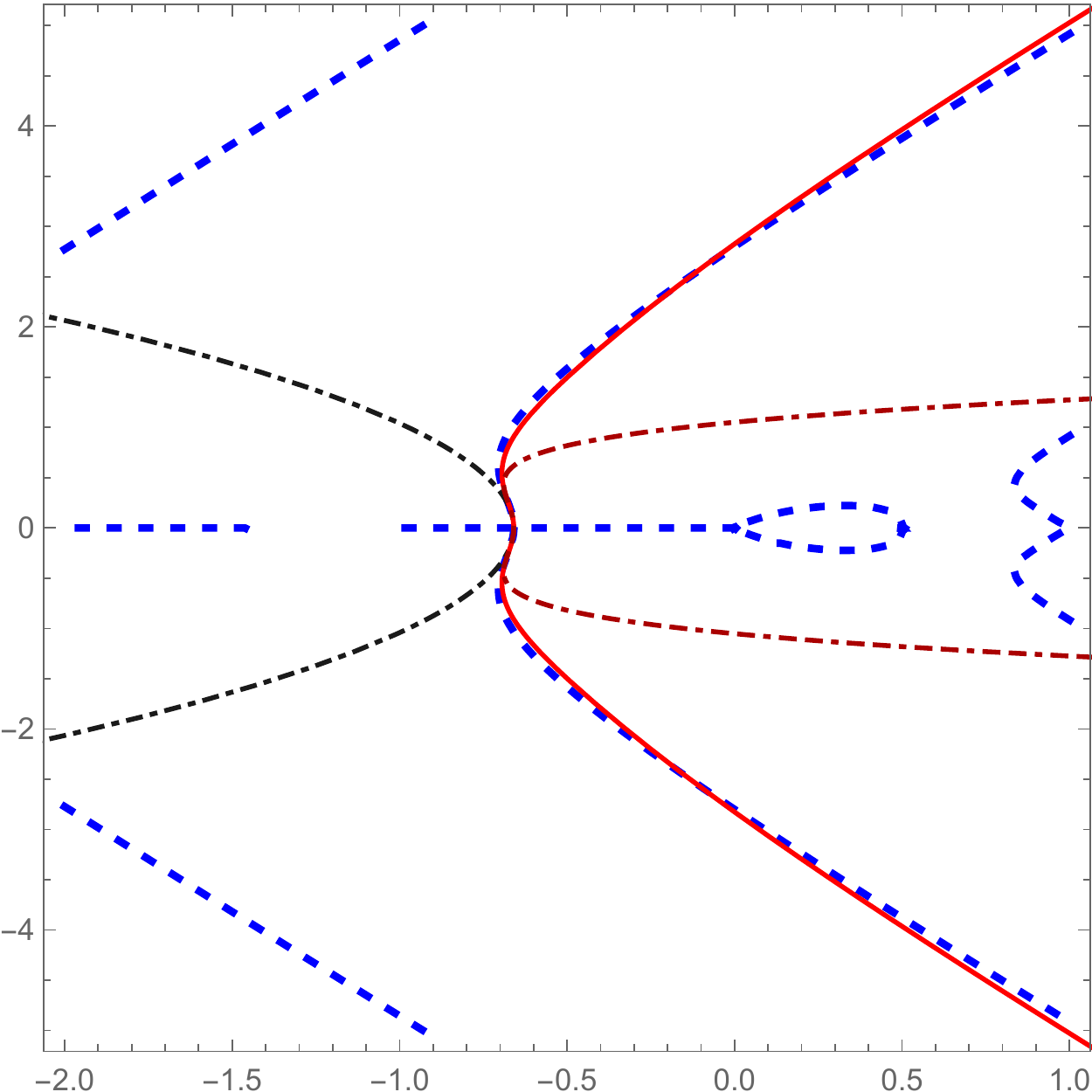}
\caption{The quadratic (dot-dashed dark gray), [3/2] Pad\'e (solid red),
and quartic (dot double-dashed dark red)
contours for the integrand $F_3(z,s)$ of
\eqn{Integrand3} with
$s=-{20}$, shown against the exact contours of zero phase (dotted blue).}
\label{Integrand3ContoursFigure}
\end{figure}

As discussed in the previous section,
we can use the asymptotic expansion for the gamma function in order
to study the large-$z$ behavior of the integrand.  For the integrand
$F_1(z,s)$ in \eqn{OneDimensionalExampleIntegrand}, 
we found~(\ref{Integrand1AsymptoticExpansion})
that it behaves like,
\begin{equation}
{\rm const}
  \,4^z (-s)^{-z} \frac{(-1+i\delta\sign\Im z)^{z+1/2}}{\sqrt{-z}}
\,.
\end{equation}
The factor $(-1+i\delta\sign\Im z)^{z}$ is convergent as $z\rightarrow\infty$
both above and below the real axis, independently of the sign of the real
part of $z$.  The first two factors do however care which direction 
the contour goes.
For $-4<s<0$, they require us to close the contour to the left; otherwise
these factors will blow up when $\Re z\rightarrow -\infty$.  Similarly,
for $s<-4$, they require us to close the contour to the right.  The 
quadratic contour~(\ref{QuadraticContourExample}),
for $s=\stdparam$, has the desired
form, veering away from the imaginary axis to the left.  This is also true
by construction, of course, for the [3/2] Pad\'e contour~(\ref{PadeContourExample}).  However,
the plain quadratic contour does not always have this property.
Consider the following integrand,
\begin{equation}
F_3(z,s) = 
  (-s)^{-z} \frac{\Gamma^3(-z)\Gamma(1+z) \psi(-z)}{\Gamma(-2z)}\,.
\label{Integrand3}
\end{equation}
The various quadratic approximations along with the exact contours
of zero phase are shown in \fig{Integrand3ContoursFigure} for $s=-20$.
In this case, the pure quadratic contour veers to the left, whereas convergence
requires the contour to veer to the right.

In contrast, the form of the [3/2] Pad\'e contour (\ref{PadeContour}) 
requires no modification; it heads
off in the correct direction, thereby solving the problem
with the simple quadratic contour.  This is one of the reasons the
use of the asymptotic contour is helpful, even though the contributions
to the integral from the asymptotic region are exponentially small.  In
this particular case, the quartic contour of \eqn{QuarticContour} would
also head off towards the correct side of the complex plane;
but one can always find examples
where a given fixed-order contour heads off in the wrong direction.
All three contours are shown in \fig{Integrand3ContoursFigure}.
(The Pad\'e contour shown has its $\zinf$ intercept shifted to
the interval $(-2,-1)$, but this does not affect the overall qualitative
features compared to having it in the interval $(-1,0)$.)

\subsection{Vanishing Curvature}

The formula~(\ref{QuadraticCoefficient}) for the coefficient of the
quadratic term in the contour assumes that the second derivative at the
local extremum does not vanish.  While this is typically true, one
encounters examples where it is false. Such an example is given
by the following integrand,
\begin{equation}
F_4(z,s) = 
(-s)^{-z} \frac{\Gamma^3(-z)\Gamma(1+z)\psi^4(1-z)}{\Gamma(-2z)}\,.
\end{equation}
In general, if the integrand blows up with the same sign at both ends
of a pole-free interval (ensuring that there is an absolute extremum on
the interval), 
and the second derivative vanishes at a given local extremum, there are
two possibilities: either the third derivative also vanishes, or it is
non-vanishing.   In the latter case, there is then another, lower minimum
(higher maximum), which is what we should pick as the base
point for the contour.  If the third derivative also vanishes, we must
modify our approach.  In the above example, the integrand vanishes
at the extremum in the interval $(-1,0)$, 
which is in fact the likeliest way for a
single-term Mellin--Barnes integrand of the type we are considering to 
have a vanishing second derivative.  We could again
handle it by shifting to a different interval along the negative
real axis, where the
extremum will be quadratic.  Evaluating it with a contour in
the interval $(-1,0)$ is also possible, but requires 
the use of techniques of the same sort considered in
 \sect{NoExtremaSection} for integrands without extrema on the real axis.

\subsection{Flat Asymptotes}
\label{FlatAsymptotesSection}

\begin{figure}[ht]
\includegraphics[clip,scale=0.66]{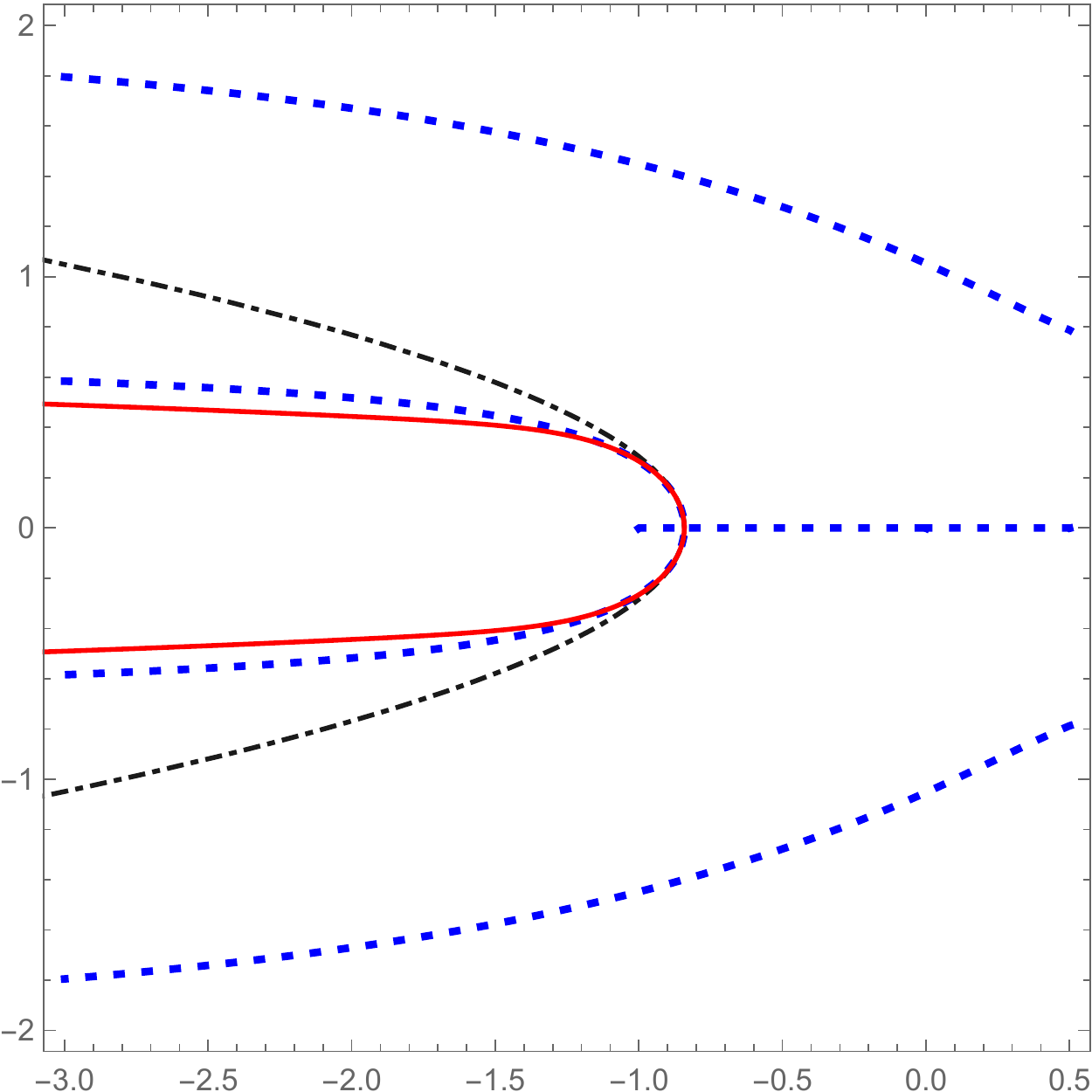}
\caption{The quadratic (dot-dashed darkk gray) and [3/2] Pad\'e (solid red)
contours for the integrand $F_5(z,s)$ of
\eqn{FlatAsymptotesIntegrand}, with
$s=\stdparam$, shown along with the exact contours of stationary phase 
(dotted blue).}
\label{FlatAsymptotesContourFigure}
\end{figure}

\begin{figure}[t]
\begin{minipage}[b]{1.03\linewidth}
\begin{tabular}{cc}
\hskip -6mm
\includegraphics[clip,scale=0.60]{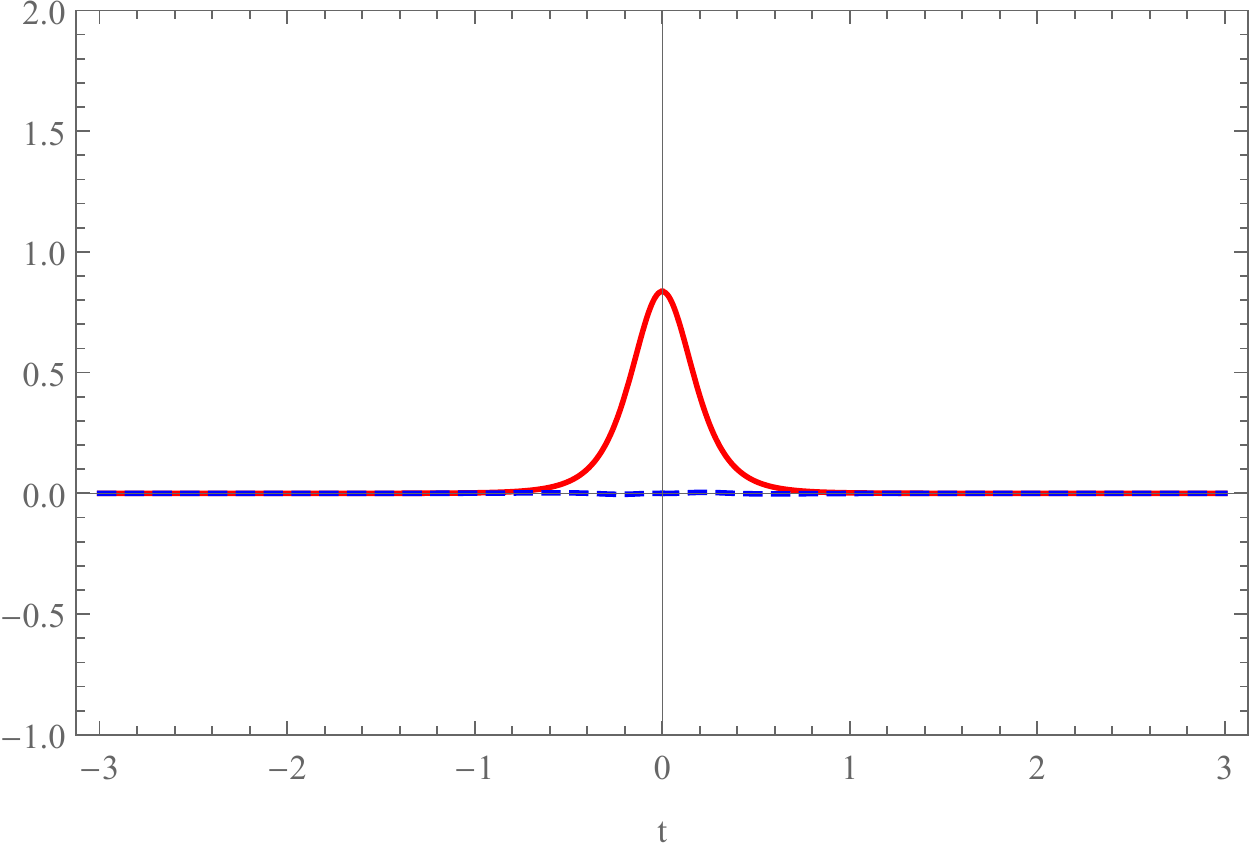}
\hskip5mm
&\includegraphics[clip,scale=0.63]{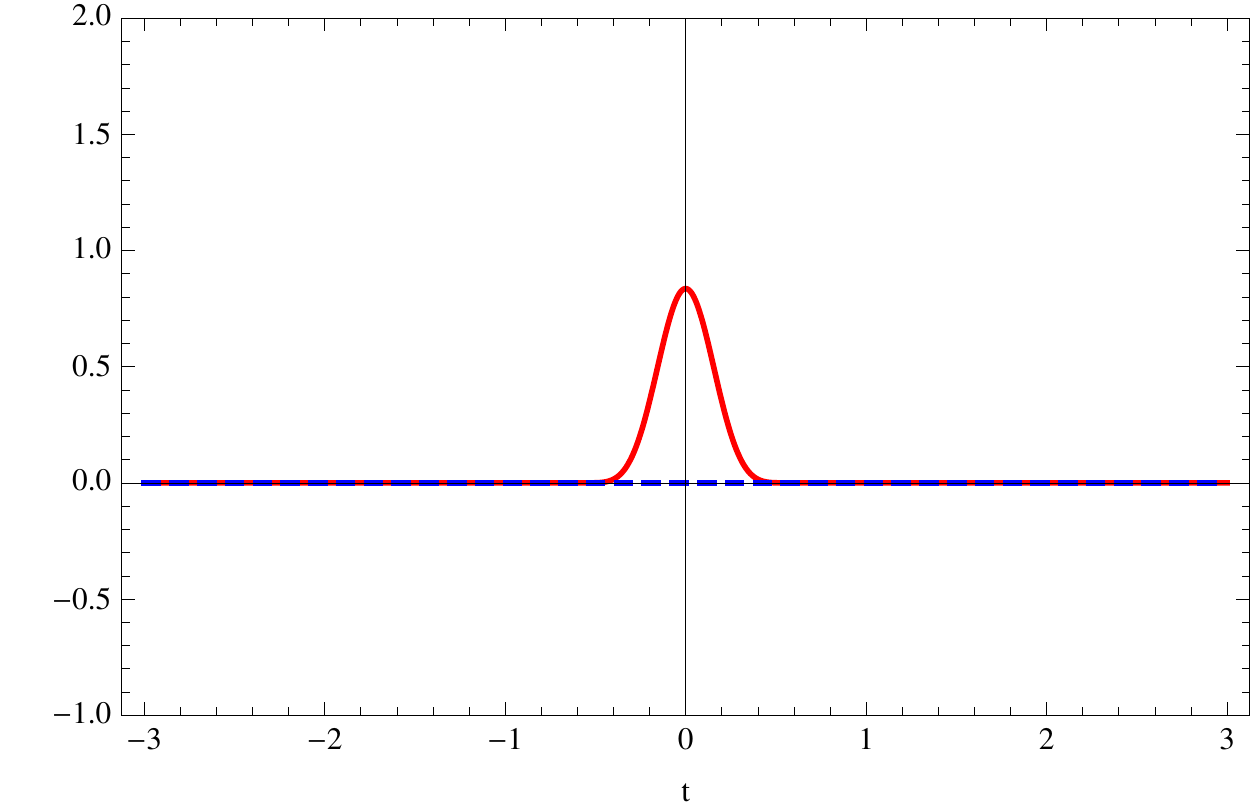}
\\[3mm]
(a)&(b)\\[3mm]
\end{tabular}
\end{minipage}
\caption{The real (red) and imaginary (dashed blue) parts of
the integrand $F_5(z,s)$ of \eqn{FlatAsymptotesIntegrand} for 
$s=\stdparam$ along (a) the [3/2] Pad\'e contour (b) the exact contour.}
\label{FlatAsymptotesIntegrandFigure}
\end{figure}

In the examples considered above, $N_{-}\neq 0$.  In contrast,
\begin{equation}
F_5(z,s) = 
(-s)^{-z} \frac{\Gamma^3(-z)\Gamma(1+z)}{\Gamma(-2z)\Gamma(1-z)\Gamma(2+z)}\,,
\label{FlatAsymptotesIntegrand}
\end{equation}
has $N_{-}=0$.  As a result,
the asymptotes of stationary-phase contours
will be parallel to the real axis.
The formul\ae{} derived in previous sections hold for this case,
but one must use the special forms for $N_{-}=0$ in \eqn{AsymptoticThetaEuclidean}.
The quadratic and [3/2] Pad\'e contours for $s=\stdparam$ are shown in 
\fig{FlatAsymptotesContourFigure}; and the behavior of the integrand
along the Pad\'e contour is shown in \fig{FlatAsymptotesIntegrandFigure}(a).
It has generic behavior, in spite of the special case needed for the contour.
In \fig{FlatAsymptotesIntegrandFigure}(b), we show the behavior along
the exact contour of stationary phase, computed using the
differential-equation approach discussed in \sect{DifferentialEquationSection};
in the latter, the imaginary part of the contour is again chosen to be $t$.
The absence of oscillations in the real part along the Pad\'e
contour, together with the
imaginary part being essentially zero in 
both parts of the figure attest to the good quality of approximation 
it furnishes.

\subsection{Closed Contours}
\label{ClosedContoursSection}

\begin{figure}[ht]
\begin{minipage}[b]{1.03\linewidth}
\begin{tabular}{cc}
\hskip -6mm
\includegraphics[clip,scale=0.5]{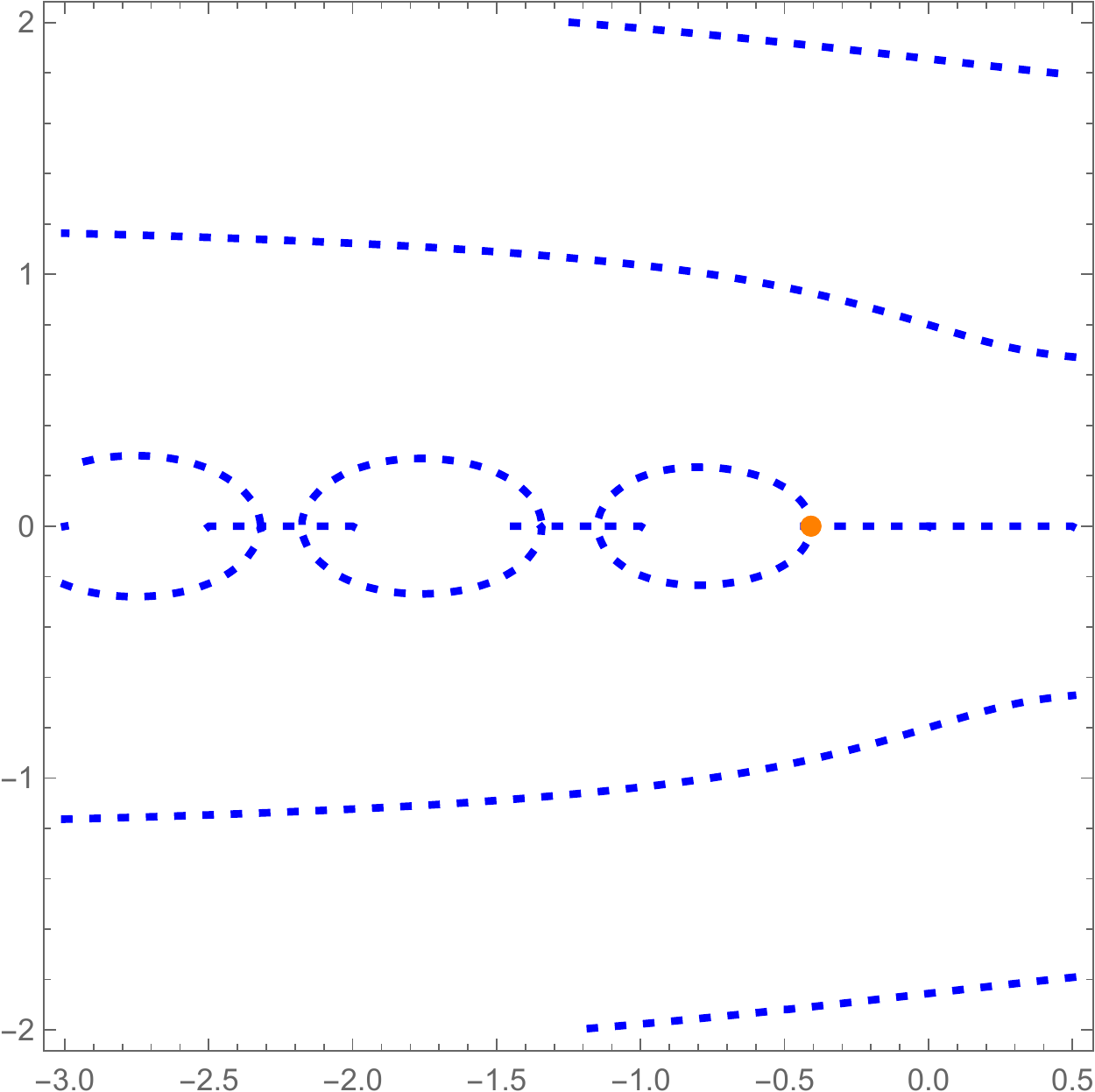}
\hspace*{10mm}
&\includegraphics[clip,scale=0.533]{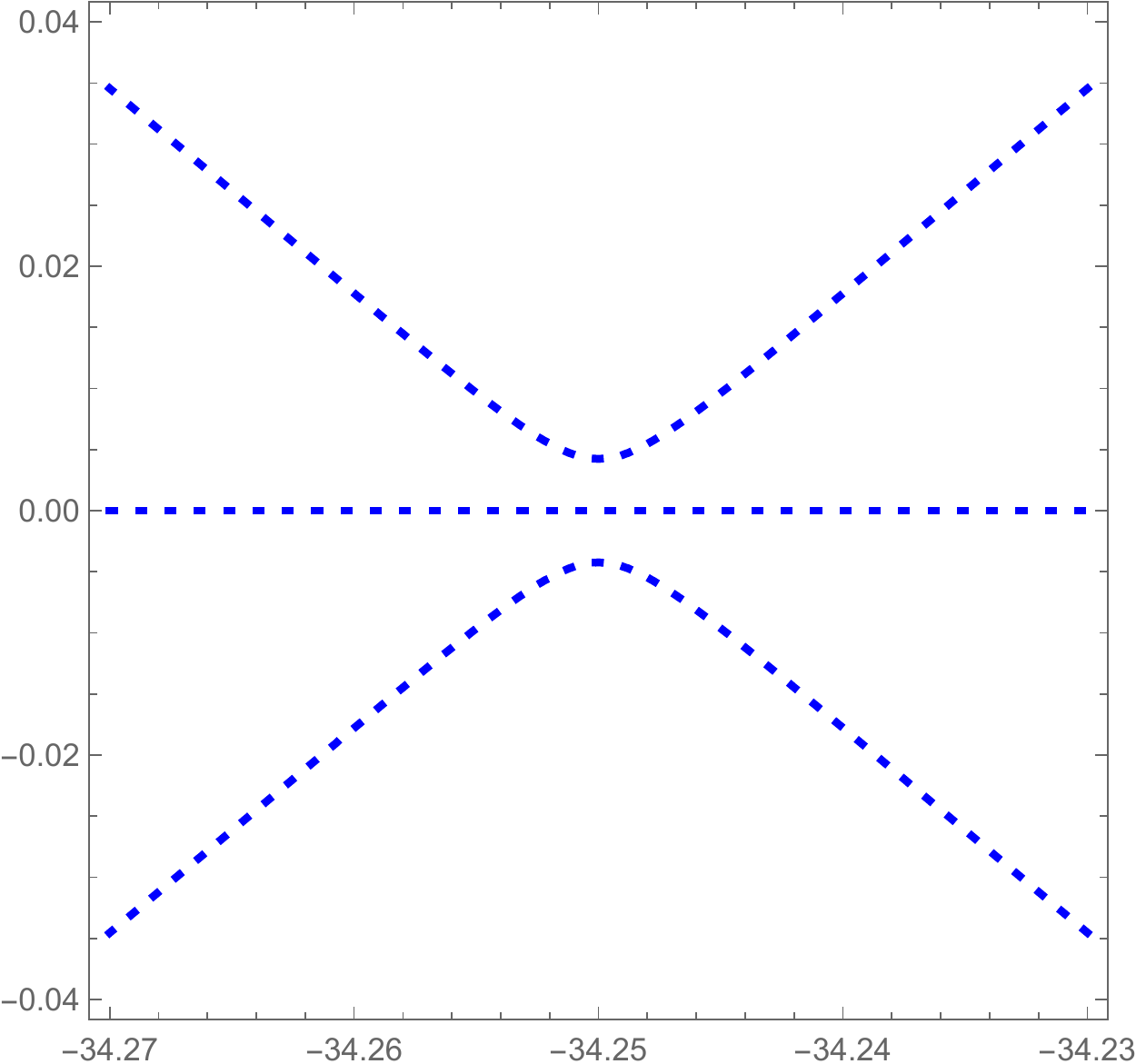}
\\[3mm]
(a)\hspace*{10mm}
&(b)\\[3mm]
\end{tabular}
\end{minipage}
\caption{Exact contours (dotted blue) of stationary phase
for the integrand
$F_6(z,s)$ of \eqn{ClosedContoursIntegrand}, with $s=-\frac18$
(a) at small $|z|$ (b) at larger $|z|$.
The saddle point in the interval $(-\frac12,0)$ is also shown (orange dot).
}
\label{ClosedContoursFigure}
\end{figure}

\begin{figure}[ht]
\includegraphics[clip,scale=0.66]{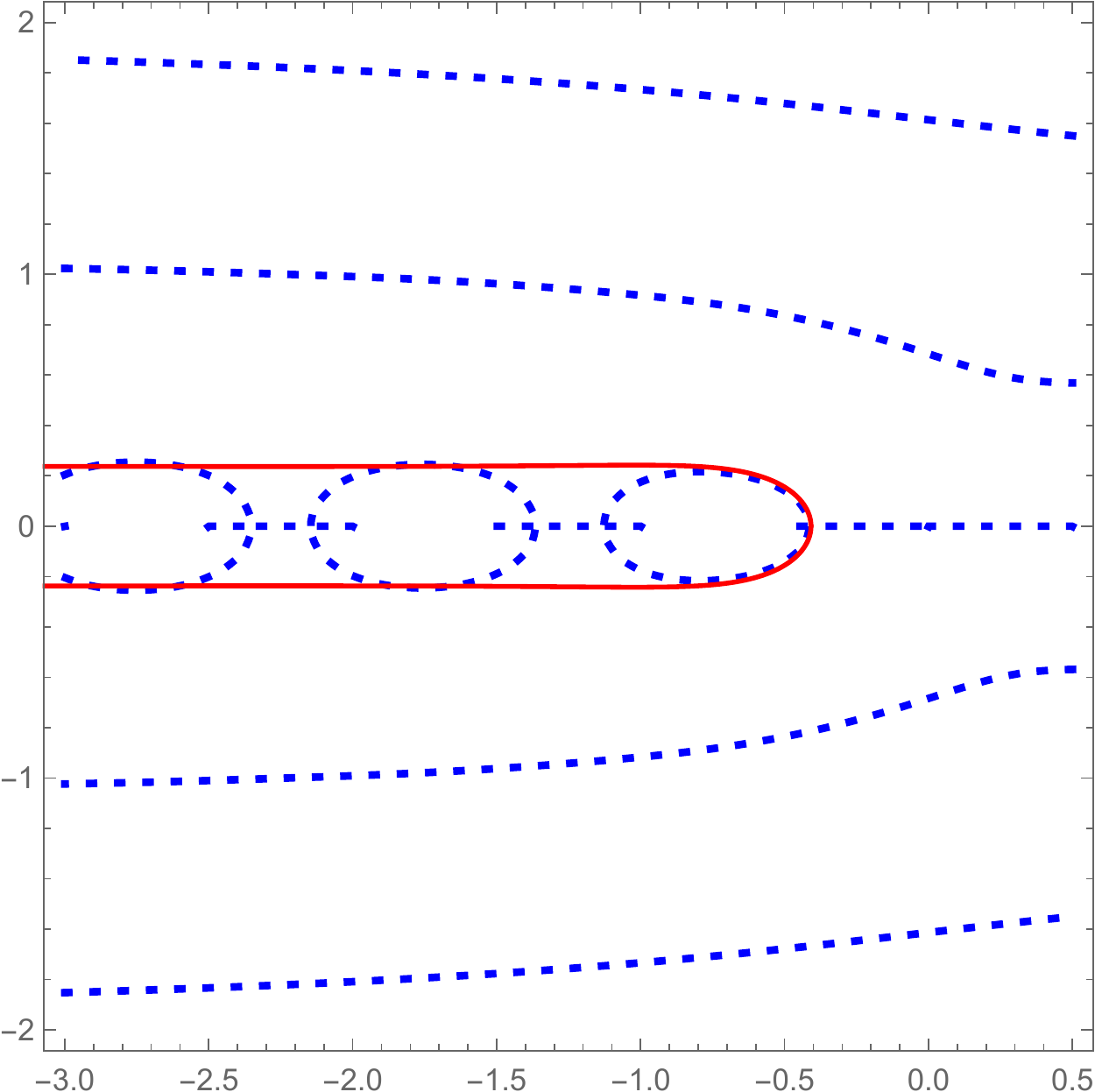}
\caption{The [3/2] Pad\'e
 (solid red)
contour for the integrand $F_6(z,s)$ of
\eqn{ClosedContoursIntegrand} with $s=-\frac18$ for 
$s=\stdparam$, shown against the exact contours of zero phase for the
integrand (dotted blue).}
\label{ClosedContourPadeFigure}
\end{figure}

\begin{figure}[t]
\includegraphics[clip,scale=0.66]{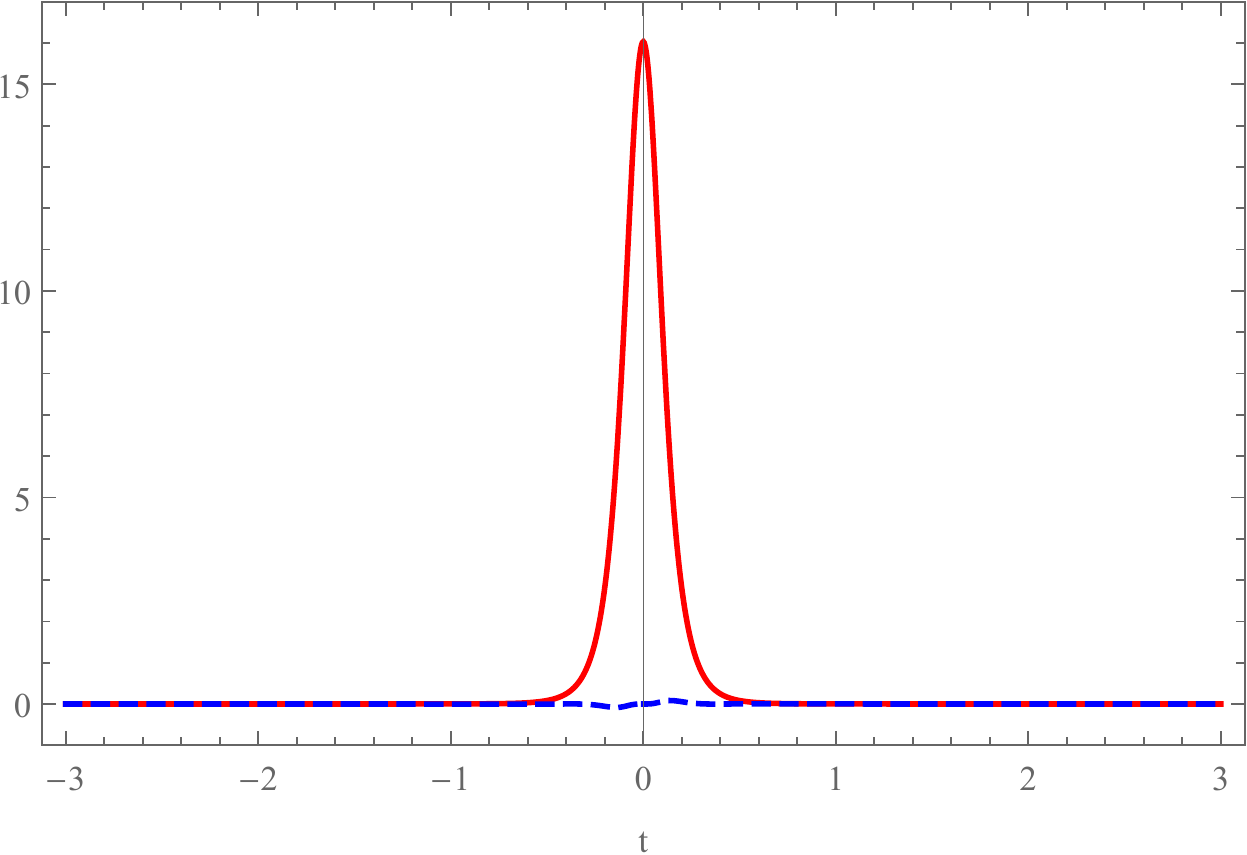}
\caption{The real (red) and imaginary (dashed blue) parts of
the integrand $F_6(z,s)$ of \eqn{ClosedContoursIntegrand} for 
$s=-\frac1{8}$ along the [3/2] Pad\'e approximation
to the contour of zero phase.}
\label{ClosedContourIntegrandFigure}
\end{figure}

The exact contour of stationary phase passing through a given
saddle point on the real axis may be closed: it may end
at a zero, or at another saddle point.  (Lines of steepest {\it ascent\/}
can also run into poles.)
The following integrand provides an example,
\begin{equation}
F_6(z,s) = (-s)^{-z}\frac{\Gamma^4(-z) \Gamma(1+2 z)}
                         {\Gamma^2(-2z)\Gamma^2(1+z)}\,.
\label{ClosedContoursIntegrand}
\end{equation}
For $s=-\frac18$, the integrand has a saddle point at,
\begin{equation}
z_s = -0.408258\,.
\end{equation}
For this integrand, saddle points come in pairs in each half-interval
$(-n-1/2,-n)$, $n\in \mathbb{Z}_+$, leading to a sequence of 
closed contours
for small $z$, as shown in \fig{ClosedContoursFigure}(a).  
Contours end on the rightmost of each pair, and begin at the leftmost.
As $n$ grows, the elements of the
pairs approach each other, and eventually move off the real axis.
At that point, we do get a single stationary-phase contour
enclosing all remaining poles.  This is illustrated for
$F_6(z,-\frac18)$ in \fig{ClosedContoursFigure}(b) (for $s=\stdparam$,
this only occurs for $z\gg 10^{12}$).
However, from a practical point
of view, using the exact contour (or ensemble of contours) is
pointless, as it would amount to summing over residues for the
bulk of the contribution.  One might as well compute the residues
analytically, and sum over them directly.  In contrast, the
approximate [3/2] Pad\'e contour interpolates between the stationary-phase
contour at the beginning, and the asymptotic contour (which in
this case is parallel to the real axis).  The Pad\'e contour obtained using
the forms described 
in \sects{ContourApproximationSection}{AsymptoticFormSection}
is shown in
\fig{ClosedContourPadeFigure}, and the value of the integrand along
this contour is shown in \fig{ClosedContourIntegrandFigure}.  It necessarily
has oscillations, but the overall fall-off, and hence the
expected convergence of integration, is rapid.

\section{Minkowski Integrals}
\label{MinkowskiSection}

\subsection{Below Threshold}
We now turn our attention to the evaluation of integrals
in the Minkowski region.  In this region, the naive textbook and
\MB{} contours yield integrals which are often only conditionally
(but not absolutely) convergent, and hence do not readily converge
numerically.  In some cases, special numerical techniques can
be brought to bear; but for most, we need to find a different contour.

\begin{figure}[ht]
\begin{minipage}[b]{1.03\linewidth}
	\begin{tabular}{cc}
		\hskip -6mm
	\includegraphics[clip,scale=0.6]{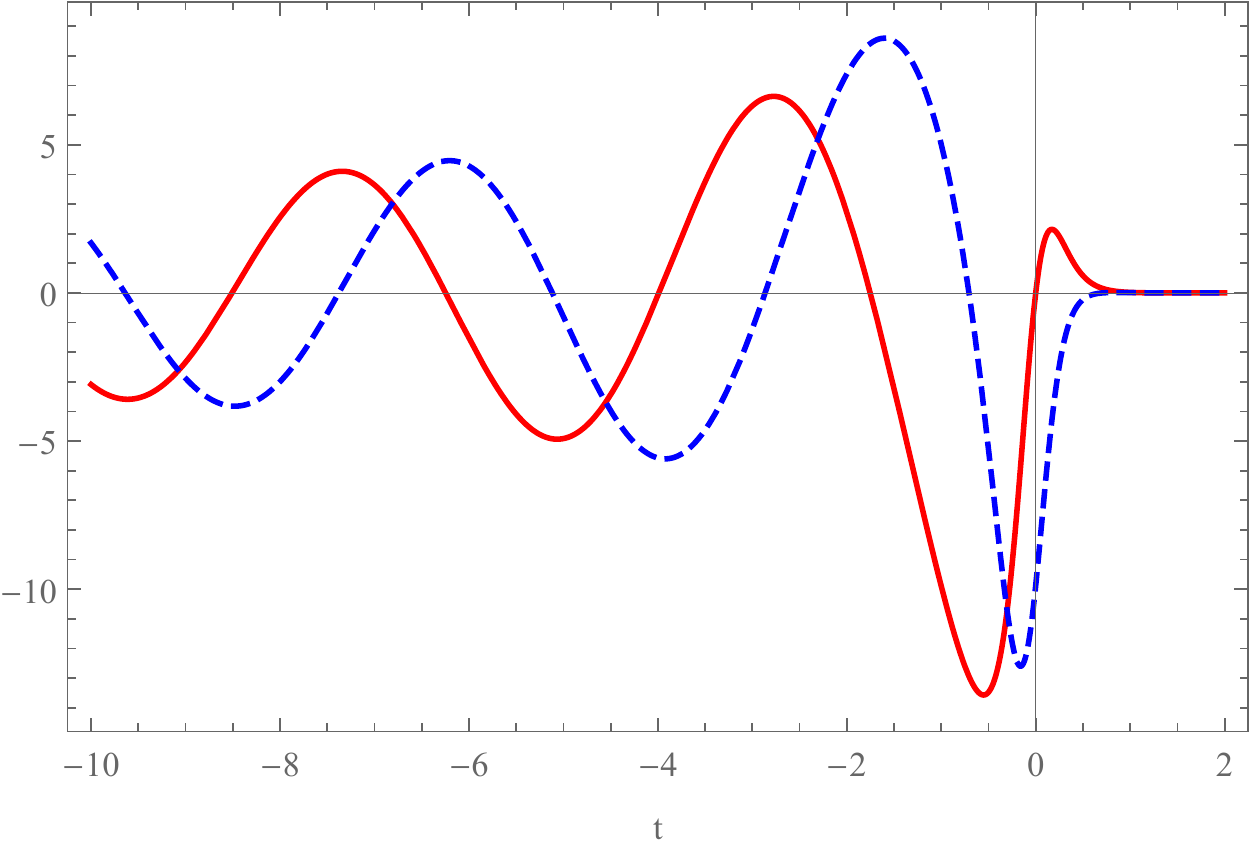}
		\hspace*{5mm}
	&\hspace*{5mm}
	 \includegraphics[clip,scale=0.6]{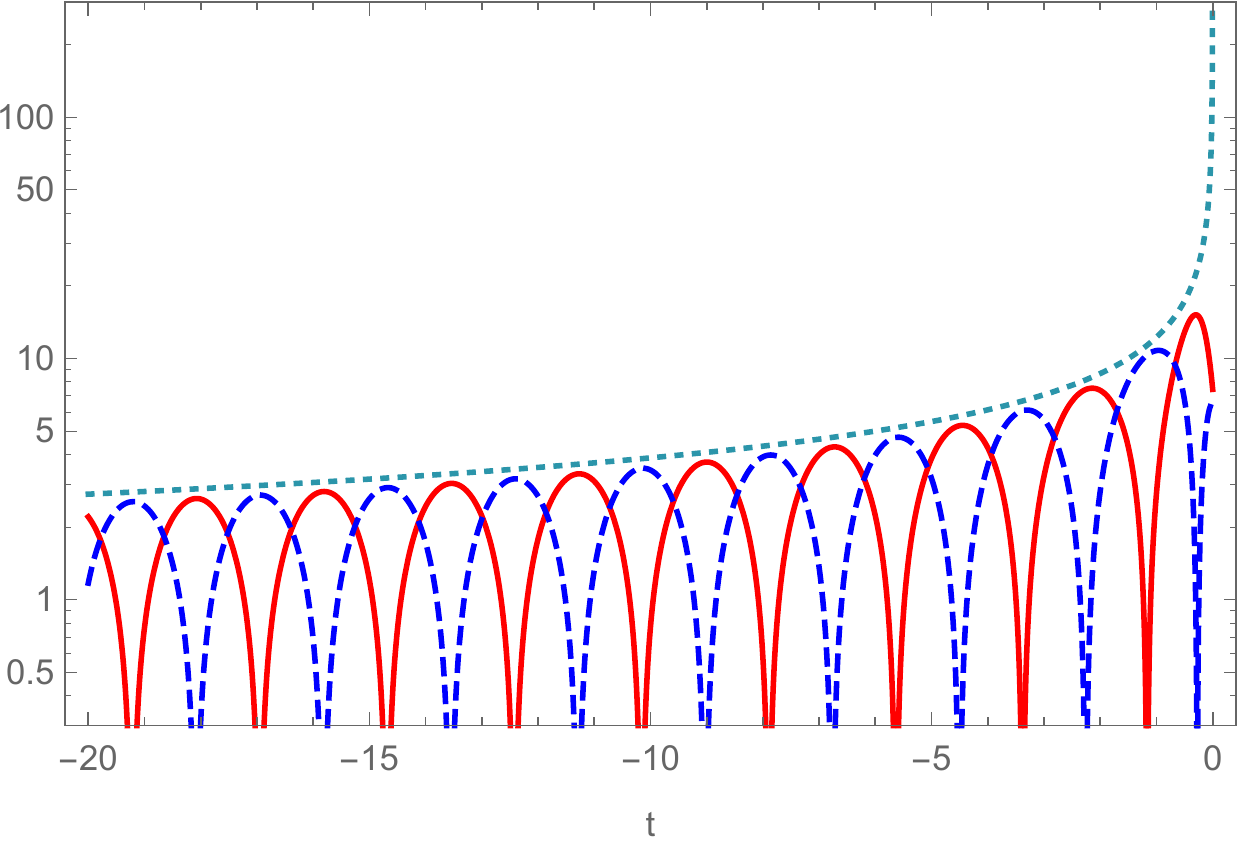}
		\\[3mm]
	(a)\hspace*{5mm}&\hspace*{5mm}(b)\\[3mm]
	\end{tabular}
\end{minipage}
\caption{The real (red) and imaginary (dashed blue) parts of
the integrand $F_1(z,s)$ of \eqn{OneDimensionalExampleIntegrand}, for 
$s=1+i\delta$, along the \MB{} contour $\Re z= \negonehalf$.  The parts are
shown on a linear scale in (a), and their absolute values
on a log scale in (b).  The dotted
(dark turquoise) curve in (b) is a curve decreasing as $t^{-1/2}$.}
\label{OneDimMinkowskiContourIntegrandA}
\end{figure}

We start with the example of \eqn{OneDimensionalExample}, but
now for $s>0$.  The integrand now has an imaginary
part even for real $z$, and correspondingly,
the integral may also have an imaginary part. 
The oscillations along
the naive contour $\Re z = \negonehalf$ are only slowly damped, as shown
in \fig{OneDimMinkowskiContourIntegrandA} for $s=1+i\delta$ (with
$\delta=10^{-10}$).  
This value is below the threshold at $s=4$, but straightforward 
numerical integration already fails to converge.

To find a better contour, we again seek a contour of 
stationary phase.  Unlike the
Euclidean case, however, this phase will not be zero; nor will
the corresponding saddle point sit on the real axis.  We seek
contours which pass from infinity (with large negative imaginary
part of $z$) through the saddle point and back to infinity (with large
positive imaginary part of $z$).  Even below threshold, we must add
an infinitesimal imaginary part to the parameter $s$ in order
to specify a contour; we will take it to be positive.
The integrand
of \eqn{OneDimensionalExampleIntegrand} has another feature which is generic,
but complicates the analysis: it vanishes for positive half-integer
values of $z$.  Zeros of the integrand complicate the analysis because
lines of stationary phase (and hence contours) can (and typically do)
end on them; this
would force us to look for half-contours, combining them with discontinuous
derivatives at the zero to obtain full contours.
We will treat that case in \sect{NoExtremaSection}.

Our first task is to find the saddle points, that is the points
where the derivative of the integrand vanishes.  When using 
{\sl Mathematica\/} to do this, it is best
to seek minima of the absolute value of
the derivative, rather than roots of the equations, as this
approach is more stable.   It is in any case helpful to bound the
search to a strip consisting of the imaginary extension of the
original integer interval containing the naive contour.

\begin{figure}[ht]
\includegraphics[clip,scale=0.66]{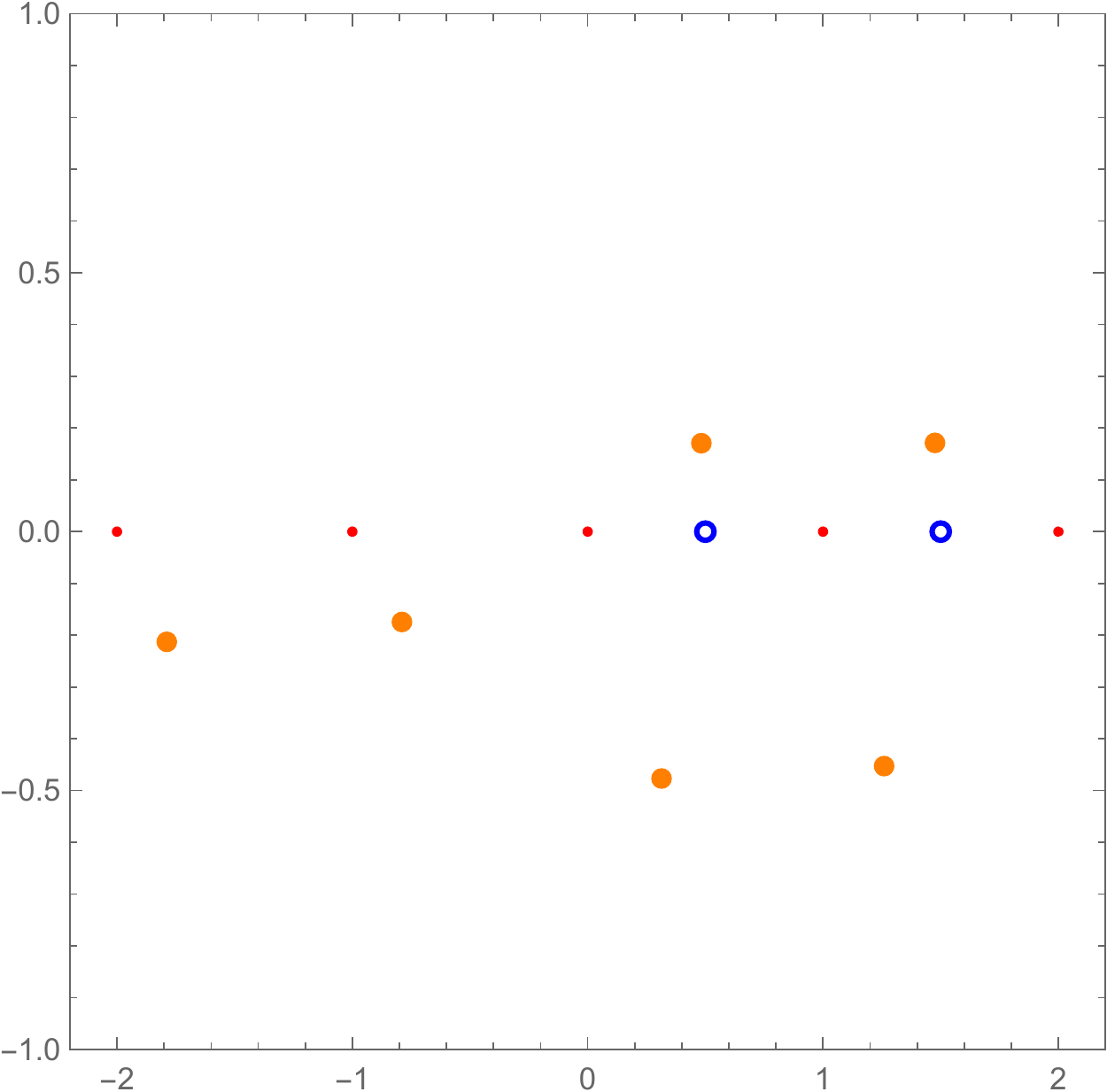}
\caption{The saddle points (large orange dots), poles (small red dots),
and zeros (blue circles)
of the integrand of \eqn{OneDimensionalExampleIntegrand} for 
$s=1+i\delta$.}
\label{SaddlePointsFigure}
\end{figure}

For $\Re z < 0$, we find a single series of solutions,
\begin{equation}
z = -0.78932 -0.174532 i\,,\qquad
-1.78841 -0.212806i\,,\ldots
\end{equation}

For $\Re z>0$, we find two series of solutions, one above the real axis,
the other below,
\begin{equation}
z = 0.313742 -0.476771i\,, 0.482908+0.17074 i\,;
1.25952 -0.452846i\,, 1.47558+ 0.171388i\,;\ldots
\end{equation}

These solutions, along with the poles and zeros of the integrand, are
shown in \fig{SaddlePointsFigure}.  The doubling of solutions for
positive $\Re z$ is directly related to the presence of nearby
real zeros at positive half-integer values.  These are stationary
points which would require
patching together two contours of stationary phase which would meet
at the associated zero.  The solutions for $\Re z<0$, in contrast,
 are associated with
a single contour running from infinity in to the stationary point,
and then back out to infinity.  Let us therefore set aside the
solutions for $\Re z>0$, and base a contour on the first solution in the
first set, $z_s = -0.78932 -0.174532 i$.  The phase at this point is
$e^{i\phi_s}$, with $\phi_s= -2.29000$.

\begin{figure}[ht]
\includegraphics[clip,scale=0.66]{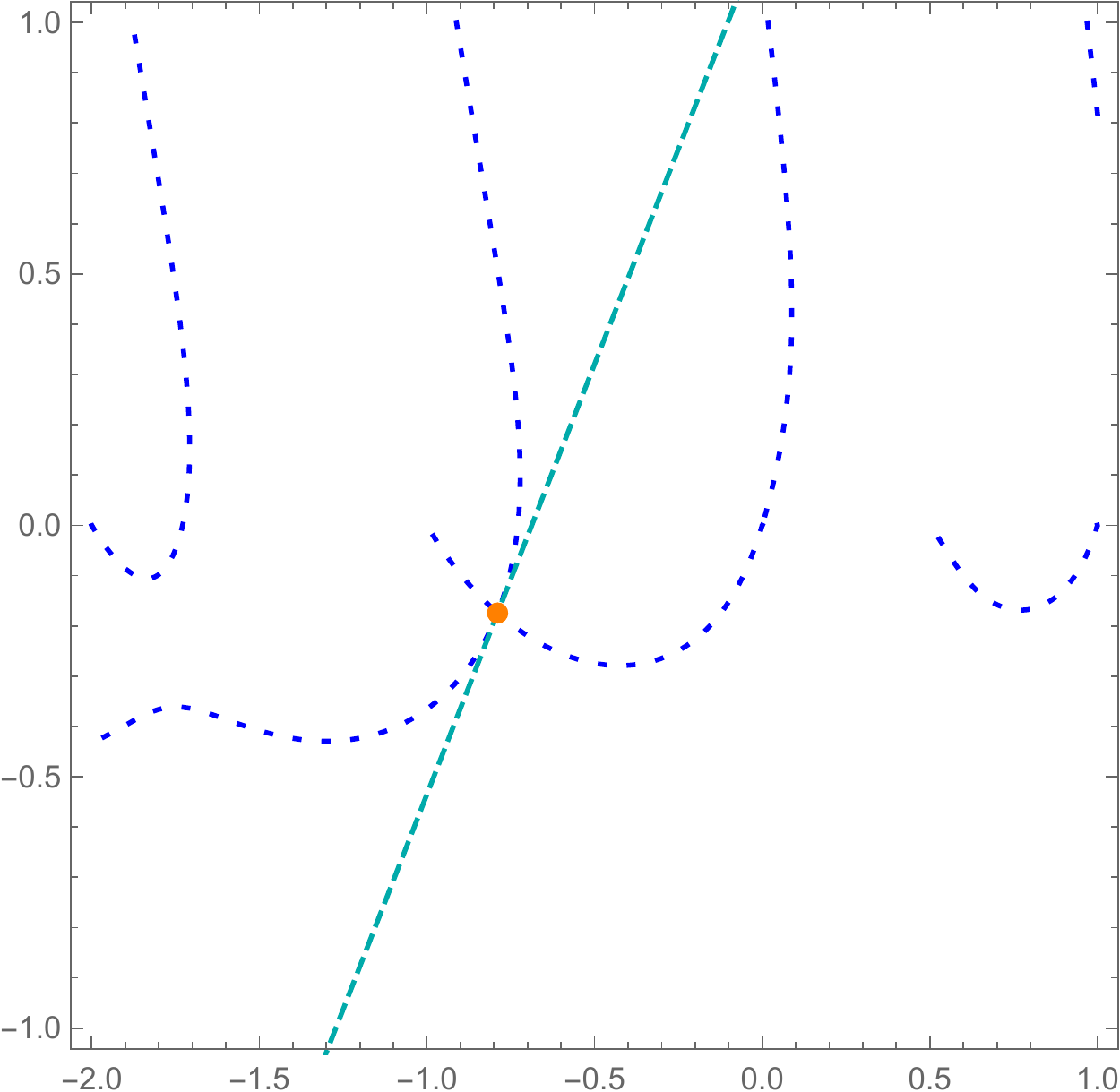}
\caption{Contours of constant phase $e^{i\phi_s}$ (dotted blue) for
the integrand $F_1(z,s)$ of \eqn{OneDimensionalExampleIntegrand} with
$s=1+i\delta$.  The saddle point is denoted by the large orange
dot.  The linear (tangent) approximation to the contour of stationary phase
is the dot-dashed (dark turquoise) line.}
\label{OneDimMinkowskiLinearContourFigure}
\end{figure}

\begin{figure}[ht]
\includegraphics[clip,scale=0.8]{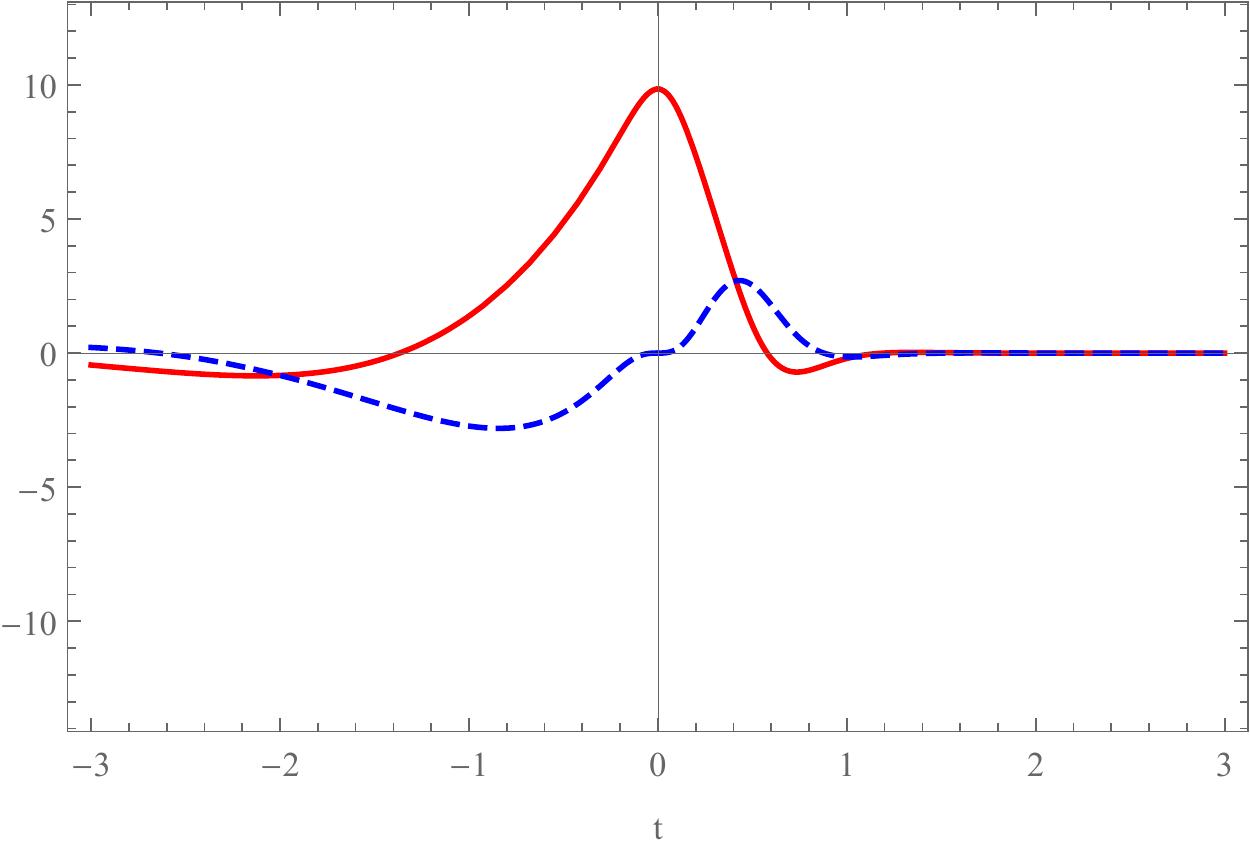}
\caption{The real (red) and imaginary (dashed blue) parts of
the integrand $F_1(z,s)$ of \eqn{OneDimensionalExampleIntegrand} with
$s=1+i\delta$, divided by the phase at the saddle point,
 along the linear contour of \eqn{OneDimMinkowskiLinearContour}.}
\label{OneDimMinkowskiContourIntegrandB}
\end{figure}

Let us again begin by finding a linear approximation to the contour of
stationary phase, given by the tangent to the contour at the
saddle point.  Writing,
\begin{equation}
z(t) = z_s + e^{i\theta_s} (x(t) + i y(t))\,,
\label{TiltedGeneralContour}
\end{equation}
where $x(0) = y(0) = 0$, without loss of generality we can
again take $x'(0) = 0$ and $y'(0) = 1$.  Because $z_s$ is a saddle
point, the expansion of the integrand $F$ around $t=0$ has no
linear term,
\begin{equation}
F = F(z_s) - \frac12 F''(z_s) e^{2 i\theta_s} t^2\,.
\end{equation}
We require the integrand to be of stationary phase along this contour,
\begin{equation}
-\frac12 \Im \biggl[\frac{F''(z_s) e^{2 i\theta_s}}{F(z_s)}\biggr] = 0\,.
\label{OneDimMinkowskiLinearAngleEquation}
\end{equation}
We can solve for $\theta_s$,
\begin{equation}
\theta_s = -\frac12 \arg\biggl[\frac{F''(z_s)}{F(z_s)}\biggr]\,.
\label{OneDimMinkowskiLinearAngleSolution}
\end{equation}
\Eqn{OneDimMinkowskiLinearAngleEquation} also allows for solutions
shifted by $\pi n/2$.  Shifting by $\pi$ is harmless; it just amounts
to exchanging $t\leftrightarrow -t$ in \eqn{TiltedGeneralContour}.  To
fix an orientation, we can
adopt the convention that $\theta_s$ should lie in the interval
$[-\frac{\pi}{2},\frac{\pi}{2}]$.  In order
to select between $\theta_s$ and $\theta_s+\pi/2$, we should pick
the direction in which the absolute value of the function decreases
in magnitude.  (This is equivalent to requiring that 
$\Re[e^{2i\theta_s} F''(z_s)/F(z_s)]$ be negative.)

In the example at hand, the this linear contour has the form,
\begin{equation}
z_t(t) = -0.78932 - 0.174532 i + (0.504583 + 0.863363 i) t\,.
\label{OneDimMinkowskiLinearContour}
\end{equation}
It is shown in \fig{OneDimMinkowskiLinearContourFigure}.
The real and imaginary parts of the integrand along this contour,
with the phase $e^{i\phi_s}$ at the saddle point divided out, are
shown in \fig{OneDimMinkowskiContourIntegrandB}; although they still
oscillate, the oscillations are damped, and the integral can be computed
numerically.  This contour is an example of the kind of contour
proposed by Freitas and Huang~\cite{FreitasHuang}, 
though their approach does not use
the criterion used here to determine $\theta_s$.  (In any case, the 
solution given here
for the tangent contour does not give a numerically stable integral
for all values of $s$ or all integrands.)

\begin{figure}[ht]
\includegraphics[clip,scale=0.66]{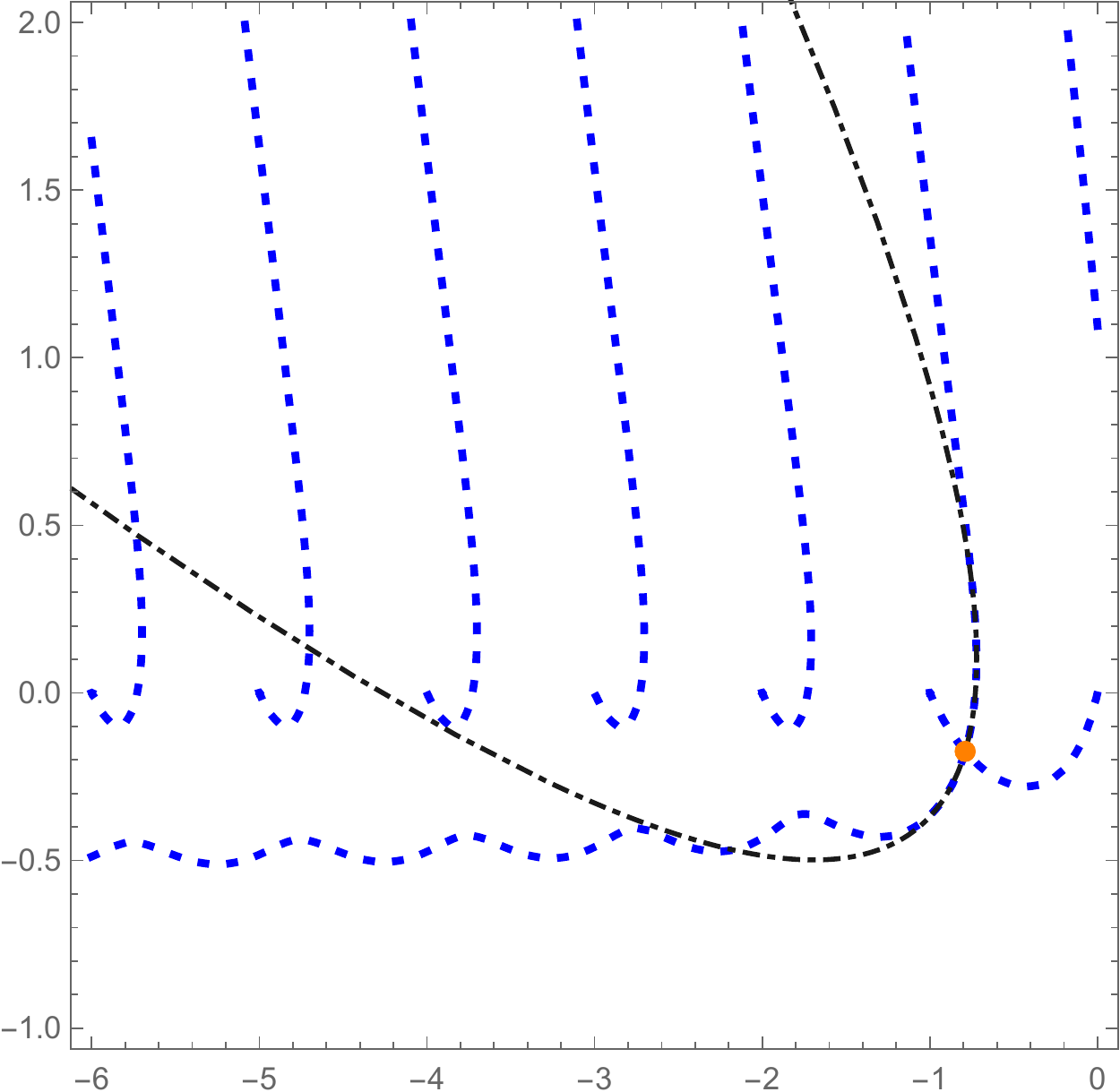}
\caption{
The quadratic approximation~(\ref{QuadraticContourM2}) 
to the stationary-phase contour (solid red)
for the integrand of \eqn{OneDimensionalExample} with
$s=1+i\delta$.
The exact contours of constant phase $e^{i\phi_s}$ are also shown
 (dotted blue).  
The saddle point is indicated by a large orange dot.
}
\label{OneDimMinkowskiQuadraticContourFigure}
\end{figure}

\def\dratioII{e^{2i\theta_s} \frac{F''(z_s)}{F(z_s)}}
\def\dratioN#1#2{e^{#1 i\theta_s} \frac{F^{(#2)}(z_s)}{F(z_s)}}
\def\tdratioII{e^{2i\theta_s} F''(z_s)/F(z_s)}
\def\tdratioN#1#2{e^{#1 i\theta_s} F^{(#2)}(z_s)/F(z_s)}
As in the Euclidean case, we can improve the contour further. 
It will be convenient to define an abbreviation,
\begin{equation}
D_n \equiv e^{i n\theta_s} \frac{F^{(n)}(z_s)}{F(z_s)}
\end{equation}
Let
us examine the stationary-phase
equation resulting from expanding the integrand to one higher order,
\begin{equation}
\begin{aligned}
&-\frac12 t^2 \Im\bigl[D_2\bigr]
-\frac16 t^3 \Re\bigl[D_3\bigr]
\\
&+\frac12 t^3 \Re\bigl[D_2\bigr] x''(0)
-\frac12 t^3 \Im\bigl[D_2\bigr] y''(0) = 0\,.
\end{aligned}
\end{equation}
Using the solution to the lower-order 
equation~(\ref{OneDimMinkowskiLinearAngleSolution}),
which forces $D_2$ to be real,
we can simplify this equation to obtain,
\begin{equation}
x''(0) = \frac13 \Re\bigl[D_3\bigr]
         \Big/  D_2\,.
\label{OneDimMinkowskiQuadraticSolution}
\end{equation}
so that
\begin{equation}
z_q(t) = z_0 + e^{i\theta_s} \bigr(i t+c_2 t^2\bigr)\,.
\label{OneDimMinkowskiQuadraticContour}
\end{equation}
with $c_2$ set to $z_2\equiv x''(0)/2$.  This generalizes 
\eqn{QuadraticCoefficient} to the Minkowski region.
In the example at hand, this would lead to the quadratic contour,
\begin{equation}
z_q(t) = (-0.78932 - 0.174532 i) 
+ (0.863363 - 0.504583 i) (i t - 1.09478 t^2)
\,.
\label{QuadraticContourM}
\end{equation}
Unlike the Euclidean case, this quadratic contour does not automatically
make the phase stationary through $\Ord(t^4)$; setting the imaginary
part of that order to zero, we find,
\begin{equation}
y''(0) = \frac{\Im\bigl[D_4\bigr]}{4 \Re\bigl[D_3\bigr]}
        -\frac{\Im\bigl[D_3\bigr]}{2 D_2 \vphantom{\bigl[\bigr]}}
\end{equation}
and set $c_2 = x''(0)/2+i y''(0)/2$.
In the example at hand, this leads to the quadratic contour,
\begin{equation}
z_q(t) = (-0.78932 - 0.174532 i) 
+ (0.863363 - 0.504583 i) (i t - (1.09478-0.026705 i) t^2)
\,,
\label{QuadraticContourM2}
\end{equation}
which is very similar to the one given in \eqn{QuadraticContourM}.
The solution for $y''(0)$ depends on truncating the contour
at the quadratic order; otherwise, $x^{(3)}(0)$ would also appear 
in the equation.

For small $t$, both of these contours provide an excellent approximation to the
true contour of steepest descent, and essentially eliminate oscillations
in the integrand.
They suffer from a problem, however, illustrated in
\fig{OneDimMinkowskiQuadraticContourFigure}: because the first is symmetric
around a line inclined to the real axis, and the second nearly so, 
at large negative $t$ they will cross the
real axis, thus failing to include the contributions of
the remaining series of poles at larger negative $z$.  The numerical
contributions of the corresponding residues are small, but this truncation
is uncontrolled: no matter how many points we throw at the integration, we
can never obtain the correct answer.  

\def\thinfp{\theta_{+\infty}}
\def\thinfm{\theta_{-\infty}}
\def\thinfpm{\theta_{\pm\infty}}
\def\mod{\mathop{\rm mod}\nolimits}
In order to solve this problem, we need to modify the behavior of the
contour at larger $t$.  One way of doing this is to match to the
asymptotic form contour for large $z$.  In this case, the two asymptotic
regions, $t\rightarrow +\infty$ and $t\rightarrow-\infty$ are no longer
complex conjugates, and so we will have two different angles, which
we denote $\thinfp$ and $\thinfm$ respectively.  

To find these angles, we generalize the discussion of
the asymptotic form of the integrand in \sect{AsymptoticFormSection}.
The only difference is in the $(-s)^{-z}$ factor; the phase now
comes from the real part of $z$ as well as the imaginary part.  We find
an additional contribution to $\arg F$ beyond that given 
in \eqn{GeneralPhaseExpression}, so that
\begin{equation}
\begin{aligned}
\arg F &= \Im z\ln\bigg|\frac{s_0}{-s}\bigg|+\pi \Re z 
(\Nm\sign\Im z+\sign\Im s)
-\frac{\pi}2 \bigl((S_{+}-2 A_{+}) \bmod 4\bigr) \sign\Im z
\\
&+\bigl[\bigl(-S_{+}/2-S_{-}/2+A_{+}+A_{-}\bigr) \arg(-z)\bigr] \bmod 2\pi\,.
\end{aligned}
\label{GeneralMinkowskiPhaseExpression}
\end{equation}

Writing the large-$z$ forms for the stationary-phase contour as follows,
\begin{equation}
\begin{aligned}
z &\simas^{t\rightarrow +\infty} \zinf + i e^{i\thinfp} t\,,
\quad t>0\,,\\
z &\simas^{t\rightarrow -\infty} \zinf + i e^{-i\thinfm} t\,,
\quad t<0\,,
\end{aligned}
\label{MinkowskiAsymptoticForm}
\end{equation}
substituting into the asymptotic form for $\arg F$, and requiring
the coefficient of $t$ in the large-$t$ expansion
to vanish, leads to the following formul\ae{}
for $\thinfpm$,
\begin{equation}
\begin{aligned}
\thinfp &= 
\atan\biggl[\frac1{\pi (\Nm+\sign\Im s)}\ln\biggl|\frac{s_0}{-s}\biggr|
\biggr]\,,&&\qquad \Nm \neq -\sign\Im s\,,\\
\thinfp &= \sign\ln\biggl|\frac{s_0}{-s}\biggr|\,\frac{\pi}2\,,
&&\qquad \Nm = -\sign\Im s\,,\\
\thinfm &= 
\atan\biggl[\frac1{\pi (\Nm-\sign\Im s)}\ln\biggl|\frac{s_0}{-s}\biggr|
\biggr]\,,&&\qquad \Nm \neq \sign\Im s\,,\\
\thinfm &= \sign\ln\biggl|\frac{s_0}{-s}\biggr|\,\frac{\pi}2\,,
&&\qquad \Nm = \sign\Im s\,.
\end{aligned}
\label{MinkowskiAsymptoticAngles}
\end{equation}
These expressions reduce to the Euclidean
results~(\ref{AsymptoticThetaEuclidean})
so long as we take $\sign\Im s$ to be 0 in the latter region.
To find $\zinf$, we must again
require that the $t^0$ term in the large-$t$ expansion
be equal to the phase at the stationary point, $\phi_s=\arg F(z_0)$. 
In the generic case,
when $\Nm\neq 0$, 
we can do this simultaneously for the $t\rightarrow \pm\infty$
limits, thereby obtaining a pair of equations and solving
for the real and imaginary parts independently,
\begin{equation}
\begin{aligned}
\Re \zinf &=
  +\frac1{2 \Nm} {(S_{+}-2 A_{+}-2\DP_{-}-2\SP_{+})\bmod 4}
\\&\hphantom{=}
 -\frac1{2 \pi \Nm}
   \bigl[\bigl(A_{+}+A_{-}-S_{+}/2-S_{-}/2-\DP_{+}-\DP_{-}\bigr) 
         \bigl(\thinfp+\thinfm-\pi\bigr)
 \bmod 2\pi\bigr]\,,\\
\Im \zinf &= \frac{\phi_s}{\ln|s_0/s|}
-\frac{\pi \sign\Im s}{2\Nm \ln |s_0/s|}\,
   (S_{+}-2 A_{+}-2\DP_{-}-2\SP_{+})\bmod 4
\\&\hphantom{=}
 +\frac1{2 \ln |s_0/s|}
\bigl[\bigl(A_{+}+A_{-}-S_{+}/2-S_{-}/2-\DP_{+}-\DP_{-}\bigr) 
     (\thinfm-\thinfp)\bmod2\pi\bigr]
\\&\hphantom{=}
 +\frac{\sign\Im s}{2 \Nm\ln |s_0/s|}
\bigl[\bigl(A_{+}+A_{-}-S_{+}/2-S_{-}/2-\DP_{+}-\DP_{-}\bigr)
      (\thinfp+\thinfm-\pi)\bmod2\pi\bigr] \,.
\end{aligned}
\end{equation}
In deriving these formul\ae{}, we have again implicitly used the
condition that $\thinfpm\in[-\frac{\pi}2,\frac{\pi}2]$.
As in the Euclidean case, these values may be shifted in order
to match the appropriate asymptote,
\begin{equation}
\delta \zinf = \frac{n_1-n_2}{N_{-}}
+i \frac{\pi(n_1+n_2)}{\ln |s_0/s|}
+i \frac{\pi(n_2-n_1)\sign\Im s}{N_{-}\ln |s_0/s|}\,,
\end{equation} 
where $n_{1,2}$ are integers.

We postpone a discussion of the $\Nm=0$ case to \sect{NoExtremaSection}.

The differing asymptotic forms require us to generalize the 
[3/2] Pad\'e form (\ref{PadeContour}) to,
\def\ah{{\hat a}}\def\bh{{\hat b}}
\begin{equation}
z_p(t) = z_s+ i e^{i\theta_s} t+\frac{e^{i\theta_s} t^2(\ah_2 + i\bh_2 \ah_3 t)}
     {1+i\bh_1 t+ \bh_2 t^2}\,.
\label{PadeContourM}
\end{equation}
Matching to a quadratic contour at small $t$ requires three coefficients,
and to a linear asymptotic contour at large $t$ an additional two coefficients
(more precisely, one complex coefficient and one phase).
A [2/1] Pad\'e approximation does not have enough free coefficients to
match both limits, so a [3/2] Pad\'e approximation is the simplest possible
one.  (A similar result is true in the Euclidean region, though the argument
is more subtle.)  However, a [3/2] Pad\'e form has one additional parameter,
that we can use to fix the cubic terms in the contour as well, so
as to make the $\Ord(t^5)$ (and in principle the $\Ord(t^6)$)
terms in the expansion of
the integrand have the phase of the stationary point as well.
As we are not truncating the contour at cubic order, however, the
equation for $y''(0)$ also involves $x^{(3)}(0)$, and additional
equations also involve higher derivatives.  In order to simplify
the structure of the equations, it is convenient to perform
a nonlinear transformation to a new set of parameters 
$\{\alpha_i,\beta_i\}$ via,
\begin{equation}
\begin{aligned}
a_s &= \alpha_3 e^{-i\theta_s}-1\,,\\
d_M &= \alpha_2 (z_s-\beta_1) + e^{i\theta_s} a_s^2
\,,\\
\ah_2 &= \alpha_2\,,\\
\ah_3 &= a_s\,,\\
\bh_1 &= i d_M^{-1} (\beta_2 (z_s-\beta_1) 
                   +i\alpha_2 e^{i\theta_s}a_s)
                   \,,\\
\bh_2 &= -d_M^{-1} e^{i\theta_s} (\alpha_2^2
                  +i \beta_2 a_s)\,,\\
\end{aligned}
\end{equation}
In these equations, $\theta_s$ is given 
by \eqn{OneDimMinkowskiLinearAngleSolution}; matching
the asymptotic behavior and taking $\thinfpm$ from 
\eqn{MinkowskiAsymptoticAngles}, we can fix $\alpha_3$ up to
an overall magnitude,
\begin{equation}
\alpha_3 = \rho_3 \bigl(e^{i\thinfp}\Theta(t)+e^{i\thinfm}\Theta(-t)\bigr)
\,,
\label{Alpha3Solution}
\end{equation}
as well as $\beta_1$,
\begin{equation}
\beta_1 = \zinf\,.
\end{equation}
The improvements from adjusting $\rho_3$ are marginal, and
trying to solve for it requires solve much higher-order polynomial
equations, so
we again simply fix it to $1$.  
Requiring the integrand to be of stationary phase through cubic
order fixes the real part of $\alpha_2$,
\begin{equation}
\Re\alpha_2 = \frac{\Re D_3}{6 D_2} = \Re c_2\,.
\label{ReAlpha2Solution}
\end{equation}
(Recall that $D_2$ is real by construction.)
Requiring the integrand to be of stationary phase through quartic
order fixes the imaginary part of $\alpha_2$ in terms of
derivatives of the integrand along with $\beta_2$,
\begin{equation}
\Im\alpha_2 = \frac{\Im D_4}{8\Re D_3}-\frac{\Im D_3}{4 D_2}
+\frac{3 D_2}{\Re D_3} \Re\beta_2\,.
\label{ImAlpha2Solution}
\end{equation}
Requiring stationarity through $\Ord(t^6)$ would give an
additional pair of equations for the real and imaginary parts
of $\beta_2$.  However, these equations are of rather high
order, and do not always admit solutions.  Furthermore,
the solutions to these equations may yield contours with
loops.  The best approach to fixing $\beta_2$ appears to be minimizing
a weighted sum of the square of the following
deviation from stationarity at quintic order,
\begin{equation}
\begin{aligned}
&-\frac{3 D_2^2}{2\Re D_3} (\Re\beta_2)^2
+\biggl(\frac{D_2\Im D_4}{4\Re D_3}-\frac12\Im D_3\biggr)\,\Re\beta_2
-\frac13\Re D_3\,\Im\beta_2
\\& -\frac{\Im D_3\Im D_4}{32 D_2}
+ \frac{5 (\Im D_4)^2}{384 \Re D_3} 
+ \frac{(\Im D_3)^2 \Re D_3}{96 D_2^2} 
+ \frac{(\Re D_3)^3}{72 D_2^2}
- \frac{\Re D_3 \Re D_4}{36 D_2} 
\\&+ \frac1{120} \Re D_5
+ D_2\,\Re\biggl[\frac{ \alpha_2^3  +
2 i \alpha_2 \beta_2 a_s + \beta_2^2 e^{-i\theta_s} (z_s-\zinf)}
{a_s^2 + \alpha_2 e^{-i\theta_s} (z_s-\zinf)}\biggr]\,,
\end{aligned}
\label{FifthOrderDeviation}
\end{equation}
the square of the relative phases of the denominator terms,
\begin{equation}
\arg(-i \bh_2/\bh_1)\,,
\end{equation}
and the square of the relative phases of the numerator terms,
\begin{equation}
\arg(-i \ah_2/(\bh_2\ah_3))\,.
\end{equation}
The minimization is over the real and imaginary parts of $\beta_2$,
after
substituting in eqs.~(\ref{Alpha3Solution},%
\ref{ReAlpha2Solution},\ref{ImAlpha2Solution}).
A good heuristic weights the quintic
significantly more than the denominator's relative phase, which
in turn is weighted more than the numerator's relative phase.
Because $\alpha_3$, which depends on the sign of $t$,
 appears implicitly on the right-hand side
of \eqn{FifthOrderDeviation} as well explicitly in the
relative phases, 
all parameters will likewise acquire a dependence on that sign.

\begin{figure}[ht]
\includegraphics[clip,scale=0.66]{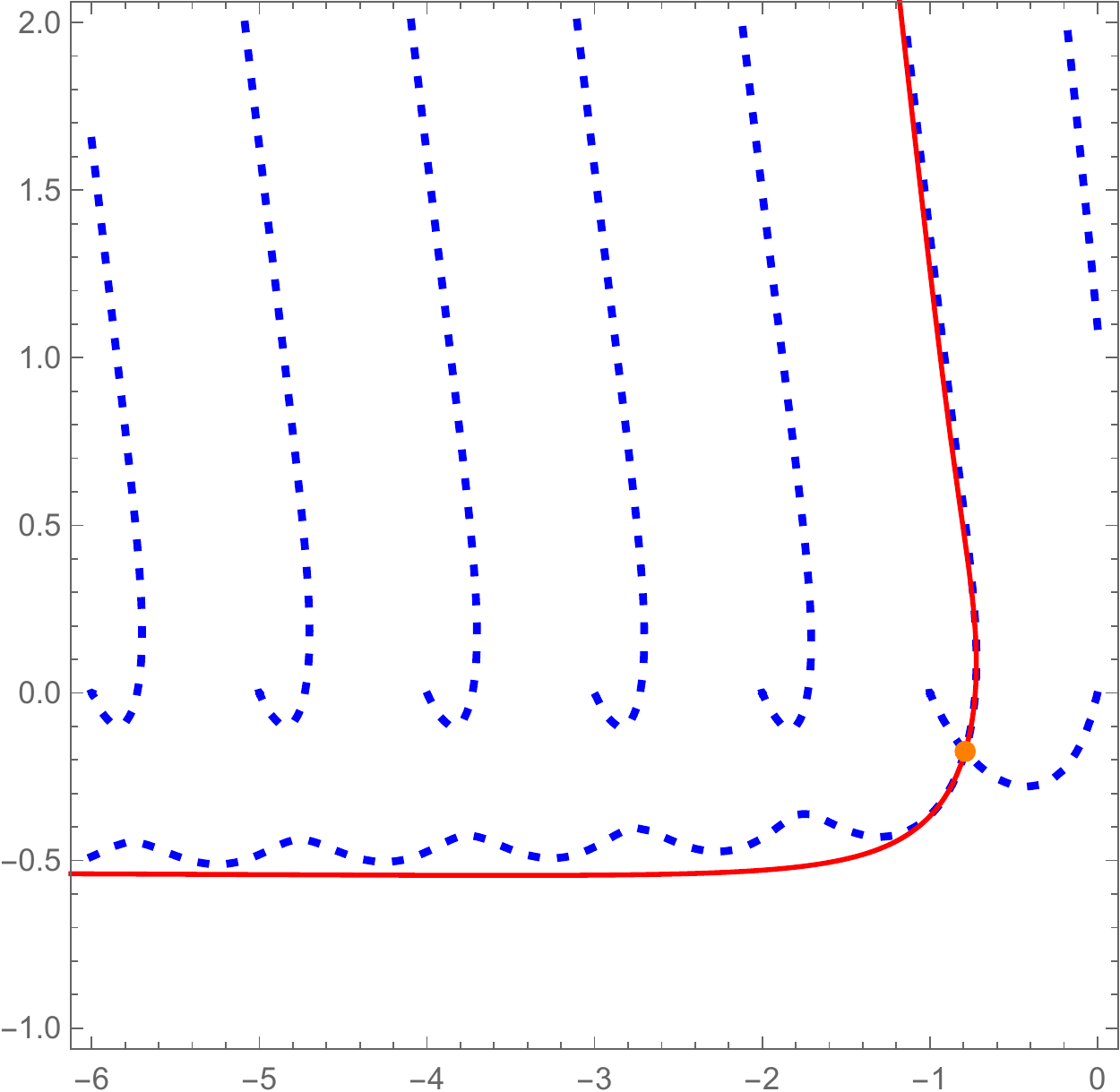}
\caption{
The [3/2] Pad\'e approximation to the contour of stationary phase $e^{i\phi_s}$
(solid red) for
the integrand $F_1(z,s)$ of \eqn{OneDimensionalExampleIntegrand} with
$s=1+i\delta$. The exact contours of this phase are
also shown (dotted blue).  
The saddle point is indicated by a large orange dot.
}
\label{OneDimMinkowskiPadeContourFigure}
\end{figure}

\begin{figure}[t]
\begin{minipage}[b]{1.03\linewidth}
\begin{tabular}{cc}
\hskip -6mm
\includegraphics[clip,scale=0.60]{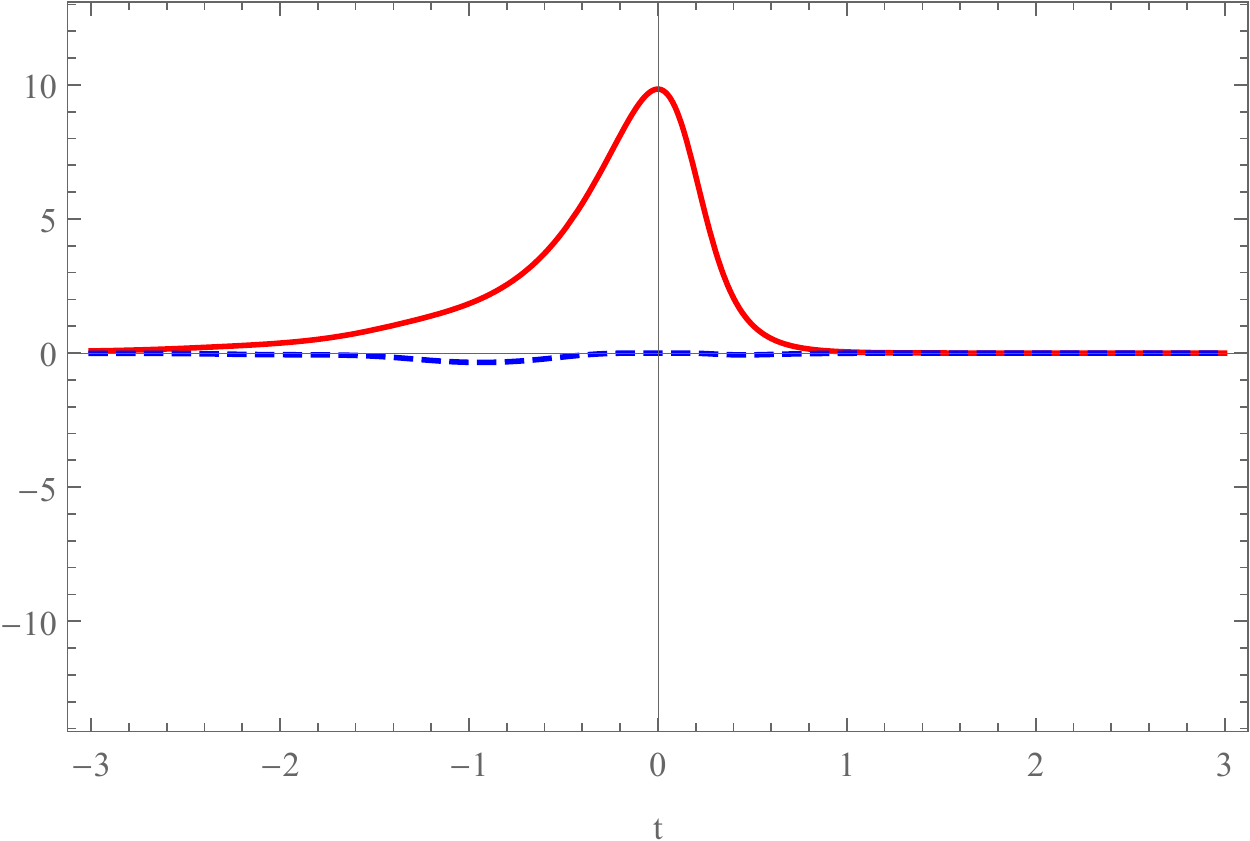}
\hspace*{6mm}
&\includegraphics[clip,scale=0.63]{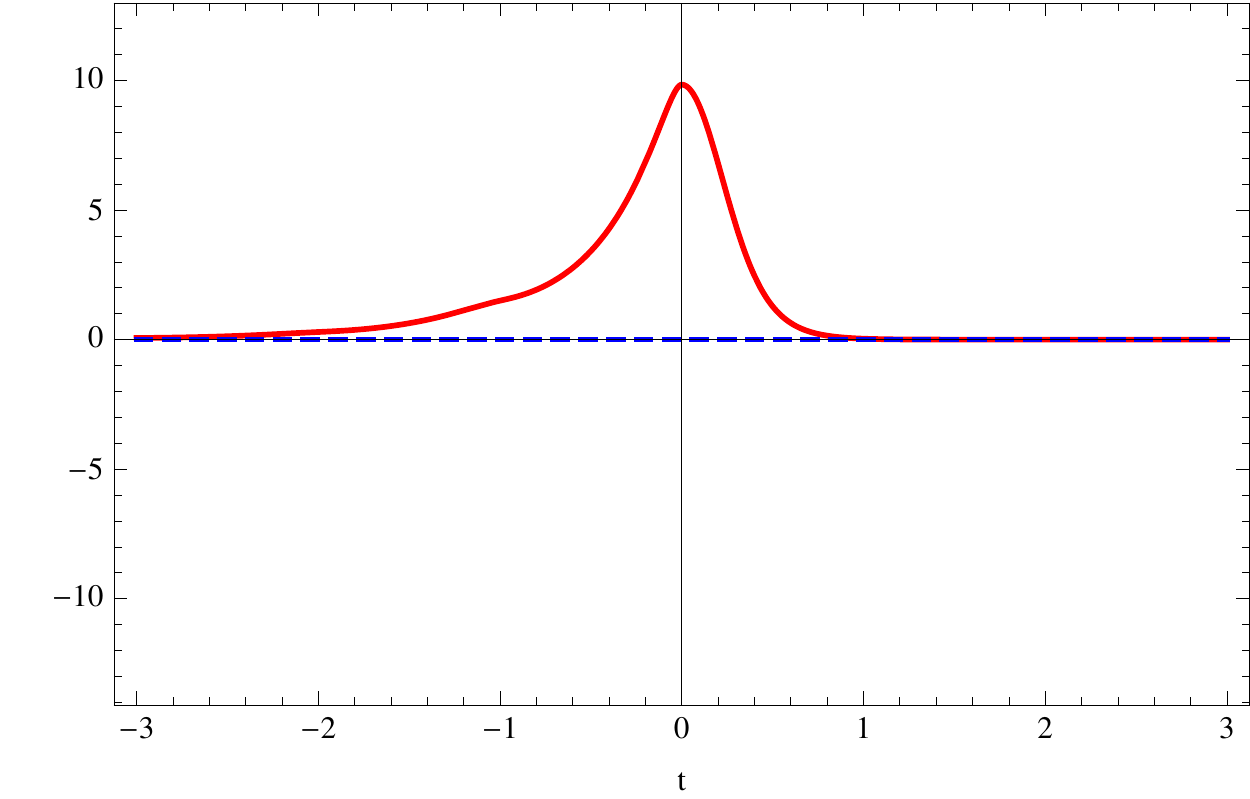}
\\[3mm]
(a)\hspace*{8mm}
&(b)\\[3mm]
\end{tabular}
\end{minipage}
\caption{The real (red) and imaginary (dashed blue) parts of
the integrand $F_1(z,s)$ of \eqn{OneDimensionalExampleIntegrand} for 
$s=1+i\delta$, after dividing out the phase at the saddle point,
 along (a) the [3/2] Pad\'e approximation 
to the contour of stationary phase passing through the saddle point (b)
the exact contour of stationary phase.}
\label{OneDimMinkowskiPadeContourIntegrand}
\end{figure}
\FloatBarrier

The resulting [3/2] Pad\'e contour for $s=1+i\delta$ is
shown in \fig{OneDimMinkowskiPadeContourFigure}.  
For negative imaginary parts of $z(t)$, the contour is asymptotically
parallel to the real axis.
Accordingly, it properly includes contributions from
all poles, and repairs the defect in the quadratic contour.
  On a linear scale, 
the values of the real and imaginary parts of the
integrand along the Pad\'e contour are quite similar to those along the
quadratic contour, because the values are very small
in regions where the contours differ.  
In \fig{OneDimMinkowskiPadeContourIntegrand}(a), we show
the real and imaginary parts of the integrand along the Pad\'e contour,
after dividing out by the phase $s^{i\phi_s}$ at the saddle point.
\Fig{OneDimMinkowskiPadeContourIntegrand}(b) shows the same parts
along the exact contour of stationary phase; the parametrization of
the exact contour is different
(for $t>0$, the imaginary part of the contour is chosen to be $t$, while
for $t<0$, the real part is chosen to be $t$), 
leading to a somewhat different shape, but the absence
of oscillations in the real part, and the small value of the imaginary
part in both parts of the figure show that the Pad\'e contour is
a very good approximation to the exact one.

\FloatBarrier
\subsection{Above Threshold}

\begin{figure}[ht]
\includegraphics[clip,scale=0.66]{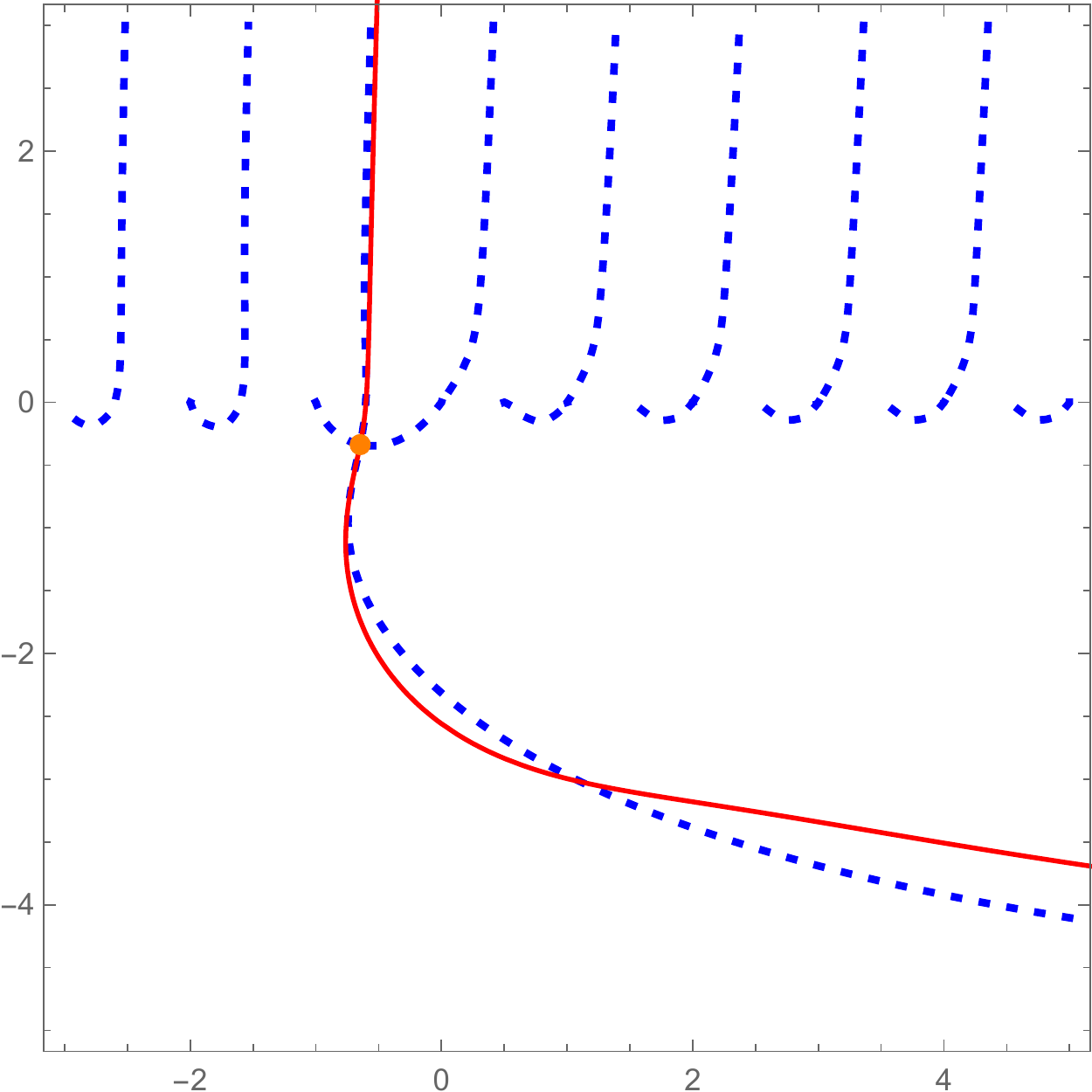}
\caption{
The [3/2] Pad\'e approximation to the contour of stationary phase $e^{i\phi_s}$
(solid red) for
the integrand $F_1(z,s)$ of \eqn{OneDimensionalExampleIntegrand} with
$s=5+i\delta$. The exact contours of this phase are
also shown (dotted blue).  
The saddle point is indicated by a large orange dot.
}
\label{OneDimMinkowskiAbovePadeContourFigure}
\end{figure}

\begin{figure}[t]
\begin{minipage}[b]{1.03\linewidth}
\begin{tabular}{cc}
\hskip -6mm
\includegraphics[clip,scale=0.60]{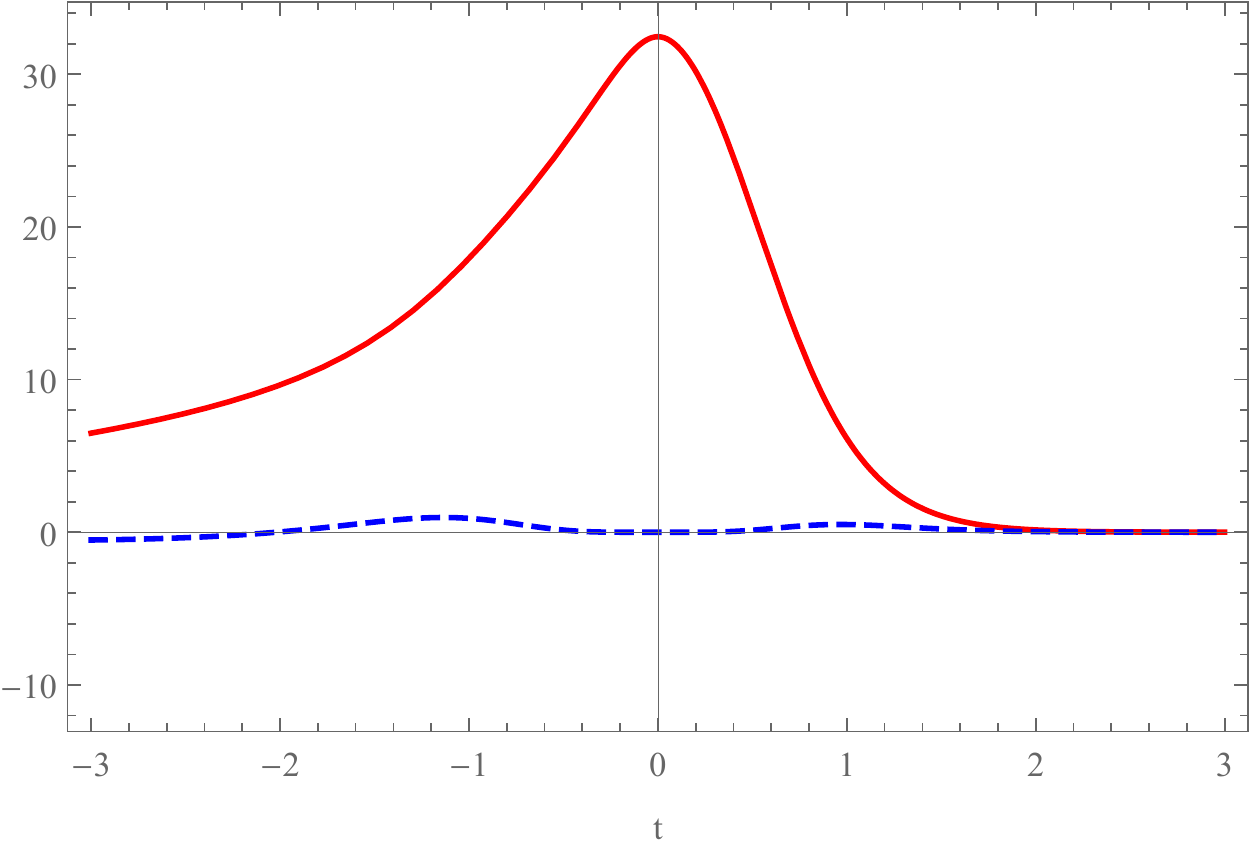}
\hspace*{6mm}
&\includegraphics[clip,scale=0.63]{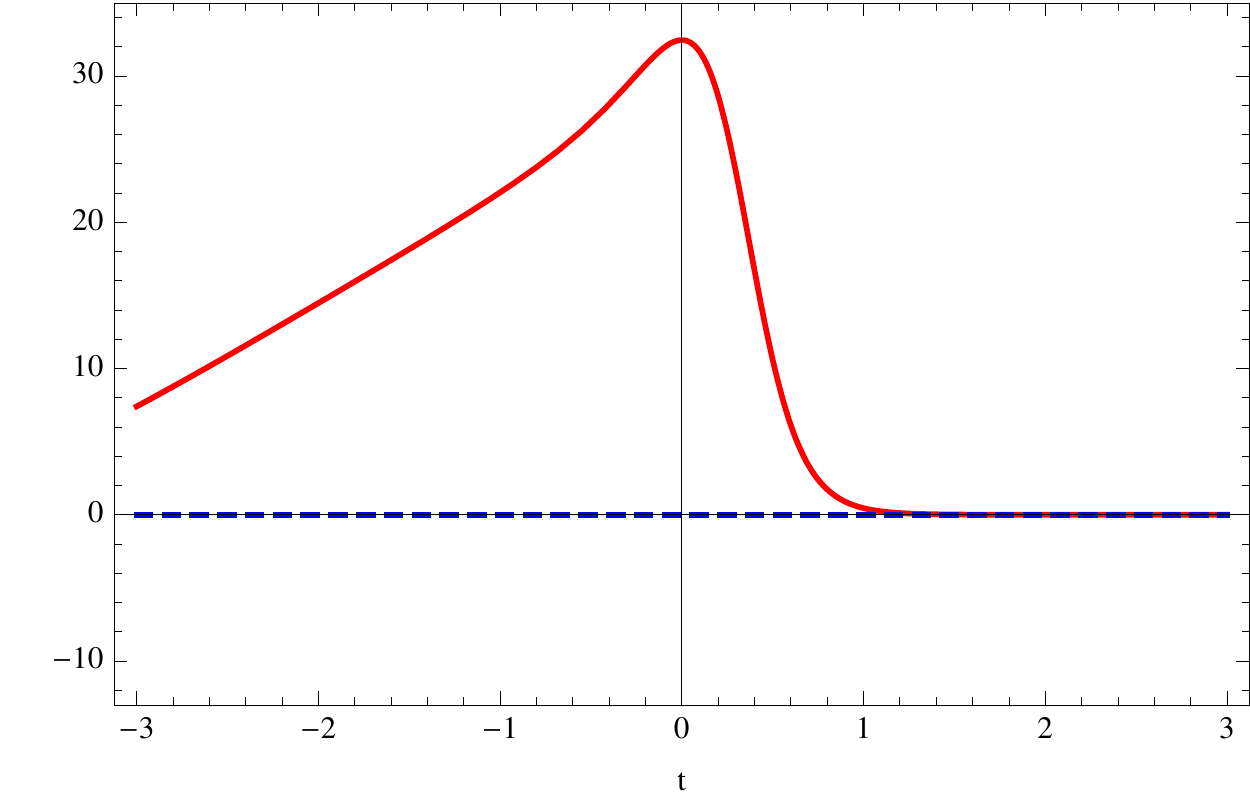}
\\[3mm]
(a)\hspace*{8mm}
&(b)\\[3mm]
\end{tabular}
\end{minipage}
\caption{The real (red) and imaginary (dashed blue) parts of
the integrand $F_1(z,s)$ of \eqn{OneDimensionalExampleIntegrand} for 
$s=5+i\delta$, with the phase at the saddle point divided out,
 along (a) the Pad\'e approximation 
to the contour of stationary phase passing through the saddle point
(b) the exact contour of stationary phase.}
\label{OneDimMinkowskiAbovePadeContourIntegrand}
\end{figure}

The approach described in the previous subsection also works above
threshold.  The integral has a branch cut starting at threshold,
so here we need to give an infinitesimal imaginary part to the
parameter $s$ in order to obtain the integral's value as well
as for finding contours.  We will again take this
imaginary part to be positive; taking it negative would complex-conjugate
the result and all the contours we find as well.  The Pad\'e approximation
contour for $s=5+i\delta$ is shown in 
\fig{OneDimMinkowskiAbovePadeContourFigure},
and the the integrand along it (with the phase at the saddle point divided out)
is shown in \fig{OneDimMinkowskiAbovePadeContourIntegrand}(a).  As
we can see, although
the contour deviates noticeably
from the exact contour in between small and
very large negative values of $t$, the integrand does not oscillate 
significantly, and hence this deviation will have 
little effect on the convergence
of the integration.  We show the integrand along the exact contour
in \fig{OneDimMinkowskiAbovePadeContourIntegrand}(b), with the
imaginary part of the contour again taken to be $t$.

\begin{figure}[ht]
	\includegraphics[clip,scale=0.66]{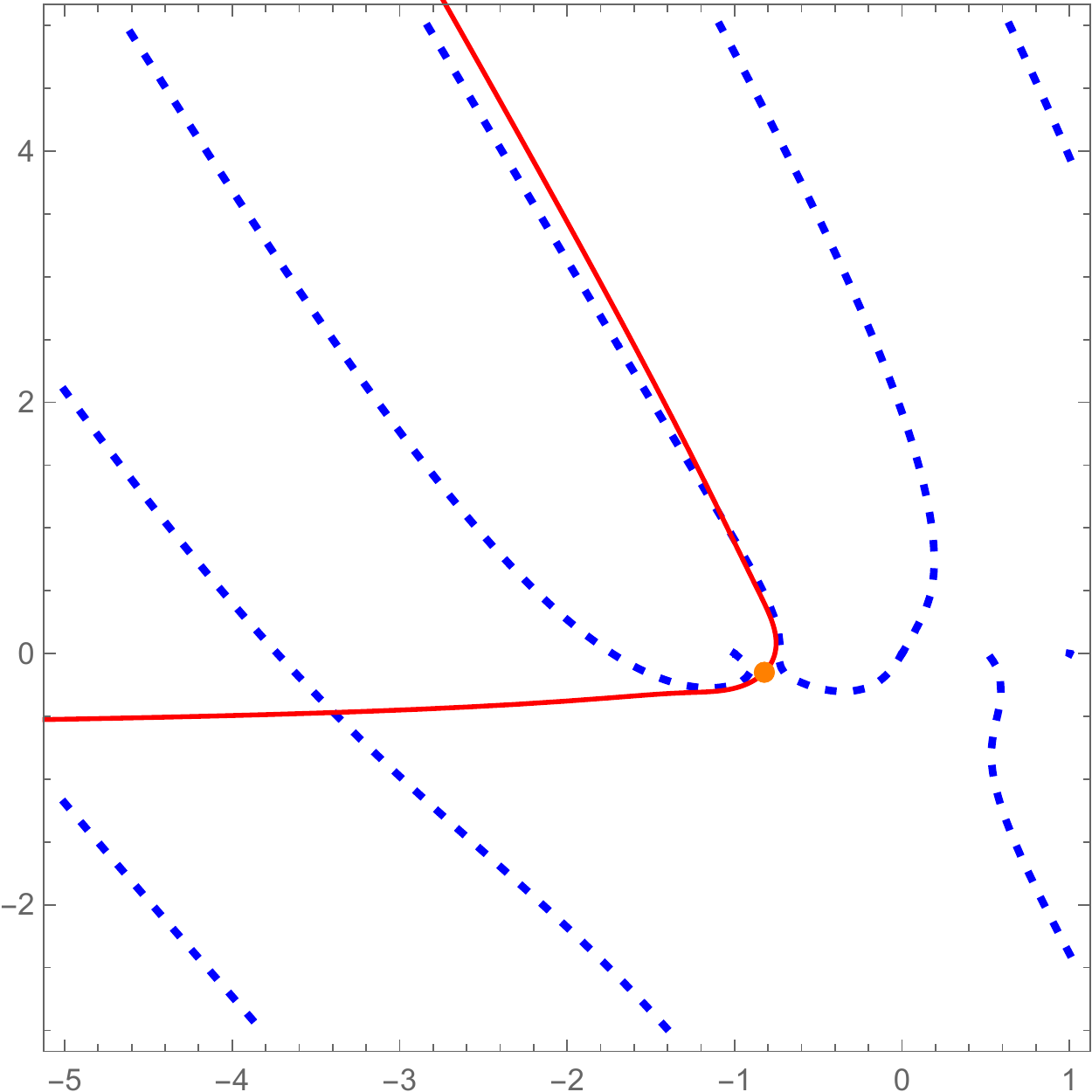}
\caption{The [3/2] Pad\'e approximation to the contour of stationary phase $e^{i\phi_s}$
	(solid red) for
	the integrand $F_5(z,s)$ of \eqn{FlatAsymptotesIntegrand} with
	$s=1+i\delta$. The exact contours of this phase are
	also shown (dotted blue).  
	The saddle point is indicated by a large orange dot.
	}
	\label{ParallelMinkowskiContoursFigure}
\end{figure}

\subsection{Parallel Asymptotes}
\label{ParallelMinkowskiAsymptotes}

As we saw in \sect{FlatAsymptotesSection}, 
the case $N_{-}=0$ requires special treatment,
because the asymptotes are parallel, and so one must require the
intercept for the asymptotic form to have an imaginary part.  Otherwise,
the Pad\'e contours are unexceptional in the Euclidean region: they
properly enclose all the poles enclosed by the `textbook' contour
selected by \MB{}.  The behavior of the integrand
along these contours is likewise unexceptional.  
In the Minkowski region, the situation is different.  As we can
see from \eqns{MinkowskiAsymptoticForm}{MinkowskiAsymptoticAngles},
the asymptotes for $t\rightarrow\pm\infty$ are parallel to each other,
but not to the real axis.  This means that a single contour cannot
enclose all poles (in general, each contour will enclose only a single
pole).  An example, the integrand
of \eqn{FlatAsymptotesIntegrand} with $s=1+i\delta$,
 is shown in \fig{ParallelMinkowskiContoursFigure}.

As in the case of closed contours considered in \sect{ClosedContoursSection},
from a practical point of view it doesn't make sense to use the
ensemble of exact contours.  (One would again be better off computing the residues
analytically and summing over them.)
The approximate Pad\'e contour can instead
be chosen to have one of the asymptotes parallel to the real axis,
thereby enclosing all poles, at the price of small oscillations in
the tail of the integrand along the contour.  This can be done
by choosing $\theta_{-\infty}$ to be $\pi/2$ (for $\sign\Im s>0$),
and then using a modified expression for $\zinf$,
\begin{equation}
\begin{aligned}
\Re \zinf &=
  \frac{\phi_s}{\pi}-\frac{\ln|s_0/s|\,\Im\zinf}{\pi}
  +\frac1{2} {(S_{+}-2 A_{+}-2\DP_{-}-2\SP)\bmod 4}\\
& -\frac1{2 \pi}
   \bigl[\bigl(A_{+}+A_{-}-S_{+}/2-S_{-}/2-\DP_{+}-\DP_{-}\bigr) 
         \bigl(2\thinfp-\pi\bigr)
 \bmod 2\pi\bigr]\,,\\
\end{aligned}
\end{equation}
where $\Im\zinf$ can be chosen with some freedom; a good heuristic
is again to take it to be of order $1/\Re c_2$.

\section{Integrands Without Extrema}
\label{NoExtremaSection}

\begin{figure}[ht]
\includegraphics[clip,scale=0.66]{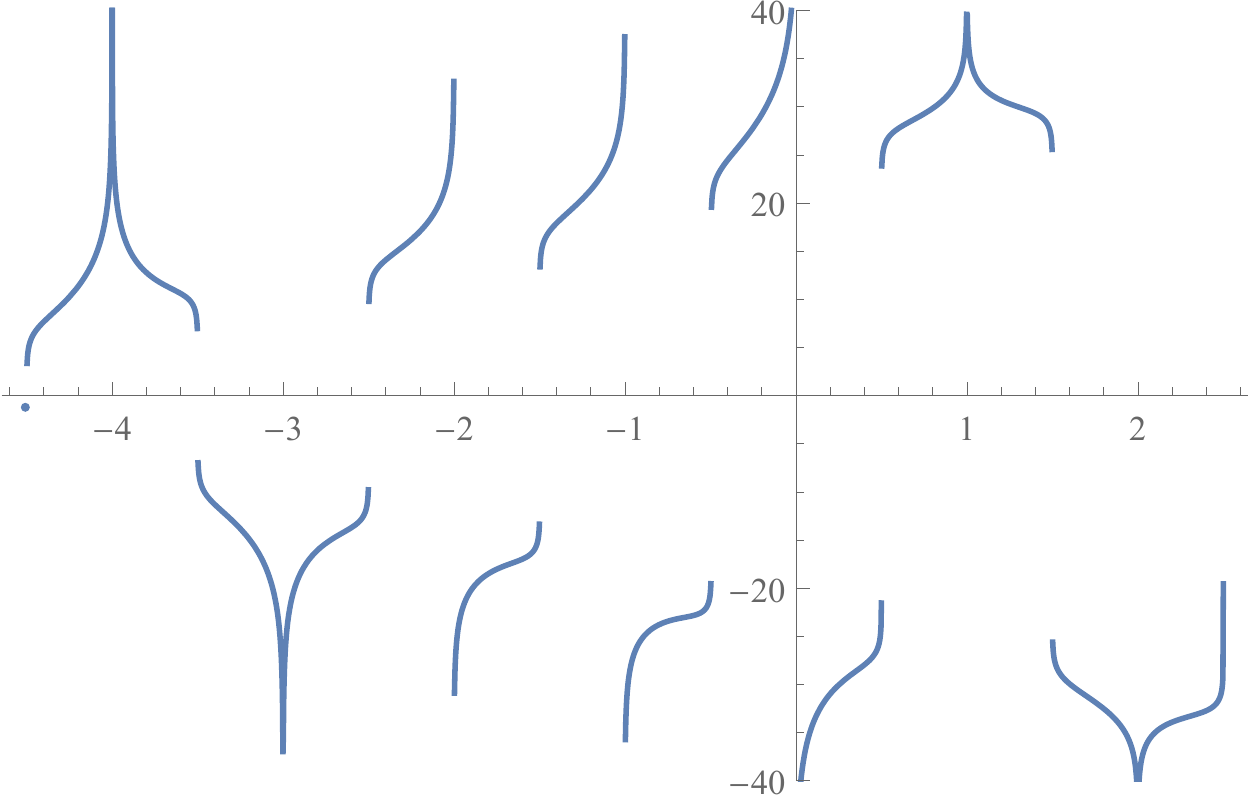}
\caption{The behavior of the integrand $F_7(z,s)$ of \eqn{OneDimensionalExampleWithZero},
	along the real axis.  The plot displays $\sln_{20}({\rm integrand})$,
	with $s=\stdparam$.}
	\label{OneDimIntegrand7}
\end{figure}

As mentioned in \sect{OtherIntegralsSection}, an integrand may
have {\it no\/} real interval in which the integrand has an extremum.
In this case, however, there is always a zero of the integrand
in any given interval between poles.  Let us consider the simplest
(and most generic) case of this type, where the integrand
has a simple zero independent of the value of the parameter $s$.
An example of such an integral is,
\begin{equation}
I_7(s) = \frac{1}{2\pi i} \int_{c_0-i\infty}^{c_0+i\infty} dz\;F_7(z,s)
\label{NoExtremumIntegral}
\end{equation}
where,
\begin{equation}
F_7(z,s) = (-s)^{-z}\frac{\Gamma^3(-z)\Gamma(3+z)\psi^{(2)}(z)}{\Gamma(-2z)}\,,
\label{OneDimensionalExampleWithZero}
\end{equation}
which is just $F_1(z,s)$ of \eqn{OneDimensionalExampleIntegrand},
multiplied by additional polygamma and polynomial factors. 

Consider first the Euclidean region; the integrand is real for real
$z$, but has no finite extrema as displayed in \fig{OneDimIntegrand7}.  
Instead, the
integrand has complex stationary points; they come in complex-conjugate
pairs.  Moreover, while there are still contours of stationary
phase, the integrand is no longer necessarily real on them.  
The integral is nonetheless real.

How is this possible?  In the generic case discussed in previous
sections, a contour of stationary phase comes in from infinity,
passing through a stationary point, and heads back out to infinity.
When the function has a zero, however, contours of stationary phase
can end there.  In the case at hand, the full contour of integration
will actually consist of two separate contours of integration, 
complex conjugates joined at the zero of the integrand.  The phase
is stationary on each separate contour, but different; indeed,
the phase of the integrand on the contour in the lower-half plane
is the complex conjugate of the phase
on the contour in the upper-half plane.

One could choose to deform the textbook contour to one passing through
the two stationary points using a quadratic approximation, 
without worrying where exactly it crosses the
real axis.  However, this will still leave substantial oscillations in
the integrand near the real axis.  To do better, we seek a contour
that passes through the stationary points and also through the zero
of the integrand on the real axis.  We join two half-contours in each
half-plane, taking the half-contour in the lower half-plane to be
the complex conjugate of that in the upper half-plane.  Each half-contour
will end at the zero.  To best approximate the (half-)contour of stationary
phase, we match the tangent at the stationary point, and also the
initial direction at the integrand's zero.  This requires six real
parameters: the real and imaginary parts of the zero location and
the stationary point, and the two angles giving the directions at those
points.  This is exactly the number of parameters available in a
quadratic curve,
\begin{equation}
z_h(t) = z_0 + e^{i\theta_0} (i a_1 t + a_2 t^2)\,,
\label{QuadraticHalfContour}
\end{equation}
where $a_1$ is real but $a_2$ is complex.  In this case, we do not
have enough parameters to match the curvature at the stationary point,
unlike contours considered in previous sections.  We give below
the formul\ae{} for a half-contour in the upper half-plane; a similar
set with appropriate replacements ($\thinfp\rightarrow\thinfm$, etc.) gives
the half-contour in the lower half-plane.

To derive formal\ae{} for the parameters in
\eqn{QuadraticHalfContour}, expand the imaginary part of the integrand
after dividing out the phase at the stationary point, obtaining,
\begin{equation}
a_1 \Im\biggl[\frac{(i F'(z_0))}{F(z_s)}e^{i\theta_0}\biggr] t+\Ord(t^2)\,.
\end{equation}
Let us restrict attention here to 
integrands with simple zeros, so that 
$F'(z_0)$ does not vanish. (The generalization to higher-order
zeros is reasonably straightforward.)
Requiring the coefficient of $t$ to vanish determines
the initial direction $\theta_0$ along the half-contour,
\begin{equation}
\theta_0 = \arg \bigl(-i F(z_s)/F'(z_0)\bigr)\,.
\label{ExtremumFreeInitialAngle}
\end{equation}
The tangent angle $\theta_s$ at the stationary point is given by 
\eqn{OneDimMinkowskiLinearAngleSolution} (up to a possible rotation by
$\pi/2$); we can solve for $a_1$ and
$a_2$ in terms of the two angles and the locations of the zero and
the stationary point,
\begin{equation}
\begin{array}{rl}
a_1 &= \displaystyle \frac{2 \Re\bigl[e^{-i\theta_s} (z_s-z_0)\bigr]}
            {\sin (\theta_s-\theta_0)}\,,\\
a_2 &= e^{-i\theta_0} (z_s-z_0) - i a_1\,.
\end{array}
\label{ExtremumFreeQuadraticContourParameters}
\end{equation}

\def\acu{\mathring{a}}
\def\bcu{\mathring{b}}
\def\zuinf{\mathring{z}_{\infty}}

To match the asymptotic behavior as well, and thereby make the
contour more robust, we again turn to a [3/2] Pad\'e approximation,
which it is here convenient to write in the form,
\begin{equation}
z_h(t) = z_0 + (z_s-z_0) t
+ t (t-1) \frac{\acu_2 + \bcu_2 \acu_3 (t-1)}
               {1+\bcu_1 (t-1) + \bcu_2 t (t-1)}\,.
\label{ExtremumFreePade}
\end{equation}
In this form, the parameter $t$ has been rescaled to
put the saddle point at $t=1$.  As
in \sect{MinkowskiSection}, it is convenient to make a nonlinear
transformation to a new set of (real) parameters $\{\rho_{2,3,b}\}$,
\begin{equation}
\begin{aligned}
\acu_2 &= i \rho_2 e^{i\theta_s}-(z_s-z_0)\,,\\
\acu_3 &= i\rho_3 e^{i\thinfp} - (z_s-z_0)\,,\\
d_Z &= \acu_3^2 - (i\rho_3 e^{i\thinfp} + 
\zuinf - z_s) (i\rho_b e^{i\theta_0} - (z_s - z_0))\,,\\
\bcu_1 &= 1-d_Z^{-1} (\acu_3^2 + (\zuinf - z_0) \acu_2)\,,\\
\bcu_2 &= d_Z^{-1} (i\rho_3 e^{i\thinfp} (\acu_2 + i\rho_b e^{i\theta_0}
 - (z_s - z_0)) + (z_s - z_0)^2 
   +\rho_2 \rho_b e^{i(\theta_s+\theta_0)})\,.
\end{aligned}
\label{ExtremumFreeChangeOfVariables}
\end{equation}
In these equations, $\theta_s$
is given by \eqn{OneDimMinkowskiLinearAngleSolution};
$\theta_0$ by \eqn{ExtremumFreeInitialAngle};
$\thinfp$ by \eqn{MinkowskiAsymptoticAngles}; and $\zinf$ by,
\begin{equation}
\begin{aligned}
\zinf &= \frac{\phi_s \sign\Im z_s}{\pi \Nm}
+\frac{(S_{+}-2 A_{+}-2\DP_{-}-2\SP)\bmod 4}{2 (\Nm+\sign\Im s)}
\\& \hphantom{=}
-\frac1{(\Nm+\sign\Im s) \pi}
   \biggl[\bigl(A_{+}+A_{-}-S_{+}/2-S_{-}/2\\
   &\hskip 43mm-\DP_{+}-\DP_{-}\bigr) 
         \biggl(\thinf-\frac{\pi}2\biggr)
 \bmod 2\pi\biggr]
\,,\quad\quad \Nm \neq -\sign\Im s\,.\\
\end{aligned}
\end{equation}
As in the generic situation, this value may be shifted in order to
match onto the desired asymptotic contour, by multiples 
of $2/(\Nm+\sign\Im s)$.
(In the case when $\Nm = -\sign\Im s$, $\zinf$ must be chosen
imaginary, and we can use the last equation in \eqn{AsymptoticThetaEuclidean},
along with possible shifts given in the text below that
equation.)
The forms in \eqn{ExtremumFreeChangeOfVariables} 
then give a contour that automatically
satisfies the correct asymptotic form; that has the correct initial
direction at $z_0$; and that has the correct tangent at $z_s$.

As in the generic case, we fix $\rho_3$ to 1.
We can fix
$\rho_{2}$ and $\rho_b$ by minimizing the square of the
following deviation from stationarity of the cubic term
in the expansion of the integrand around the stationary point,
\begin{equation}
+\Re\biggl[\frac{(\rho_2 e^{i\theta_s}-\rho_3 e^{i\thinfp})^2}
                {i\rho_3 e^{i\thinfp}+\zuinf-z_s}\biggr]
+\Re\biggl[d_Z^{-1}\frac{(\acu_3^2+\acu_2 (\zuinf-z_0))^2}
             {i\rho_3 e^{i\thinfp}+\zuinf-z_s}\biggr]
-\frac{\Re[D_3]}{6 D_2}\,.
\label{NoExtremumCubic}
\end{equation}
along with 
the square of the relative phases of the denominator terms,
\begin{equation}
\arg(\bcu_2/\bcu_1)\,,
\end{equation}
and the square of the relative phases of the numerator terms,
\begin{equation}
\arg(\acu_2/(\bcu_2\acu_3))\,.
\end{equation}
The minimization is after
substituting the expressions in \eqn{ExtremumFreeChangeOfVariables}.
For certain integrands or values of the parameter $s$, this approach
appears to avoid looping contours that would otherwise arise.
A good heuristic weights the cubic~(\ref{NoExtremumCubic})
significantly more than the denominator's relative phase, which
in turn is weighted more than the numerator's relative phase.
As with the reparametrizations and the fixing of $\theta$s, the
minimization should be carried out independently for the upper-
and lower-half planes.

\begin{figure}[ht]
\includegraphics[clip,scale=0.66]{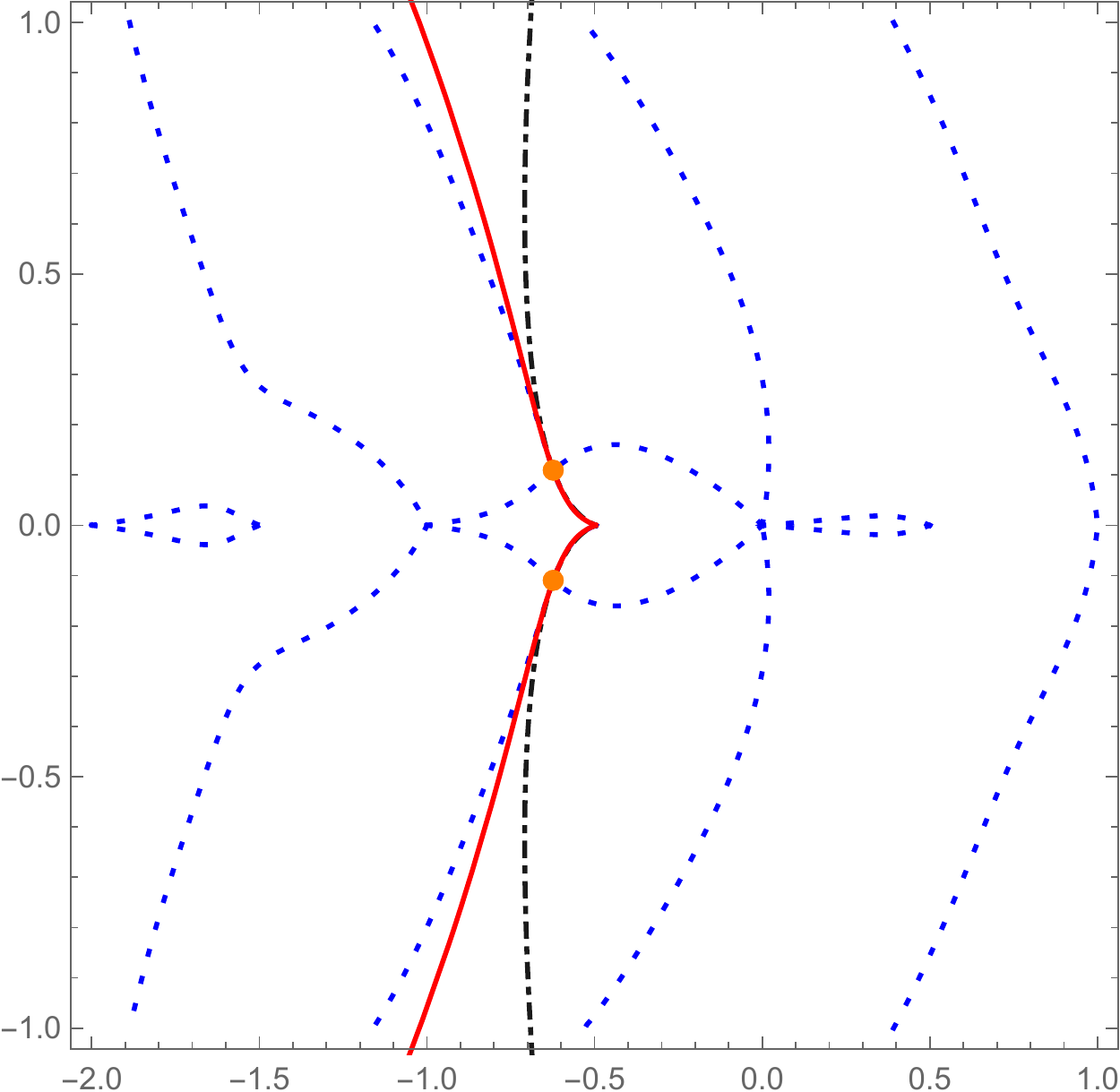}
\caption{
The joined quadratic (dot-dashed dark gray) and
joined [3/2] Pad\'e approximations (solid red) to the contour of 
stationary phase for
the integrand $F_7(z,s)$ of \eqn{OneDimensionalExampleWithZero} with
$s=\stdparam$. The exact contours of constant phase are
also shown (dotted blue).  
The saddle points are indicated by large orange dots.
}
\label{OneDimEuclideanNoExtremumContourFigure}
\end{figure}

\begin{figure}[t]
\begin{minipage}[b]{1.03\linewidth}
\begin{tabular}{cc}
\hskip -6mm
\includegraphics[clip,scale=0.60]{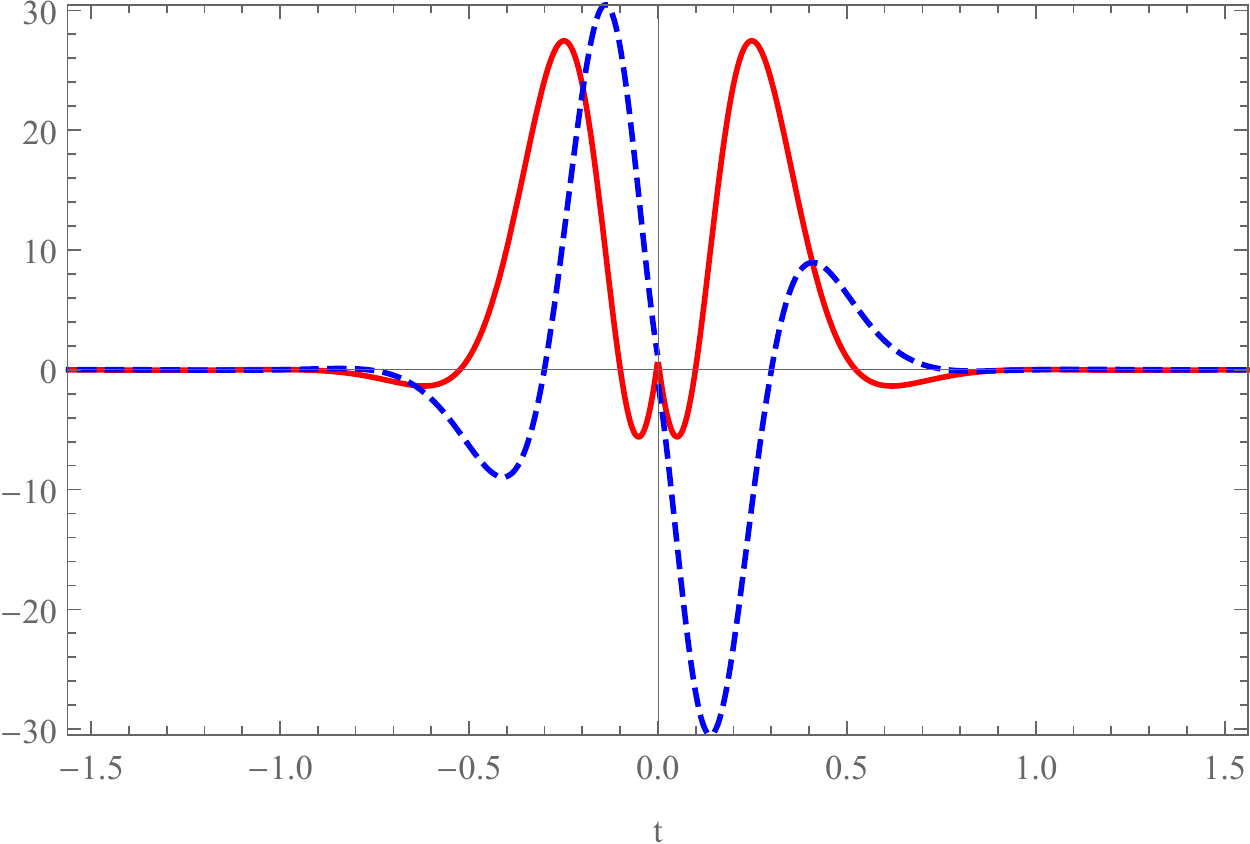}
\hspace*{6mm}
&\includegraphics[clip,scale=0.60]{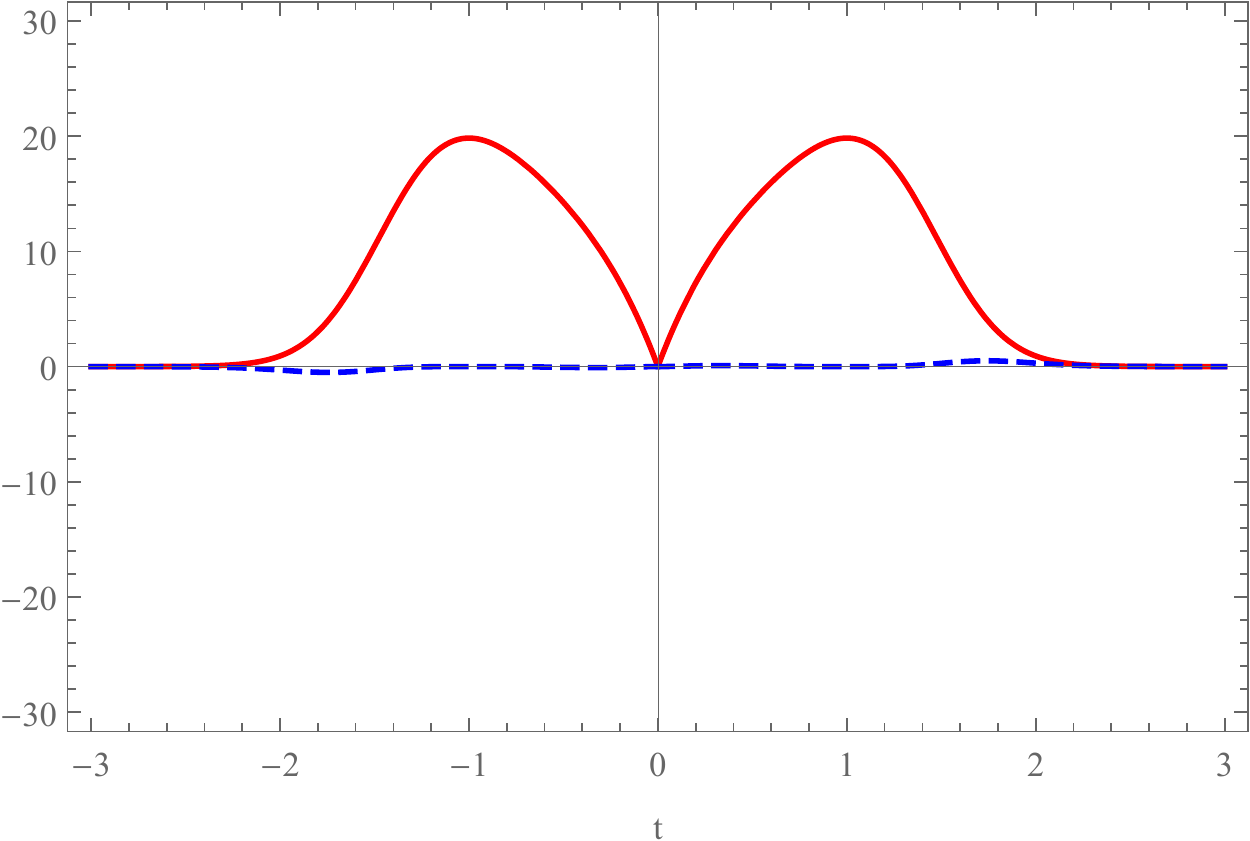}
\\[3mm]
(a)\hspace*{8mm}
&(b)\\[3mm]
\end{tabular}
\end{minipage}
\caption{The real (red) and imaginary (dashed blue) parts of
the integrand $F_7(z,s)$ of \eqn{OneDimensionalExampleWithZero} for 
$s=\stdparam$ (a) along the simple contour $\Re z=\negonehalf$ (b) 
 along the joined Pad\'e approximations,
to the contour of stationary phase,
with the phases at the saddle points divided out.}
\label{OneDimEuclideanNoExtremumPadeContourIntegrand}
\end{figure}

\begin{figure}[ht]
\includegraphics[clip,scale=0.66]{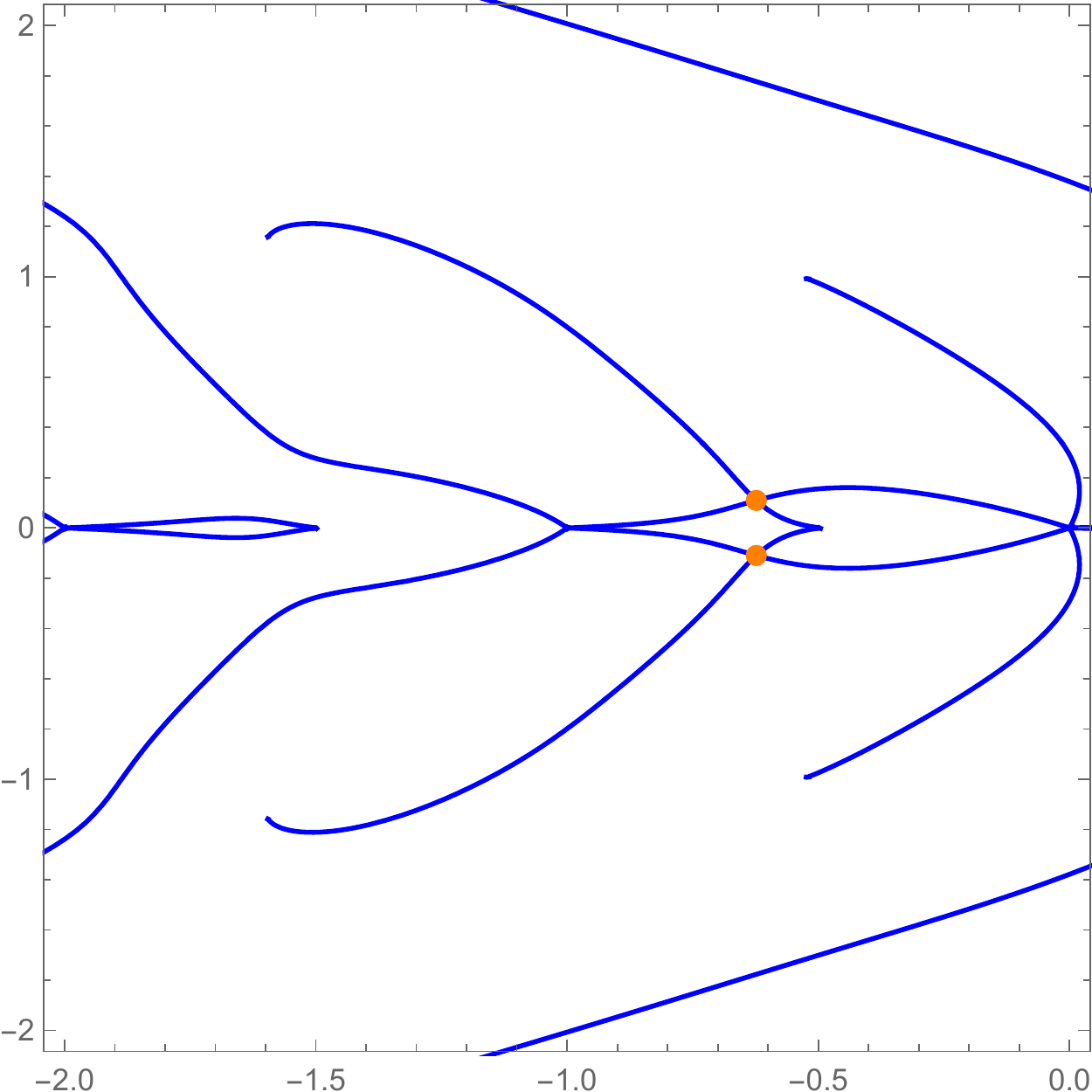}
\caption{Exact contours (solid blue) of stationary phase for
	the integrand $F_7(z,s)$ of \eqn{OneDimensionalExampleWithZero} with
	$s=\stdparam$. 
	The saddle points are indicated by large orange dots.
}
\label{OneDimEuclideanNoExtremumExactContourFigure}
\end{figure}

The joined quadratic
and joined [3/2] Pad\'e contours for $F_7(z,s=\stdparam)$ are shown in
\fig{OneDimEuclideanNoExtremumContourFigure}, 
while the integrand along the joint Pad\'e contour is
shown in \fig{OneDimEuclideanNoExtremumPadeContourIntegrand}, 
and contrasted with the behavior along the `textbook'
\MB{} contour $\Re z=\negonehalf$.  The phase at the upper saddle point
is divided out for $t>0$, and that at the lower saddle point for $t<0$.

The exact contours are shown in \fig{OneDimEuclideanNoExtremumExactContourFigure}.
The contours passing through the saddle points at $-0.623407 \pm 0.109501 i$
illustrate another potential complication with exact contours: in addition
to ending at zeros on the real axis, they can end at zeros off in the complex
plane without ever making it out to infinity.  Their use would then necessitate
finding the zeros and gluing on additional contours starting there.  In
contrast, the Pad\'e contour smoothly interpolates to a curve reaching infinity,
at the price of very small oscillations in the tail of the integrand.  These
oscillations do not disturb our ability to use the contour to calculate the
integral precisely and efficiently.

\begin{figure}[ht]
\begin{minipage}[b]{1.03\linewidth}
\begin{tabular}{cc}
\hskip -6mm
\includegraphics[clip,scale=0.5]{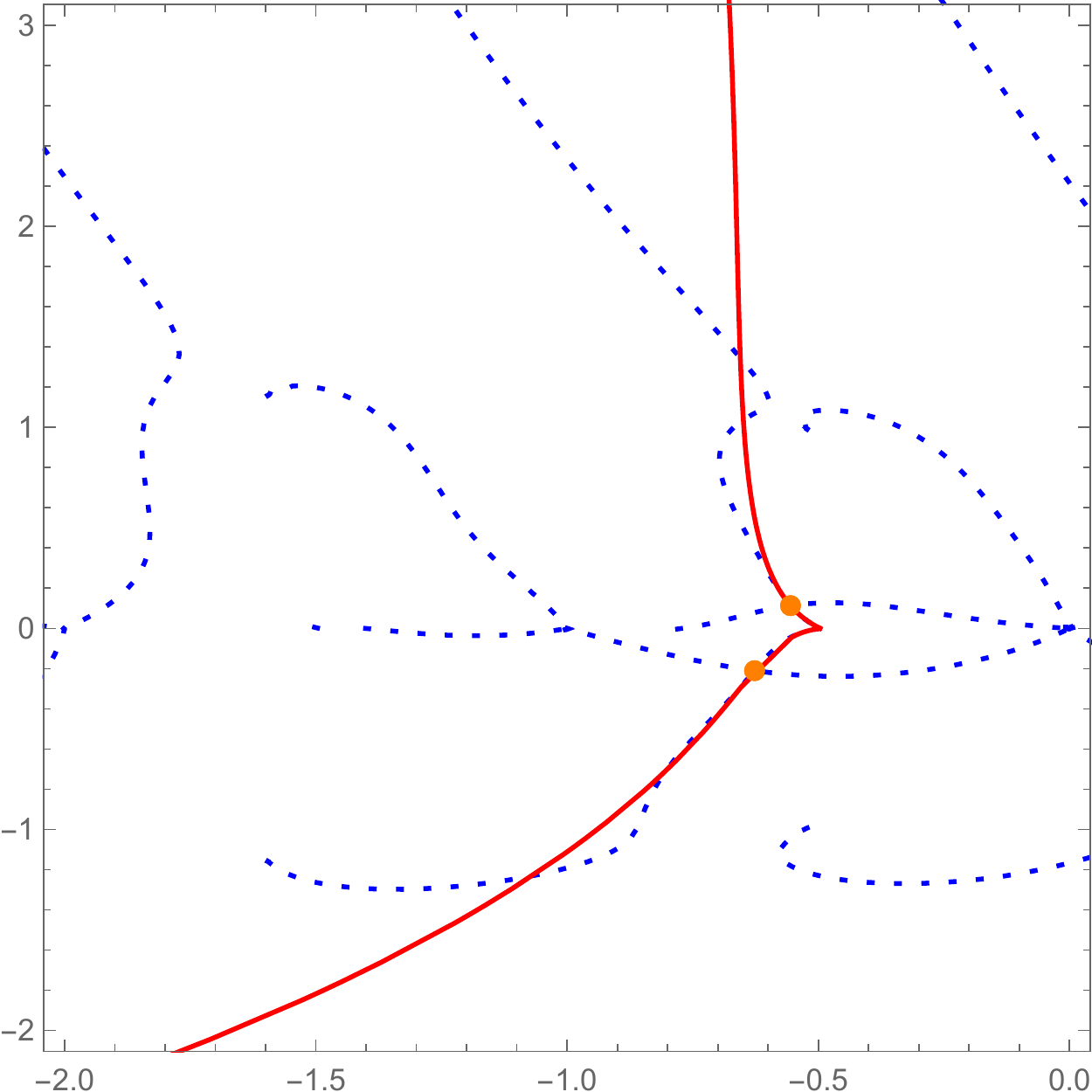}
\hskip 5mm
&\includegraphics[clip,scale=0.5]{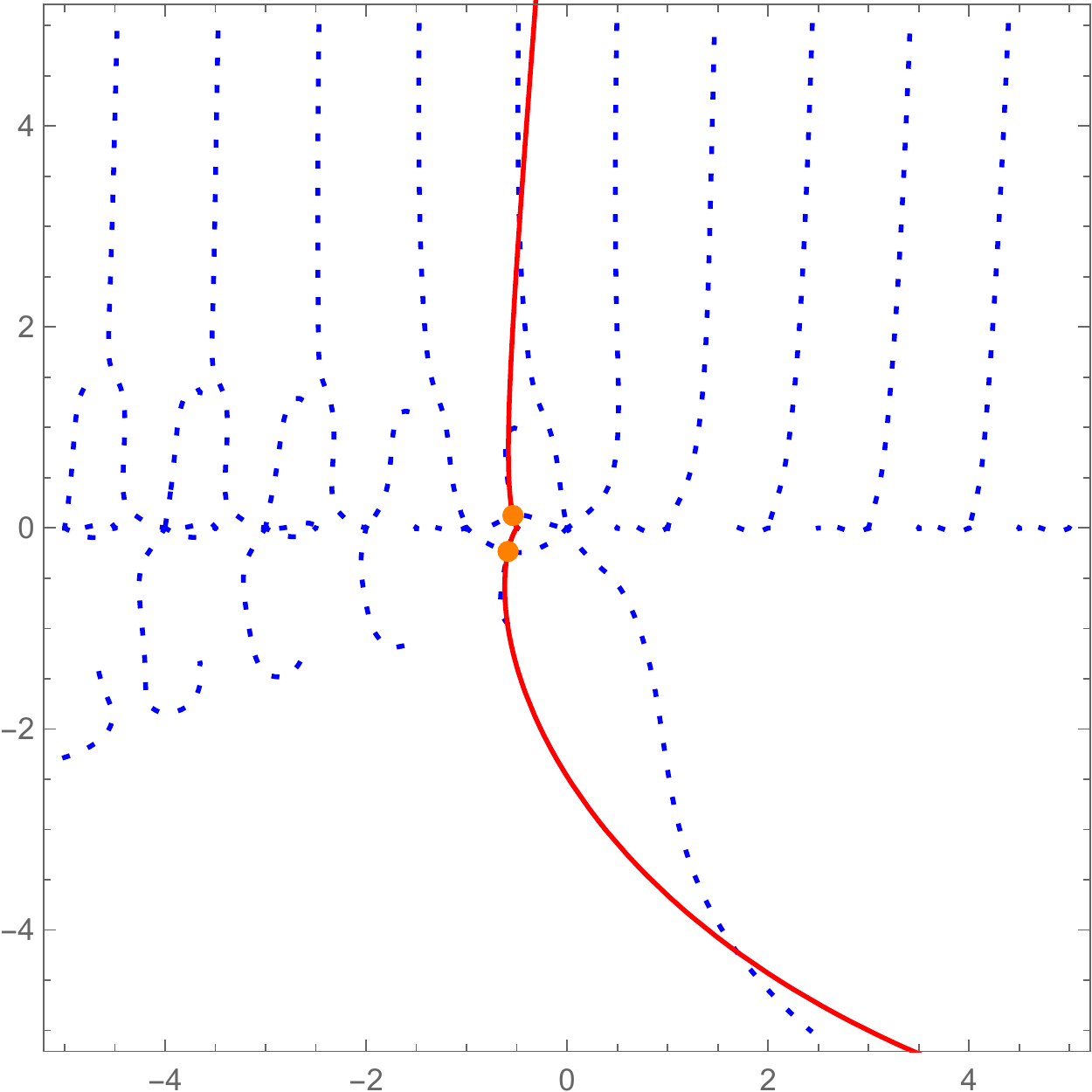}
\\[3mm]
(a)&(b)\\[3mm]
\end{tabular}
\end{minipage}
\caption{
The 
joined [3/2] Pad\'e approximations (solid red) to the contour of 
stationary phase for
the integrand $F_7(z,s)$ of \eqn{OneDimensionalExampleWithZero} with
(a) $s=1+i\delta$ (b) $s=5+i\delta$. The exact contours of constant phase are
also shown (dotted blue).  
The saddle points are indicated by large orange dots.
}
\label{OneDimMinkowskiNoExtremumContourFigure}
\end{figure}

\begin{figure}[ht]
\begin{minipage}[b]{1.03\linewidth}
\begin{tabular}{cc}
\hskip -6mm
\includegraphics[clip,scale=0.5]{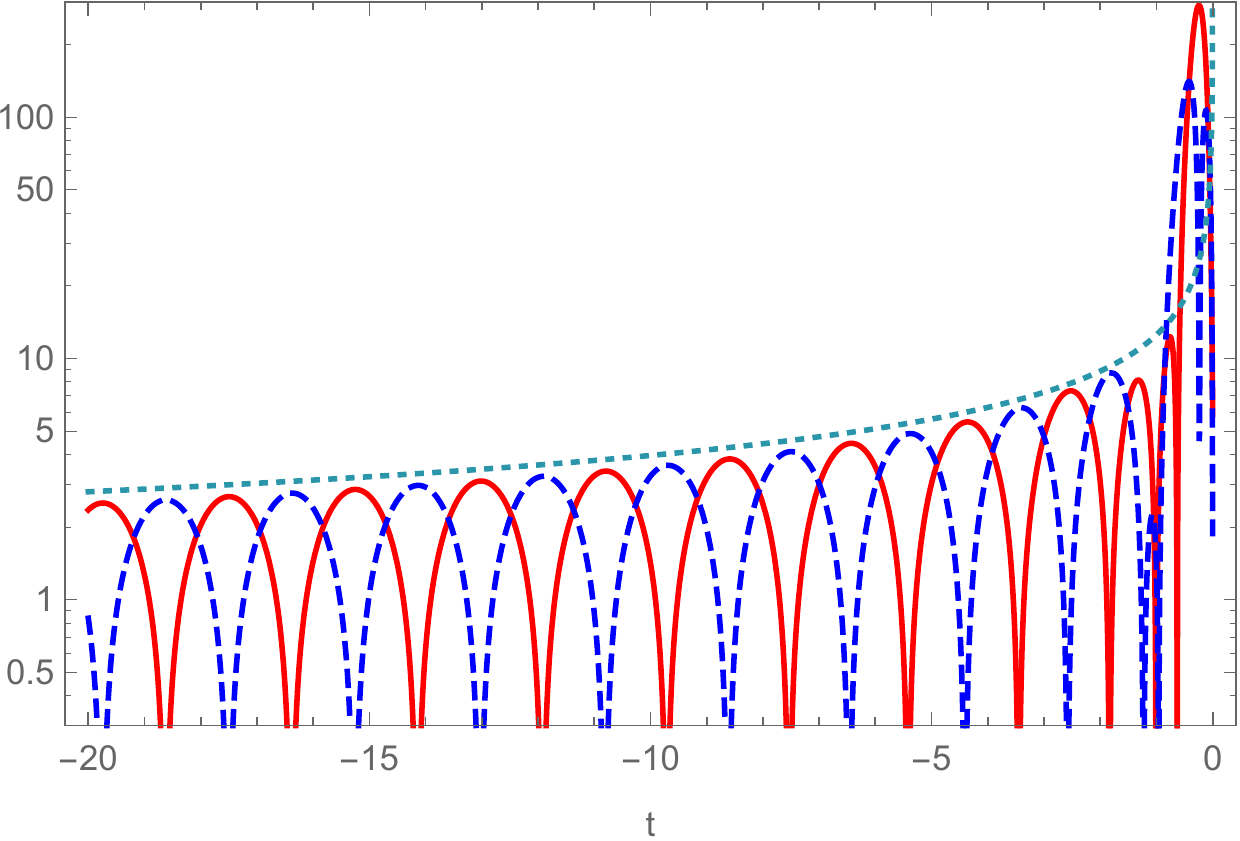}
\hspace*{5mm}
&\hspace*{5mm}
\includegraphics[clip,scale=0.5]{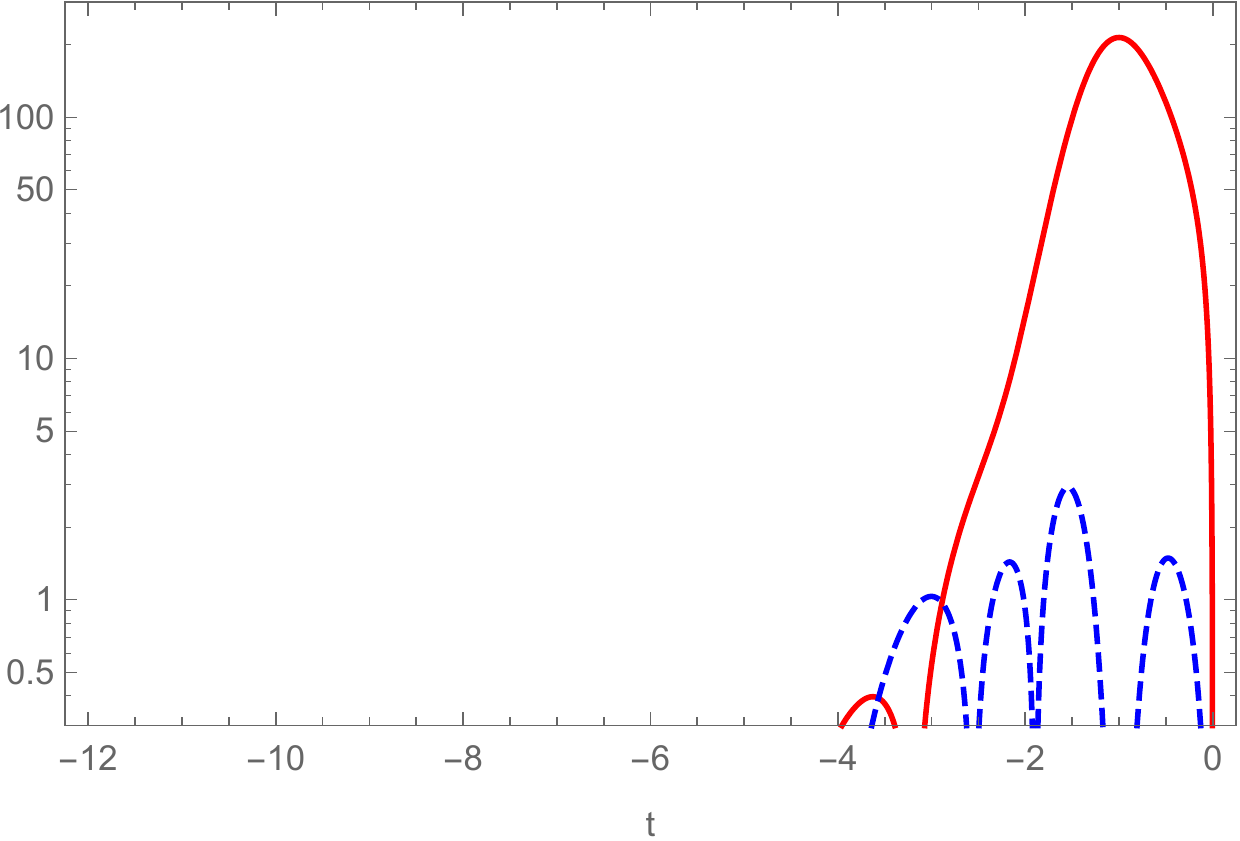}
\\[3mm]
(a)\hspace*{5mm}&\hspace*{5mm}(b)\\[3mm]
\hskip -6mm
\includegraphics[clip,scale=0.5]{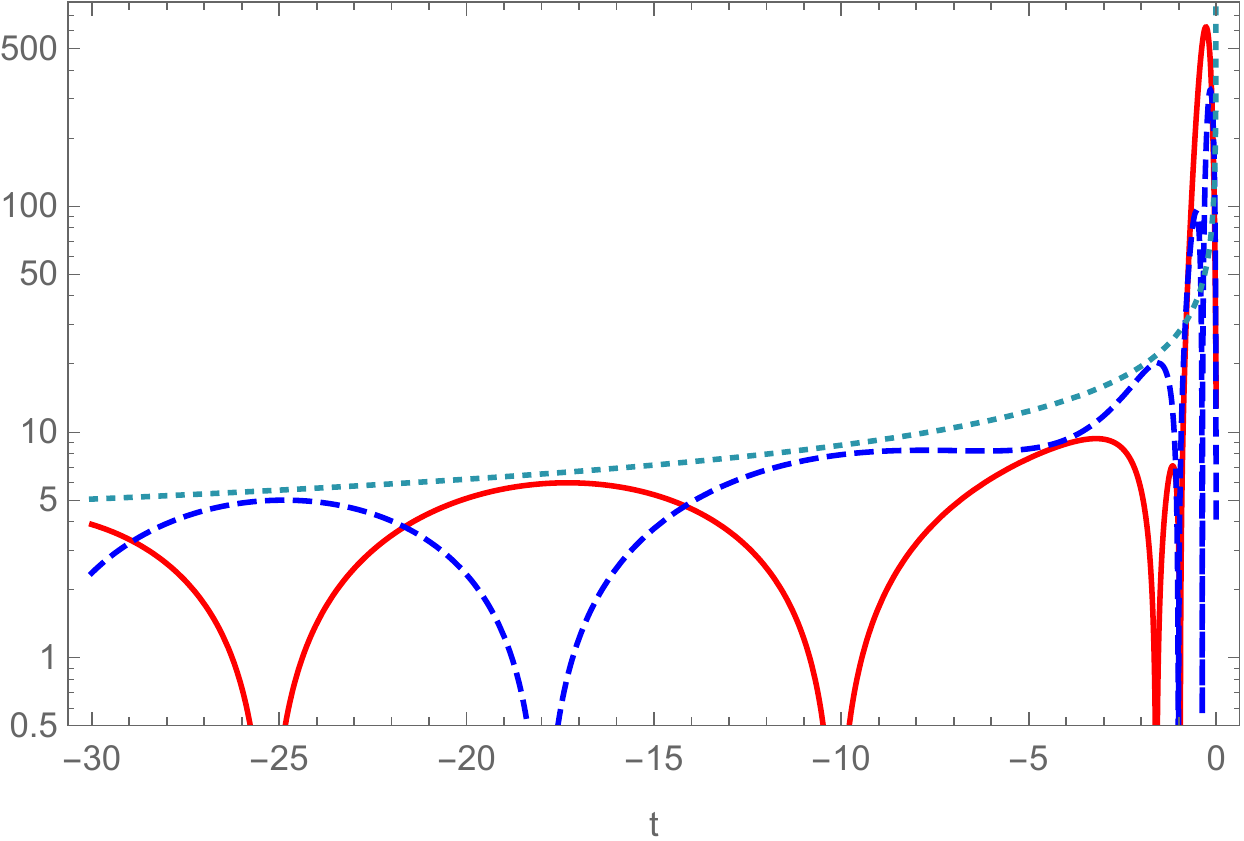}
\hspace*{5mm}
&\hspace*{5mm}
\includegraphics[clip,scale=0.5]{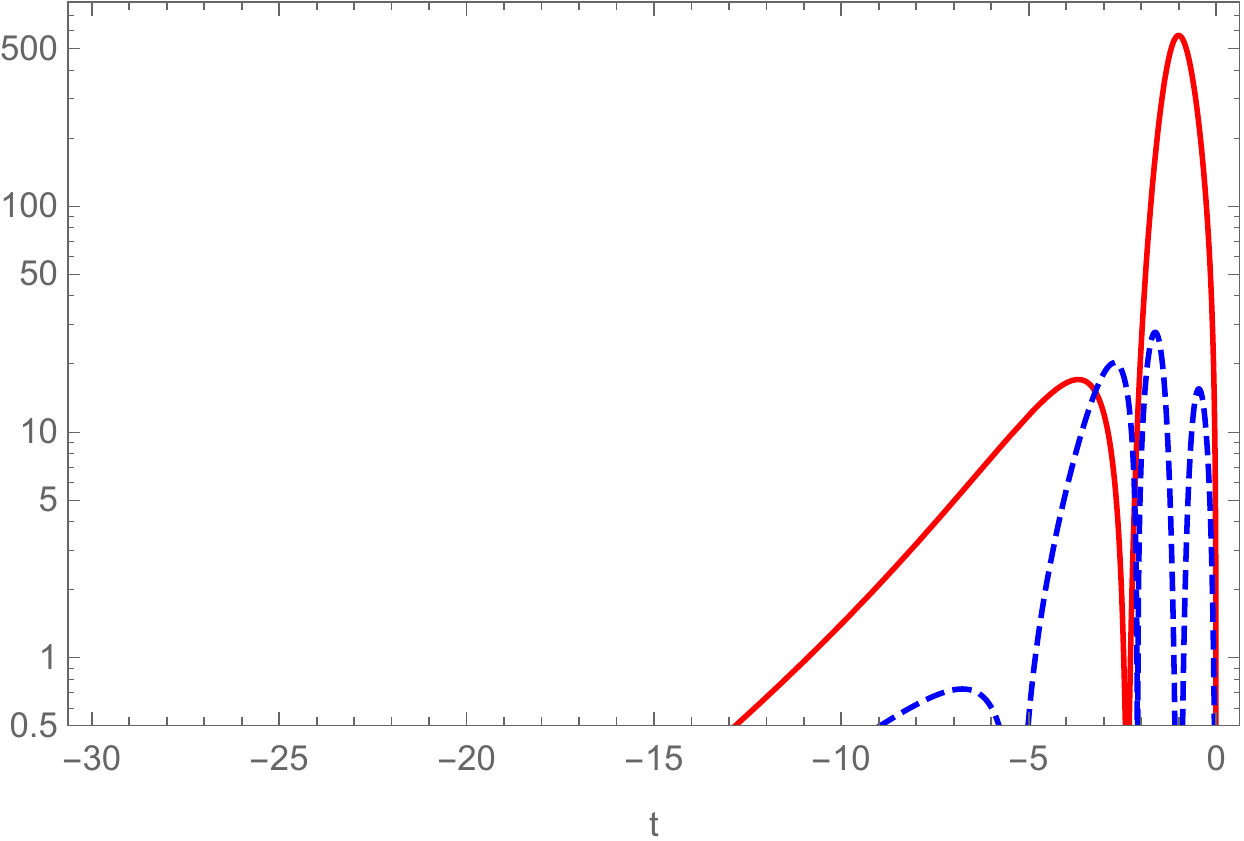}
\\[3mm]
(c)\hspace*{5mm}&\hspace*{5mm}(d)\\[3mm]
\end{tabular}
\end{minipage}
\caption{The absolute values of the
real (red) and imaginary (dashed blue) parts of
the integrand of \eqn{OneDimensionalExampleWithZero}, shown on
a log scale, for 
(a) $s=1+i\delta$ along the simple contour $\Re z=\negonehalf$
(b) $s=1+i\delta$  along the joined Pad\'e approximations
to the contour of stationary phase
(c) $s=5+i\delta$ along the simple contour $\Re z=\negonehalf$ 
(d) $s=5+i\delta$  along the joined Pad\'e approximations
to the contour of stationary phase.  The phases at the saddle points are 
divided out before taking real and
imaginary parts.  In (a) and (c), the dotted (dark turquoise) curve shows
a curve decreasing as $t^{-1/2}$.}
\label{OneDimMinkowskiNoExtremumPadeContourIntegrand}
\end{figure}
\FloatBarrier

The same approach works in the Minkowski region as well; here of course,
the two half-contours will no longer be complex conjugates.  The
[3/2] Pad\'e contours for $s=1+i\delta$ and $s=5+i\delta$ are shown in
\fig{OneDimMinkowskiNoExtremumContourFigure}, 
while the behavior of the integrands along these contours
are shown in \fig{OneDimMinkowskiNoExtremumPadeContourIntegrand}.  
They are contrasted with the behavior along the 
`textbook' \MB{} contour $\Re z=\negonehalf$.  
The integrand along the latter contour is again not absolutely convergent,
and hence not numerically stable.  A different contour
is again required
for a convergent numerical integration, and the Pad\'e contour provides
an efficient one.  (Note that $I_7$ has an imaginary part already starting
at $s=0$, and not just at $s=4$.)

\section{Evaluating Integrals}
\label{EvaluationSection}

In the previous sections, we have described simple approximations
to the exact contours of stationary phase in both the Euclidean
and Minkowski regions.  We turn now to a brief discussion of how to
evaluate the integrals along these contours, postponing a more
complete investigation to future work.

The simplest approach to evaluating the integral using any
of the contours is with an adaptive numerical routine,
such as {\tt gsl\_integration\_qagiu\/} and
{\tt gsl\_integration\_qagil\/}
from the GNU Scientific Library (GSL)~\cite{GNUScientificLibrary}.

\def\finfty{f_\infty}
\def\sech{\mathop{\textrm{sech}}\nolimits}
\def\atanh{\mathop{\textrm{atanh}}\nolimits}
Another approach to evaluating the integral along the [3/2] Pad\'e
contour takes advantage of the exponential decay in the
integrand from the saddle point out to $t\rightarrow \pm\infty$,
and uses Gaussian quadrature with a finite number of evaluation points based on
orthogonal polynomials for an appropriate weight function.
At small $t$, the integrand behaves like $e^{-c t^2}$, while at
large $t$, it behaves like $e^{-c' t}$, so the classical weight functions
are not optimal for us.  Instead, motivated by the observation that,
\begin{equation}
\Gamma(-z)\Gamma(1+z)= -\frac{\pi}{\sin\pi z}\,,
\end{equation}
so that,
\begin{equation}
\Gamma\biggl(\frac12-iy\biggr)\Gamma\biggl(\frac12+iy\biggr) = \pi\sech \pi y\,,
\end{equation}
we take $\sech u$ as our weight function.  It has the required
behavior at small and large $u$.  

Because the coefficients $c$ and $c'$ governing small-$t$ and large-$t$ behavior
are not the same, in principle we ought to
interpolate between different linear
arguments for small and large $t$.  We could do this, for example,
via,
\begin{equation}
f_i(t) = \finfty t + \biggl({\sqrt{2f_2}}-{\finfty}\biggr) 
                 \frac{t}{1+b_3 t}\,,
\end{equation}
where $\finfty$ and $f_2$ will be extracted from our integrand below,
and $b_3$ is an additional parameter.
(There is no need to have an analytic form for the inverse function,
so other forms could be used.)  

The required 
coefficients $\finfty$ and $f_2$ describe the large- and small-$t$ behavior
of the integrand $F(z(t))$, respectively:
\begin{equation}
\begin{aligned}
f_2 &= -\frac{F''(z_s)}{2F(z_s)}\,,
\\
\finfty &= \bigg|\ln\bigg|\frac{s_0}{-s}\bigg|\,\sin\thinf 
                         +(N_{-}+\sign\Im s)\pi\,\cos\thinf \bigg|\,.
\end{aligned}
\end{equation}
(Recall that with our conventions, $z(0)$ is the saddle point $z_s$.)
However, it turns out (surprisingly) that the integration is ultimately
{\it much\/} more efficient if we do not interpolate, but rather take,
\begin{equation}
\sech \finfty t/4\,,
\end{equation}
as our weight function.

We can next perform two changes of
variables: first, the change of variables $u=\atanh v$ takes us from
integrating over $[0,\infty)$ to integrating over the interval $[0,1]$.
The new weight function (including the Jacobian) is $(1-v^2)^{-1/2}$,
which would suggest the use of Gauss-Chebyshev quadrature, were it not
for the region of integration failing to match the required $[-1,1]$.
  Instead, we can make another change of
variables, $v=\cos\omega$, to arrive at the integral,
\begin{equation}
I_n=\frac4{\finfty}\int_0^{\pi/2} d\omega\;F((4\atanh\cos\omega)/\finfty)
z_p'(4(\atanh\cos\omega)/\finfty)\,,
\end{equation}
in the upper half-plane (and a similar integral with $\cos\omega$ replaced
by $-\cos\omega$ for the lower half-plane) for the original
Mellin--Barnes integrand $F(z)$.  The integral $I_n$ can be computed
efficiently via Gauss-Legendre quadrature, with an $n$-point
evaluation at the roots of the $n^{\rm th}$ Legendre polynomial $P_n(x)$, and
the weight for the $j{}^{\rm th}$ root given by~\cite{GaussLegendreQuadrature},
\begin{equation}
\frac{2}{(1-x_j^2)[P_n'(x_j)]^2}\,.
\end{equation}
Other techniques, such as recursive subdivision, may also be appropriate, but
we have not explored them.

\begin{table}
	\renewcommand{\arraystretch}{0.85} 
	\begin{tabular}{||l|l|c|c|c|c||}
		\hline &
		& \multicolumn{4}{|c||}{Integration Contour and Method}\\
		\cline{3-6}
		\hfil Integral &\hfil Value & 
                      \MB{} & Tangent & Pad\'e & Pad\'e \\
		& & GSL & GSL & GSL & Gauss-Legendre\\
		\hline
        $I_1(\stdparam)$ & $0.04958745585$ & 195 & 195 & 75 & 16\\
        \hline
        $I_1(-20)$ & $5.639661654$ & 135 & 135 & 105 & 19\\
        \hline
        $I_1(1+i\delta)$ & $ -1.2091995762$ & --- & 930 & 1080 & 80\\
        \hline
        $I_1(5+i\delta)$ & $4.30408941-14.04962946 i$ & --- & --- & 660 & 160\\
	\hline
	$I_7(\stdparam)$ & $-1.954168464$ & 225 & 165 & 195 & 26\\
	\hline
	$I_7(-20)$ & $-35.72650854$ & 135 & 165 & 165 & 19\\
	\hline
	$I_7(1+i\delta)$ & $2.831441537+17.99925456 i$ 
                 & --- & 1140 & 780 & 159\\
	\hline
	$I_7(5+i\delta)$ & $-27.40504335+37.26381174 i$ 
                 & --- & --- & 720 & 200\\
        \hline
	\end{tabular}
\caption{Number of evaluations required to obtain a relative error of $10^{-8}$
	for a variety of integrals, using several integration methods.  A missing
		entry indicates that the integral is not convergent numerically using the given method.}		
	\label{EvaluationPointsTable}
\end{table}
\FloatBarrier

\Tab{EvaluationPointsTable} gives examples of evaluating 
the Mellin--Barnes integrals $I_1$ (\ref{OneDimensionalExample}) 
and $I_7$ (\ref{NoExtremumIntegral}) using
both the contour chosen by \MB{}, as well as the tangent and Pad\'e contours,
all using the GSL routines mentioned earlier, as well as an evaluation using
the Pad\'e contour and the Gauss-Legendre approach.  In the Euclidean
region, the Pad\'e contour provides
a more efficient evaluation, especially within the Gauss-Legendre approach.
In the Minkowski region, it is again more efficient where other contours can
be used, and provides a reliable means of evaluating the integrals even when
linear contours fail to provide a numerically convergent result.

\section{Conclusions}
\label{Conclusions}

In this paper,
we have re-examined the numerical evaluation of Mellin--Barnes integrals.  
Contours chosen by the \MB{} or \MBresolve{} packages are not always
suitable for numerical evaluation.  Using contours of stationary
phase, or approximations thereto, resolves problems that arise
in numerical evaluation.  We discussed the computation of exact contours
of stationary descent for one-dimensional
integrals, as well as several approximations which 
are likely of greater practical importance.   The [3/2] Pad\'e
approximations~(\ref{PadeContour},\ref{PadeContourM},\ref{ExtremumFreePade}) 
to contours of stationary phase are likely to be
the most robust and widely useful of these approximations.
A remapping and Gauss-Legendre quadrature appears to be an efficient
means of evaluating integrals using the Pad\'e contour.  
We hope to extend these ideas to the more practically important
multidimensional case in future work.

\section*{Acknowledgments}

We thank Valery Yundin for collaboration in early stages of this research.
JG and TJ also thank Tord Riemann, Ievgen Dubovyk, and Johann Usovitsch
for discussions.
  JG's
research is supported by 
the Polish National Science Center under 
Grant Agreement no.~DEC--2013/11/B/ST2/04023.
DAK thanks the Ambrose Monell Foundation for its support of his stay
at the Institute for Advanced Study.


\end{document}